\documentclass[usenatbib]{mn2e}

\usepackage{graphicx}
\usepackage{url}
\usepackage[T1]{fontenc} 
\usepackage{aecompl}
\title[ISM in NGC\,1569 with {\it Herschel\/}]{Probing the interstellar medium of NGC\,1569 with {\it Herschel\/}\thanks{{\it Herschel} is an ESA space observatory with science instruments provided by European-led Principal Investigator consortia and with important participation from NASA.}}
\author[S. Lianou, P. Barmby, A. R{\'e}my-Ruyer, et al.]{S. Lianou$^{1}$\thanks{E-mail:
slianou@uwo.ca (SL); pbarmby@uwo.ca (PB)}, 
P. Barmby$^{1}$, 
A. R{\'e}my-Ruyer$^{2,3}$, 
S. C. Madden$^{2}$\thanks{E-mail:
suzanne.madden@cea.fr (SM); aurelie.remy@cea.fr (ARR); frederic.galliano@cea.fr (FG); vianney.lebouteiller@cea.fr (VL)}, 
F. Galliano$^{2}$ and 
\newauthor % starts a new line in the
V. Lebouteiller$^{2}$\\
$^{1}$Department of Physics \& Astronomy, University of Western Ontario, London, ON N6A 3K7, Canada\\
$^{2}$Laboratoire AIM, CEA/IRFU/Service d'Astrophysique, Universit\'{e} Paris Diderot, Bat. 709, 91191 Gif-sur-Yvette, France\\
$^{3}$Institut d'Astrophysique Spatiale, CNRS, UMR8617, 91405, Orsay, France}
\begin{document}

\date{Accepted 2014 August 27. Received 2014 August 27; in original form 2014 May 29}

\pagerange{\pageref{firstpage}--\pageref{lastpage}} \pubyear{2014}

\maketitle

\label{firstpage}

\begin{abstract}
NGC\,1569 has some of the most vigorous star formation among nearby galaxies. It hosts two super star clusters (SSCs) and has a higher star formation rate (SFR) per unit area than other starburst dwarf galaxies. Extended emission beyond the galaxy's optical body is observed in warm and hot ionised and atomic hydrogen gas; a cavity surrounds the SSCs. 
We aim to understand the impact of the massive star formation on the surrounding interstellar medium in NGC\,1569 through a study of its stellar and dust properties.
We use {\it Herschel} and ancillary multiwavelength observations, from the ultraviolet to the submillimeter regime, to construct its spectral energy distribution, which we model with {\it magphys} on $\sim$300\,pc scales at the SPIRE\,250\,$\mu$m resolution. 
The multiwavelength morphology shows low levels of dust emission in the cavity, and a concentration of several dust knots in its periphery. The extended emission seen in the ionised and neutral hydrogen observations is also present in the far--infrared emission. The dust mass is higher in the periphery of the cavity, driven by ongoing star formation and dust emission knots. The SFR is highest in the central region, while the specific SFR is more sensitive to the ongoing star formation. The region encompassing the cavity and SSCs contains only 12 per cent of the dust mass of the central starburst, in accord with other tracers of the interstellar medium. The gas--to--dust mass ratio is lower in the cavity and fluctuates to higher values in its periphery.
\end{abstract}


\section{Introduction}
\label{sec:introduction}

   The most extreme form of massive star formation occurring in galaxies is in the form of super star clusters (SSCs). These are very luminous, with L$_{V}$ between 10$^{6}$--10$^{8}$\,L$_{\odot}$, massive, with stellar masses higher than 10$^5$M$_{\odot}$, and very compact, with effective radii less than 5\,pc \citep[][and references therein]{Billett02,O'Connell04}. The star formation conditions during gravitational interactions favour the formation of SSCs \citep{Larsen10,Billett02}. SSCs are important as they have been considered as the younger analogues to the old Galactic globular clusters (GCs) and show that the formation of very massive stellar clusters is continuous as a function of time \citep[][]{Larsen10,deGrijs05,O'Connell04}. 

   The most prominent example of a nearby dwarf galaxy that contains one SSC is the Large Magellanic Cloud (LMC; with R\,136 in the 30\,Doradus \mbox{H\,{\sc ii}} region; \citealt{Massey98}). The next nearest extragalactic resolved SSCs in a dwarf galaxy are hosted in NGC\,1569 \citep[e.g.,][]{O'Connell94,Grocholski12} and NGC\,4214 \citep[e.g.,][]{Leitherer96,Drozdovsky02,Dopita10}, at similar distances ($\sim$3\,Mpc). Our focus in the current work is placed on NGC\,1569, a starburst low--metallicity dwarf irregular galaxy, with an oxygen abundance of 8.02$\pm$0.02 \citep[in the \citealt{Pilyugin05} scale adapted from \citealt{Kobulnicky97};][]{Madden13}, and a stellar metallicity of Z\,=\,0.1--0.2\,Z$_{\odot}$ \citep{Grocholski08,Aloisi01}. At a distance of 2.96$\pm$0.22\,Mpc \citep{Grocholski12}, it is a member of the IC\,342 group of galaxies, and the intense starburst may have been triggered by the gravitational interactions with other galaxy group members or with a nearby low-mass \mbox{H\,{\sc i}} companion \citep{Johnson13,Hunter12,Holwerda13,Jackson11,Stil98}.

   NGC\,1569 hosts two SSCs, one of which consists of two components \citep{deMarchi97,O'Connell94,Hunter00}. The star formation in the central starburst is concentrated in three distinct events within the last 1--2\,Gyr, with the most recent between 8--27\,Myr ago \citep{Angeretti05}. In the outer parts of the galaxy, the star formation peaked around 500\,Myr ago along with subsequent recent bursts of lower intensity and duration \citep[]{Grocholski12}. SSC\,A and SSC\,B formed during the last bursts of star formation, having ages of the order of 7\,Myr and 10-20\,Myr \citep{Hunter00,Larsen08}. NGC\,1569 contains, apart from the two SSCs, a large number of young ($\sim$30\,Myr), compact and massive star clusters \citep{Hunter00,Anders04}, similar to the ones found in the LMC \citep[e.g.,][]{Hunter03}. Even though NGC\,1569 and LMC are galaxies that are gravitationally interacting \citep{Besla12,Putman98,Johnson13,Stil98}, the formation of compact star clusters in galaxies is not unique to interacting\,/\,merging galaxies \citep{Larsen10}. 

   In NGC\,1569, the winds of the SSCs have profoundly impacted the interstellar medium (ISM) around them leading to a complex morphology \citep[]{Westmoquette07a}, including a cavity evident in \mbox{H\,{\sc i}} and H\,$\alpha$ \citep{Israel90,Hunter00}. Studies of the hot and warm ionised gas reveal filaments of ionised gas, expanding shells, superbubble kinematics and galactic scale outflows \citep[][and references therein]{Heckman95,Martin02,Westmoquette08}. Moreover, \citet{Waller91} detects \mbox{H\,{\sc ii}} regions in the periphery of the cavity, especially in the East and West of SSC\,A, while \citet{Taylor99} detect giant molecular clouds in the West and North--West direction of the SSC\,A. With a diameter of $\sim$200\,pc, the \mbox{H\,{\sc i}}\,/\,H\,$\alpha$ cavity is a strong manifestation of the impact of the massive star formation on the ISM. This cavity is reminiscent of the giant \mbox{H\,{\sc ii}} region NGC\,604 in M\,33, which also presents a complex ISM morphology powered by a cluster of massive stars in its centre \citep[][and references therein]{MartinezGalarza12}.  

   Several feedback mechanisms have been proposed to explain the discrepancy between observed and theoretically predicted numbers of low-mass galaxies \citep[][and references therein]{Benson03,Sawala10}. Attributed to feedback processes is also the mixing of metals with the metal--poor ISM, while galactic-scale outflows further drive metals to the intergalactic medium to enrich it \citep[][and references therein]{Recchi14,Kawata14}. The intriguing observation of NGC1569's SSCs clearing out the ISM and forming the cavity seen in \mbox{H\,{\sc i}} and H\,$\alpha$ leaves us with the following questions: how are the dust properties distributed across the central starburst, and especially in the cavity region? how do the dust properties relate to the ongoing and embedded star formation? how does the dust emission relate to the other ISM tracers of the warm ionised and atomic hydrogen gas? Dust is an important component of a galaxy's ISM, and a significant amount of a galaxy's bolometric luminosity is emitted in the infrared, through reprocessing by dust of the radiation emitted by massive stars. The sensitivity and spatial resolution offered with {\it Herschel}, together with the new wavelength coverage in the submillimeter (submm), allow a new study to investigate these questions, in much greater detail than before. Our study primarily focuses on understanding the stellar and dust properties in the cavity and the interplay between them, as well as across the central starburst.

   \citet{Galliano03} and \citet{Lisenfeld02} focus on the global ISM properties of NGC\,1569, consistently modelling the mid-infrared (MIR) to millimetre observations of ISO, IRAS, SCUBA/JCMT and MAMBO/IRAM multiwavelength dust spectral energy distribution (SED), finding an emission in excess compared to the model predictions at submm wavelengths, often referred to as submm excess. \citet[][]{Remy13} and R\'{e}my--Ruyer et al.\,(in prep.) use {\it Herschel} observations along with both modified black--body fits to the SED and detailed SED modelling to study the global dust properties, as well as the gas--to--dust mass ratio \citep{Remy14}, for a sample of low--metallicity dwarf galaxies \citep{Madden13}, including NGC\,1569. 

%% TABLE 1 - Basic properties %%%%%%%%%%%%%%%%%%%%%%%%%%%%% 
\begin{table}
      \begin{minipage}[t]{\columnwidth}
      \caption[]{Properties of NGC\,1569.}
      \label{sl_table1} 
      \renewcommand{\footnoterule}{}
      \begin{tabular}{lc}
\hline\hline
  Quantity                            & Value (ref.)\footnote{{\bf References.--} (1) \citet{Burstein82}; (2) \citet{Grocholski12}; (3) \citet{Hunter06}; (4) \citet{Hunter12}; (5) \citet{Johnson12}; (6) \citet{Hunter10}; (7) \citet{Greve96}; (8) \citet{Remy13}; (9) \citet{Madden13}.} \\
\hline
  Type                           & IBm                    \\
  RA~(J2000.0)                   & $04^h30^m49.0^s$        \\
  Dec~(J2000.0)                  & $+64^\circ 50' 53.0''$   \\
  $\rm E(B-V)$~(mag)             & 0.51\,(1)               \\
  $(m-M)_0$~(mag)                & 27.52$\pm$0.14\,(2)     \\
  Distance~(Mpc)                 & 2.96$\pm$0.22\,(2)      \\
  D$_{25}$($\arcsec$)             & 216\,(3)                \\
  M$_{\rm V}$~(mag)                &$-18.2$\,(4)                 \\
  M$_{\star}$~(M$_{\sun}$)          &2.8$\times$10$^{8}$\,(5)  \\
  M$_{\mbox{H\,{\sc i}}}$~(M$_{\sun}$) &2.5$\times$10$^{8}$\,(4)  \\
  M$_{H\alpha}$~(M$_{\sun}$)         &1.3$\times$10$^5$\,(6)    \\
  M$_{H_{2}}$~(M$_{\sun}$)           &2$\times$10$^6$\,(7)       \\
  M$_{D}$~(M$_{\sun}$)              &2.8$^{+1.5}_{-1.4}\times$10$^5$\,(8)  \\
  12+log(O\,/\,H)                 &8.02$\pm$0.02\,(9)          \\
\hline
\end{tabular} 
\end{minipage}
\end{table}
%%%%%%%%%%%%%%%%%%%%%%%%%%%%%%%%%%%%%%%%%%%%%%%%%%%%%%%%%%%%%%%%%%
%
   In the present study, we consider the star and dust properties on a pixel--by--pixel basis and in two regions via modelling the observed 0.15\,$\mu$m to 500\,$\mu$m SED using {\it magphys} \citep{dacunha08}, and we compare our findings with other ISM tracers, such as the warm ionised and atomic hydrogen gas and the CO--traced molecular gas. A detailed examination of the ISM properties in the central region of the starburst of NGC\,1569 will provide us with valuable insights to lend interpretation to the properties of more distant starburst galaxies. We assume a distance modulus of 27.52$\pm$0.14\,mag or a distance of 2.96$\pm$0.22\,Mpc \citep[based on the tip of the red giant branch and reddening consistent with the \citet{Burstein82} value;][]{Grocholski12}. Table~\ref{sl_table1} lists the basic properties of NGC\,1569. With the SED modelling on a pixel--by--pixel basis, the smallest spatial scale we are probing is 21$\arcsec$ or 294\,pc. The results of the SED modelling on the dust and star properties are model--dependent and thus tied to the scale defined by {\it magphys}. The flux densities of the observations considered here are in the Vega photometric system, except from the {\it GALEX} data which are in the AB photometric system.
%__________________________________________________________________

\section{Observations \& analysis}
\label{sec:observations}

%% TABLE 2 -- Multi--data \& Calibration errors %%%%%%%%%%%%%%%%%%%
\begin{table}
      \begin{minipage}[t]{\columnwidth}
      \caption[]{Wavelength coverage and adopted calibration errors.}
      \label{sl_table2} 
      \renewcommand{\footnoterule}{}
      \begin{tabular}{llclclc}
\hline\hline
Telescope\,/\,Filter\footnote{Filter here is defined to mean either the instrument used, or the camera used, or the filter used, depending on the telescope facility.}       
                           &$\lambda$ ($\mu$m)     &FWHM            &Ref.\footnote{{\bf References for literature data.--} (1) \citet{GilDePaz07}; (2) \citet{Hunter06}; (3) \citet{Skrutskie06}; (4) \citet{Wright10}; (5) \citet{Fazio04}; (6) \citet{Bendo12}; (7) \citet{Remy13}.}
                                                                            &Error     &Ref.\footnote{{\bf References for calibration errors.--} (1) \citet{Morrissey07}; (2) \citet{Hunter06}; (3) \citet{Skrutskie06}; (4) \citet{Jarrett11}; (5) \citet{Aniano12}; (6) \citet{Bendo12}; (7) \citet{Remy13}.}                                                                                    \\
\hline                                  
GALEX\,/\,FUV              &0.15                   &4.2$\arcsec$    &1      &5\%        &1           \\
GALEX\,/\,NUV              &0.23                   &5.3$\arcsec$    &1      &3\%        &1           \\
KPNO\,/\,B                 &0.44                   &18.2$\arcsec$   &2      &3\%        &2           \\
KPNO\,/\,V                 &0.54                   &18.2$\arcsec$   &2      &3\%        &2           \\
2MASS\,/\,J                &1.25                   &2.5$\arcsec$    &3      &3\%        &3           \\
2MASS\,/\,H                &1.65                   &2.5$\arcsec$    &3      &3\%        &3           \\
2MASS\,/\,K$_{S}$          &2.17                   &2.5$\arcsec$    &3      &3\%        &3           \\
{\it WISE}\,/\,W1          &3.4                    &8.4$\arcsec$    &4      &2.4\%      &4            \\
{\it WISE}\,/\,W2          &4.6                    &9.2$\arcsec$    &4      &2.8\%      &4            \\
{\it WISE}\,/\,W3          &12                     &11.4$\arcsec$   &4      &4.5\%      &4            \\
{\it WISE}\,/\,W4          &22                     &18.6$\arcsec$   &4      &5.7\%      &4            \\
{\it Spitzer}\,/\,IRAC     &3.6                    &1.7$\arcsec$    &5      &8.3\%      &5            \\
{\it Spitzer}\,/\,IRAC     &4.5                    &1.7$\arcsec$    &5      &7.1\%      &5            \\
{\it Spitzer}\,/\,IRAC     &5.8                    &1.9$\arcsec$    &5      &22.1\%     &5            \\
{\it Spitzer}\,/\,IRAC     &8.0                    &2.0$\arcsec$    &5      &16.7\%     &5            \\
{\it Spitzer}\,/\,MIPS     &24                     &6.0$\arcsec$    &6      &4\%        &6            \\
{\it Herschel}\,/\,PACS    &70                     &5.8$\arcsec$    &7      &5\%        &7            \\
{\it Herschel}\,/\,PACS    &100                    &7.1$\arcsec$    &7      &5\%        &7            \\
{\it Herschel}\,/\,PACS    &160                    &11.2$\arcsec$   &7      &5\%        &7            \\
{\it Herschel}\,/\,SPIRE   &250                    &18.2$\arcsec$   &7      &7\%        &7            \\
{\it Herschel}\,/\,SPIRE   &350                    &25.0$\arcsec$   &7      &7\%        &7            \\
{\it Herschel}\,/\,SPIRE   &500                    &36.4$\arcsec$   &7      &7\%        &7            \\
\hline
\end{tabular} 
\end{minipage}
\end{table}
%%%%%%%%%%%%%%%%%%%%%%%%%%%%%%%%%%%%%%%%%%%%%%%%%%%%%%%%%%%%%%%%%%
%
   While this study is motivated by {\it Herschel} observations, we use numerous archival data sets to cover the ultraviolet (UV) to the infrared (IR) and submm wavelength regime. We use {\it Herschel} PACS \citep{Poglitsch10} and SPIRE \citep{Griffin10} imaging observations to probe the far--IR (FIR) and submm part of the spectrum, from 70\,$\mu$m to 500\,$\mu$m; these data are part of the Dwarf Galaxy Survey (PI S. Madden; \citealt{Madden13}), and we refer to \citet{Remy13} for more information on the data reduction techniques. The ancillary observations that we use for our study are taken from publicly available archives. Table~\ref{sl_table2} lists the available multiwavelength data set coverage.

   The far-UV (FUV; 1350--1750 \AA) and near-UV (NUV; 1750--2750 \AA) images come from GALEX \citep{Martin05}, as part of the GALEX Nearby Galaxies Survey \citep{GilDePaz07}, and are retrieved through the archive\footnote{\url{http://galex.stsci.edu/GR6/}} as background subtracted intensity maps. The optical part of the data consist of broadband B and V data adopted from \citet{Hunter06}\footnote{\url{http://www2.lowell.edu/users/dah/littlethings/n1569.html}}. The near--IR part of the spectrum is covered with data from the 2MASS \citep{Skrutskie06} archive in the J, H, Ks filters. For the MIR and FIR wavelength regime, we use {\it WISE} data \citep[in the W1, W2, W3, W4 filters;][]{Wright10}, as well as {\it Spitzer} data \citep{Werner04}, both IRAC \citep[in the 3.6, 4.5, 5.8 and 8.0\,$\mu$m bands;][]{Fazio04} and MIPS \citep[only 24\,$\mu$m;][]{Rieke04}\footnote{As the MIPS\,70\,$\mu$m and MIPS\,160\,$\mu$m images cover only a small part of the galaxy, we make use of only the MIPS\,24\,$\mu$m image. The field of view covered by the MIPS\,24\,$\mu$m image places a limit on the field of view used from the remaining imaging data set.}. The {\it WISE} and IRAC data sets are retrieved through the NASA\,/\,IPAC Infrared Science Archive (IRSA)\footnote{\url{http://irsa.ipac.caltech.edu/index.html}}, while the MIPS data are from \citet[][]{Bendo12}. 

   All images are treated initially in their native pixel scales, where the unit conversion of the flux density takes place from the native unit to Jy\,/\,pixel. Subsequently, we use SExtractor \citep{Bertin96} in each image to estimate the background flux density and subtract it. We use the segmentation image created by SExtractor, in order to mask the images from the detected sources. The segmentation image is the input image with all detected point and extended sources assigned to positive pixel values and the remaining background structure set to zero pixel values. We then estimate the background flux density from the masked image. The segmentation image is also used to mask the foreground stellar contamination. The GALEX, 2MASS and MIPS images are already background--subtracted. 

   We then correct the UV--to--MIR images for Galactic foreground extinction: 
\begin{equation}
F_{\rm i} = F_{\rm obs} 10^ {0.4 E(B-V) [{\rm \kappa}({\rm \lambda} - V) + R_{V}]},
\end{equation}
where {\it F$_{\rm i}$} is the flux density corrected for foreground extinction, {\it F$_{\rm obs}$} is the observed flux density, {\it E(B$-$V)} is the colour excess, {\it R$_{V}$}\,=\,3.1, and {\it ${\rm \kappa}({\rm \lambda}$-V)} is the extinction curve. In order to estimate the {\it ${\rm \kappa}({\rm \lambda}$-V)} values at the optical bands, B and V, we use the Galactic extinction curve of \citet{Fitzpatrick99}, while for the near--IR to MIR bands (2MASS bands; {\it WISE} W1, W2, \& W3 bands; IRAC 3.6, 4.5, 5.8, \& 8.0\,$\mu$m bands) we use the one from \citet{Indebetouw05}\footnote{\url{http://svo2.cab.inta-csic.es/theory/fps/index.php}}. The GALEX data are corrected for the Galactic foreground extinction using the selective extinction {\it A$_{FUV}$}\,/\,{\it E(B$-$V)}\,=\,8.29 and {\it A$_{NUV}$}\,/\,{\it E(B$-$V)}\,=\,8.18 \citep{Seibert05}. We adopt for NGC\,1569 a Galactic foreground reddening {\it E(B$-$V)}\,=\,0.51\,mag \citep{Burstein82}, which gives a Galactic foreground extinction {\it A$_{V}$}\,=\,1.58\,mag. The motivation for this choice of reddening is from past studies on the internal extinction across the galaxy, which found the total reddening, i.e. Galactic foreground plus internal, to vary between 0.56 to 0.80 mag \citep[][]{Pasquali11,Relano06,Devost97,Kobulnicky97,Calzetti96}. We do not perform internal extinction corrections to the data set for the SED modelling procedure, while in the occasion where these are performed then this is explicitly noted each time.  

%% TABLE 3 -- Colour Corrections %%%%%%%%%%%%%%%%%%%%%%%%%%%%%%
\begin{table}
      \caption[]{Applied colour corrections.}
      \label{sl_table3} 
      \begin{tabular}{lcc}
\hline\hline
  Camera                &Power law index    &Colour correction   \\
\hline
  W1\,3.4\,$\mu$m       &1                  &0.9961              \\
  W2\,4.6\,$\mu$m       &1                  &0.9976              \\
  W3\,12\,$\mu$m        &$-$2               &1                   \\
  W4\,22\,$\mu$m        &$-$2               &1                   \\
  IRAC\,3.6\,$\mu$m     &1                  &1.0037              \\
  IRAC\,4.5\,$\mu$m     &1                  &1.0040              \\
  IRAC\,5.8\,$\mu$m     &$-$2               &1.0052              \\
  IRAC\,8\,$\mu$m       &$-$2               &1.0111              \\
  MIPS\,24\,$\mu$m      &$-$2               &0.960               \\
  PACS\,70\,$\mu$m      &$-$2               &1.016               \\
  PACS\,100\,$\mu$m     &2                  &1.034               \\
  PACS\,160\,$\mu$m     &2                  &1.075               \\
  SPIRE\,250\,$\mu$m    &2                  &0.9878              \\
  SPIRE\,350\,$\mu$m    &2                  &0.9890              \\
  SPIRE\,250\,$\mu$m    &2                  &1.0128              \\
\hline
\end{tabular} 
\end{table}
%%%%%%%%%%%%%%%%%%%%%%%%%%%%%%%%%%%%%%%%%%%%%%%%%%%%%%%%%%%%%%%%%%
%
   We additionally perform colour corrections to the {\it WISE}, {\it Spitzer}, and {\it Herschel} images. Colour corrections are necessary in the case of wide filters where the spectral shapes of the sources used in the calibration process are different than the spectral shape of the targets under study. We assume power law indexes and colour corrections, as listed in Table~\ref{sl_table3}. The power law indexes were determined using the galaxy's SED (in an aperture similar as the one used in Sec.~\ref{sec:regions} for the central starburst region), and solving the equation 
F$_{\nu}\,\propto\,\nu ^{\alpha}$
for $\alpha$, where F$_{\nu}$ is the specific flux density, $\nu$ is the frequency, and $\alpha$ is the power law index. The power law index was determined: for W1, W2, IRAC\,3.6\,$\mu$m, and IRAC\,4.5\,$\mu$m solving for $\alpha$ the above equation at the wavelengths in the 2MASS K$_{S}$ band and in the W2 band; for IRAC\,5.8\,$\mu$m, IRAC\,8\,$\mu$m, W3, W4, MIPS\,24\,$\mu$m, and PACS\,70\,$\mu$m solving for $\alpha$ at the wavelengths in the W2 and PACS\,70\,$\mu$m; for the remaining bands, solving for $\alpha$ at the wavelengths in the PACS\,100\,$\mu$m and SPIRE\,500\,$\mu$m. The colour corrections listed in Table~\ref{sl_table3} are taken from \citet{Wright10} for {\it WISE}, from the IRAC Instrument Handbook (version 2.0.3; 2013) for IRAC\footnote{\url{http://irsa.ipac.caltech.edu/data/SPITZER/docs/irac/iracinstrumenthandbook/18/}}, from the MIPS Instrument Handbook (version 3.0; 2011) for MIPS\footnote{\url{http://irsa.ipac.caltech.edu/data/SPITZER/docs/mips/mipsinstrumenthandbook/51/}}, from \citet{Poglitsch10} for PACS, and from the SPIRE Data Reduction Guide (version 2.3; 2013) for SPIRE\footnote{\url{http://herschel.esac.esa.int/hcss-doc-11.0/load/spire_drg/html/ch05s07.html}}. The colour correction of the SPIRE\,data additionally includes beam effect corrections which also depend on the spectral shape of the source of the emission. Finally, we apply extended source corrections to the IRAC bands, as recommended in the IRAC Instrument Handbook (version 2.0.3; 2013), using the surface brightness corrections listed in Table~4.9.

   After these corrections are applied, we register the -- background, galactic foreground extinction, and colour--corrected -- images to a common world coordinate system, centred on the galaxy and covering the region of 294$\arcsec$\,$\times$\,294$\arcsec$ ($\sim$\,4116\,pc $\times$\,4116\,pc, at our adopted distance). The final step of the image processing is the convolution of all images to a common point-spread function (PSF), using the convolution kernels of \citet{Aniano11}. For bands without a convolution kernel in the \citet{Aniano11} database (the 2MASS and B, V data sets), we use a Gaussian kernel with FWHM taken from \citet[][Column 7 of their Table 6]{Aniano11}. 

   In order to retain the best spatial resolution possible for the purposes of our study, as well as to fully exploit the whole {\it Herschel} imaging data set, we convolve the available imaging to two different PSFs. First, to the PSF of the SPIRE 250\,$\mu$m band, a FWHM of 18.2$\arcsec$, thus keeping all imaging up to this band for our analysis. Second, to the PSF of the SPIRE 500\,$\mu$m band, which corresponds to 36.4$\arcsec$. 

   Each set of the convolved images is, then, re-sampled to a 21$\arcsec$ pixel size for the set of images convolved to the SPIRE\,250\,$\mu$m PSF, and 42$\arcsec$ for the set of images convolved to the SPIRE\,500\,$\mu$m PSF. The re--sampled pixel sizes are larger than the common PSF, in order to derive statistically independent properties. In the former case, the pixel size probes the ISM properties in spatial scales equivalent to 294\,pc, which is larger than the size of the \mbox{H\,{\sc i}}\,/\,H\,$\alpha$ cavity (a diameter of 198\,pc at our adopted distance; \citealt{Hunter00}). The data set convolved to the SPIRE\,500\,$\mu$m PSF consists of a total of 22 images, exploiting the full {\it Herschel} imaging, while the data set of images convolved to the SPIRE 250\,$\mu$m PSF consists of a total of 20 images, i.e., dropping the two longer wavelength SPIRE bands. Several steps of the image processing have been performed using Imagecube \citep*[Taylor, Lianou, \& Barmby 2013;][]{Lianou13a}.

   The photometric uncertainties are derived directly on the convolved and re--sampled images, on a pixel--by--pixel basis, using the approach of \citet{Aniano12}. These authors model the SEDs of two spiral galaxies (NGC\,628 \& NGC\,6946) drawn from the KINGFISH sample \citep{Kennicutt11} using the dust model of \citet{Draine07} and deriving the dust properties both on global and local scales. Here, we assume background variations and calibration errors as the main source of uncertainties. The background variations have been estimated on the final images, after masking the galaxy with the SExtractor segmentation image. In this way, background pixels are defined, and eq.\,D2 of \citet{Aniano12} has been used to estimate the background noise on a pixel--by--pixel basis. The calibration errors adopted are listed in Table~\ref{sl_table2}. The two uncertainties of the background variation and calibration errors are then added in quadrature. 
%______________________________________________________________

\section{Multiwavelength emission morphology}
\label{sec:morphology}

%%%%% FIGURE 1 - SSCs with HST \& 100um %%%%%%%%%% 
 \begin{figure}
%  \centering
      \includegraphics[width=8cm]{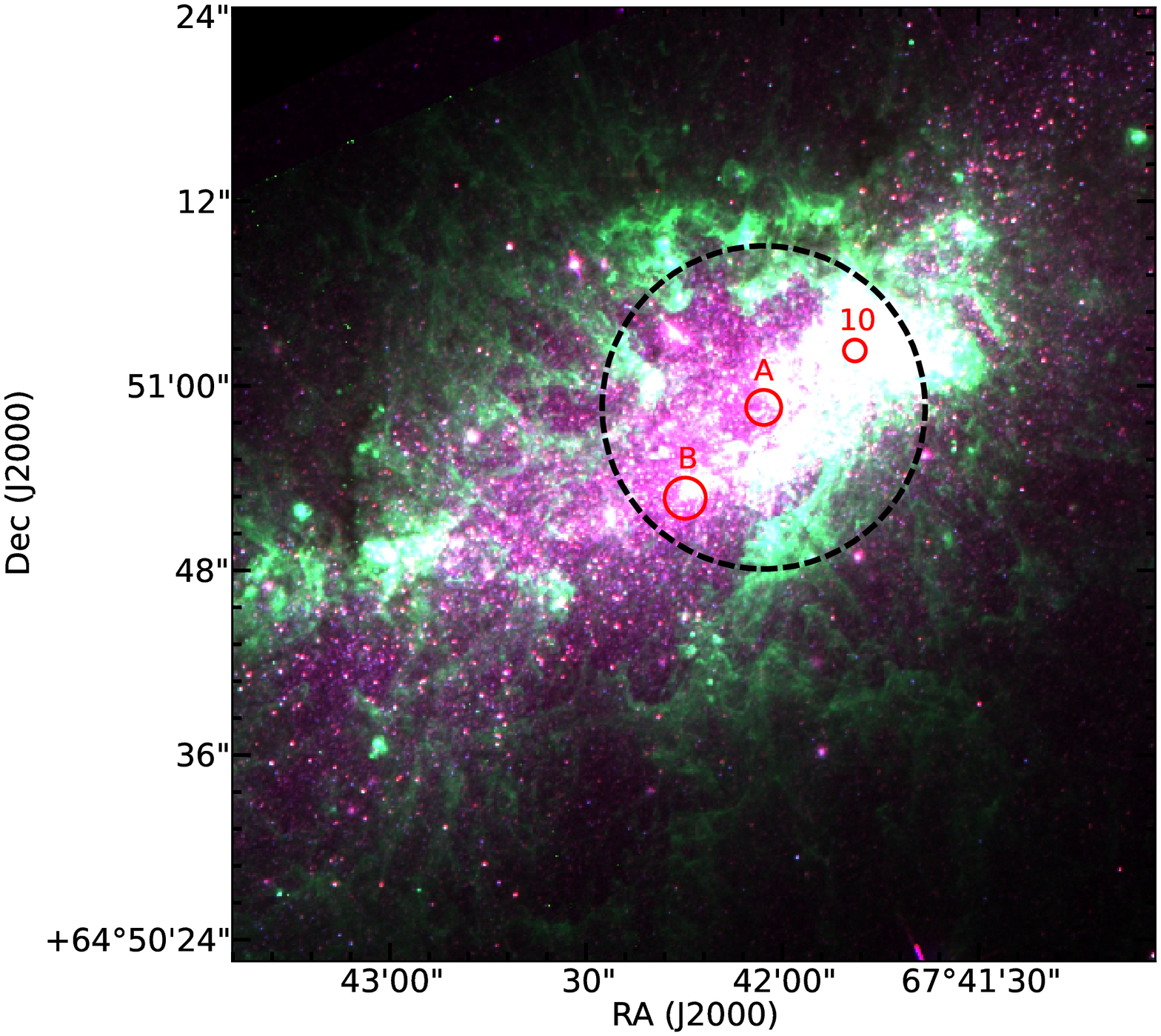}
      \includegraphics[width=8cm]{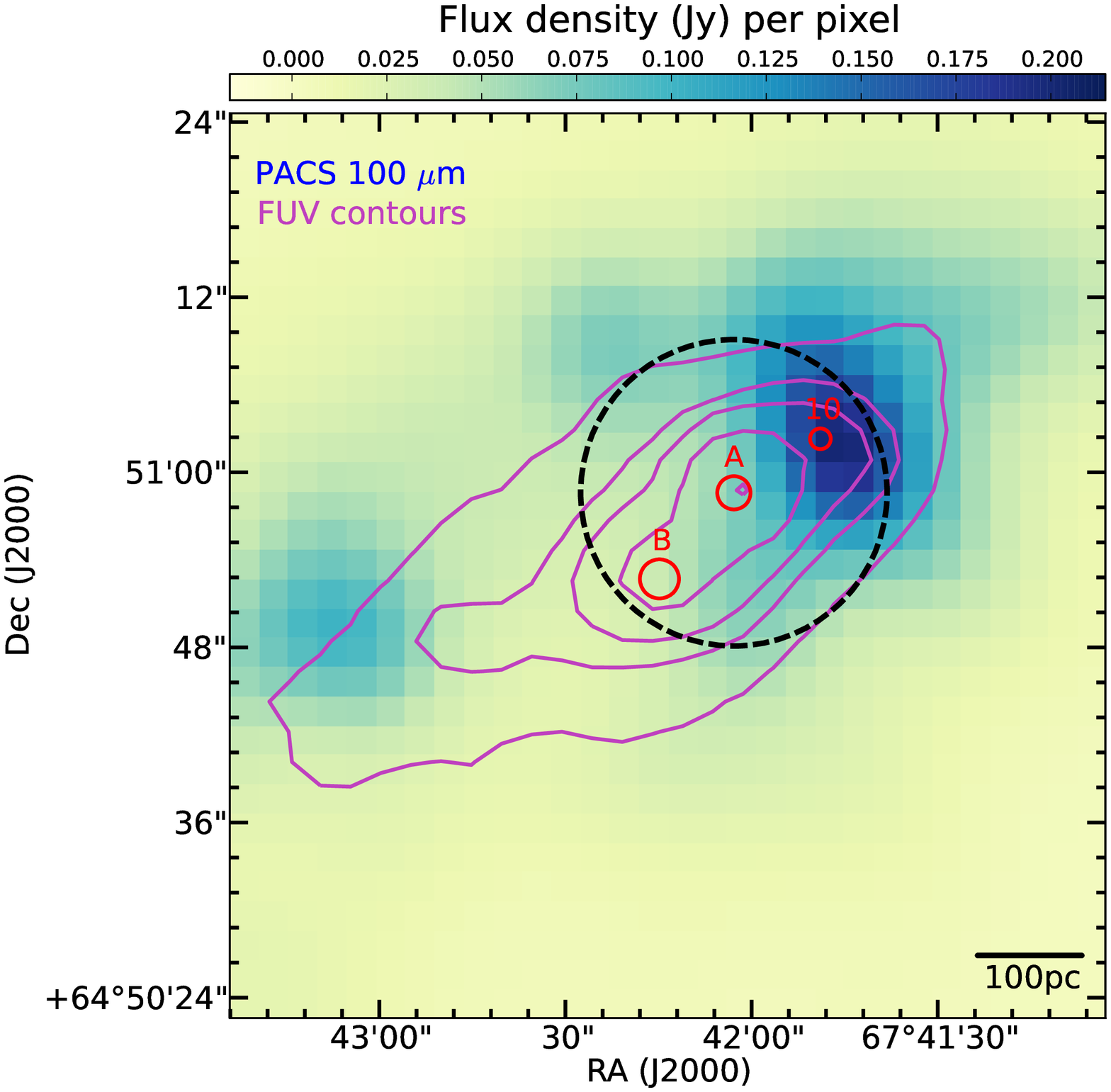}
  \caption{{\it Upper panel}: Three-colour image of NGC\,1569 showing the inner 60$\arcsec$\,$\times$\,60$\arcsec$ region, and composed of HST\,/\,ACS images in the following filters: F606W (blue), H\,$\alpha$ (green), and F814W (red). The two SSCs and SC\,10 are shown with the red solid circles, while the cavity region is shown with the black dashed circle (diameter of 21$\arcsec$ to match the one used in Sec.~7). The pixel scale of the three-colour image is 0.05$\arcsec$.  
{\it Lower panel}: PACS\,100\,$\mu$m image of NGC\,1569 shown in its native angular resolution (7.1$\arcsec$ FWHM) and covering the same region as in the upper panel. The contours are shown at the GALEX FUV native angular resolution (4.2$\arcsec$ FWHM), at the levels of 2.5\,mJy, 1\,mJy, 0.5\,mJy, 0.3\,mJy, and 0.1\,mJy, going from the inner to the outer contour, respectively. The pixel scale is 2$\arcsec$. At the adopted distance of NGC\,1569, 1$\arcsec$ corresponds to 14\,pc.}
  \label{sl_figure1}
 \end{figure}
%%%%%%%%%%%%%%%%%%%%%%%%%%%%%%%%%%%
%
   The upper panel of Fig.~\ref{sl_figure1} shows a three-colour image of NGC\,1569 based on HST images (GO\,10885; PI A.\,Aloisi; \citealt{Grocholski12}) and focusing on the central starbursting area of 840\,pc\,$\times$\,840\,pc (60$\arcsec$\,$\times$\,60$\arcsec$). SSC\,A, SSC\,B, and SC\,10 are shown with their true angular sizes, 1.14$\arcsec$, 1.34$\arcsec$, and 0.71$\arcsec$, respectively \citep{Hunter00}. The H\,$\alpha$ emission, which traces the ongoing star formation in timescales of about 10\,Myr \citep{Lee11,Kennicutt12}, reveals the cavity \citep{Hunter00,Pasquali11} that has been initially observed in \mbox{H\,{\sc i}} \citep{Israel90}. 

   The cavity is centred on SSC\,A and has a diameter of 198\,pc \citep[][]{Hunter00}, thus it extends to also contain SSC\,B.  SSC\,A and SSC\,B have ages of 7\,Myr and 10-25\,Myr \citep{Hunter00,Larsen08}, respectively, while their masses range between 6--7$\times$10$^5$\,M$_{\odot}$ \citep{Grocholski12,Larsen08}. SC\,10 is located at the edges of the cavity and towards its North--West direction, where giant molecular clouds have been observed together with ongoing star formation in an embedded phase \citep{Tokura06,Taylor99,Waller91}. SC\,10, which is not considered an SSC, has a stellar mass of $\sim$10$^{3}$\,M$_{\odot}$ and consists of two components, both with young ages: one of 5--7\,Myr, and another with less than 5\,Myr \citep{Westmoquette07b}. 

  The lower panel of Fig.~\ref{sl_figure1} shows contours of the GALEX FUV emission overlaid on the PACS\,100\,$\mu$m map. The PACS\,100\,$\mu$m emission traces the warm dust, while the FUV emission reveals the recent star formation occurring at a timescale of less than 100\,--\,300\,Myr \citep[e.g.,][]{Lee11,Calzetti13}. The emission from these young stars dominates the region inside the \mbox{H\,{\sc i}}\,/\,H\,$\alpha$ cavity, where the dust emission at 100\,$\mu$m seems less prominent. There are several dust knots revealed at 100\,$\mu$m, which are located outside the very central starbursting region and mark the periphery of the cavity. At the location of these knots, several \mbox{H\,{\sc ii}} regions and giant molecular clouds are found \citep{Waller91,Taylor99,Pasquali11,Johnson12}. These molecular clouds host embedded star formation \citep{Tokura06,Westmoquette07b,Clark13}. The brightest of these dust knots spatially coincides with SC\,10 and extends to its north, northwest, and southwest. At the location of the brightest dust knot, the onset of an H\,$\alpha$ feature has been observed by \citet{Waller91}, while \citet{Taylor99} report that the detected molecular clouds are located in the interface region between the cold atomic gas phase and the hot gas, as due to the shocks of the outflow. \citet{Hong13} and \citet{Westmoquette07a} identify at the same location shocks due to stellar feedback and the outflow. These reports from the literature may suggest that the bright dust knot in the northwest to the SSC\,A is due to the embedded and ongoing star formation and the complex ISM dynamics due to the outflow.

%
%%%%% FIGURE 2 - Flux density maps %%%%%%%%%% 
 \begin{figure*}
%  \centering
      \includegraphics[width=6.3cm,clip]{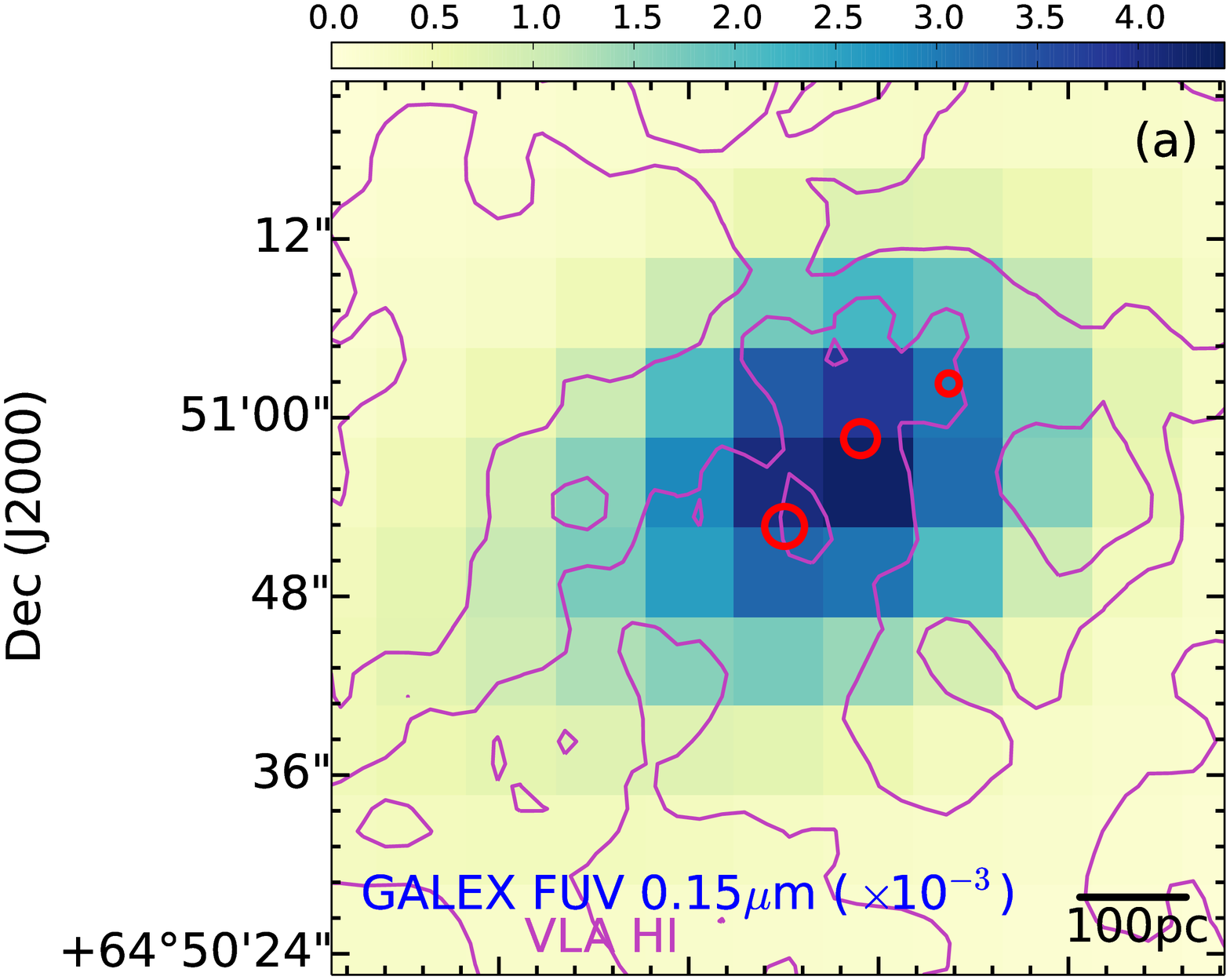}
      \includegraphics[width=4.7cm,clip]{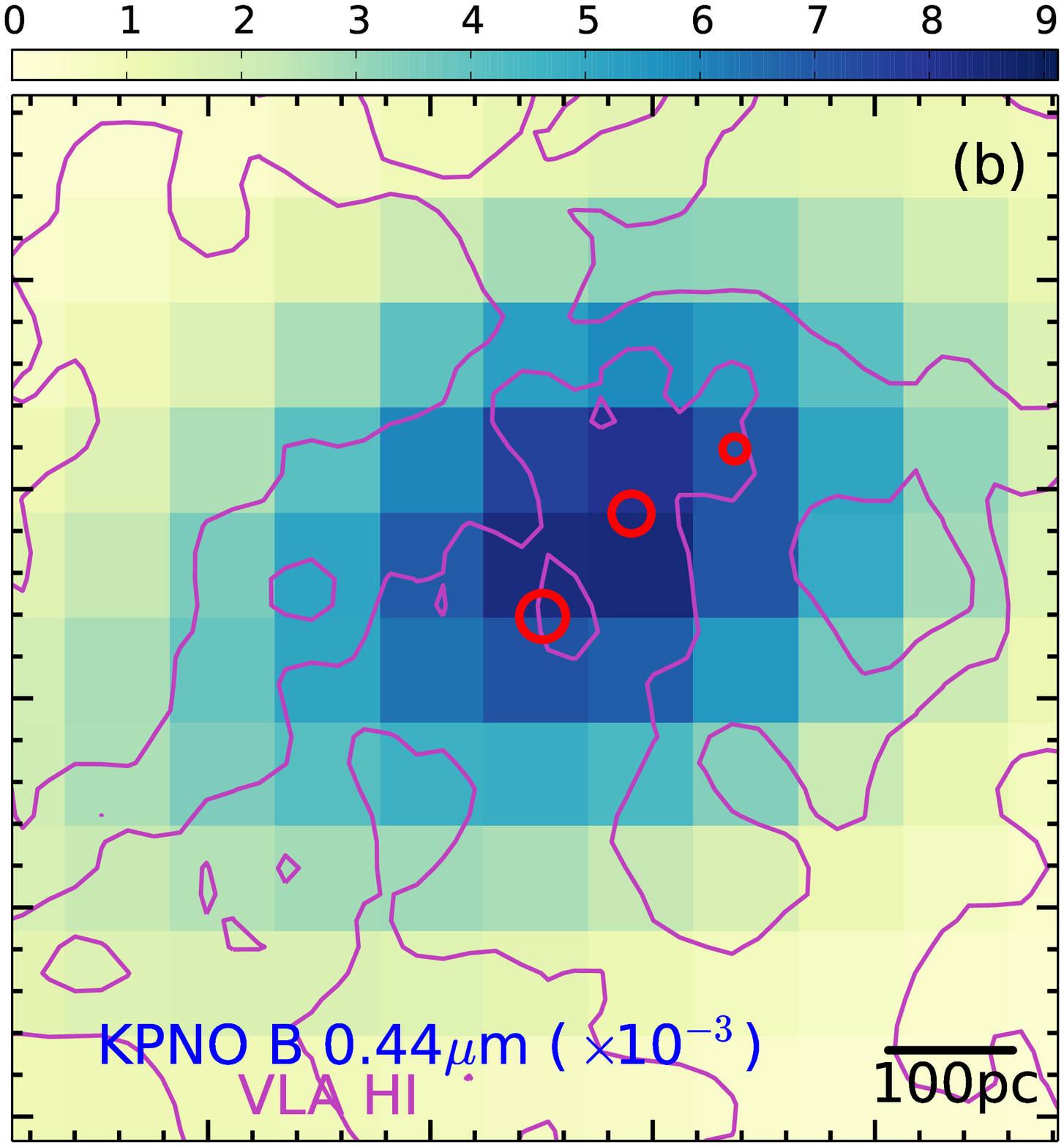}
      \includegraphics[width=4.8cm,clip]{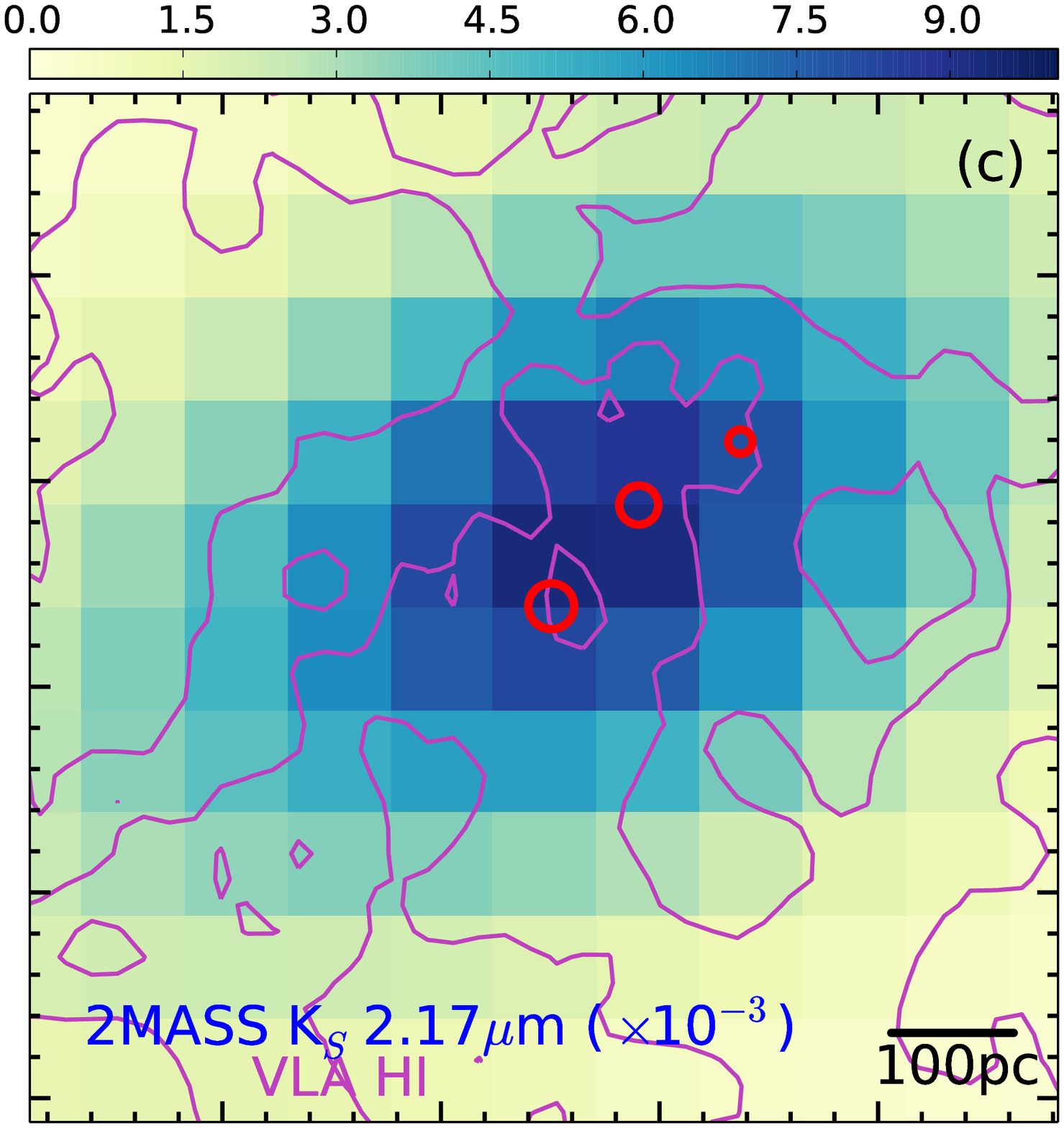}  
      \includegraphics[width=6.3cm,clip]{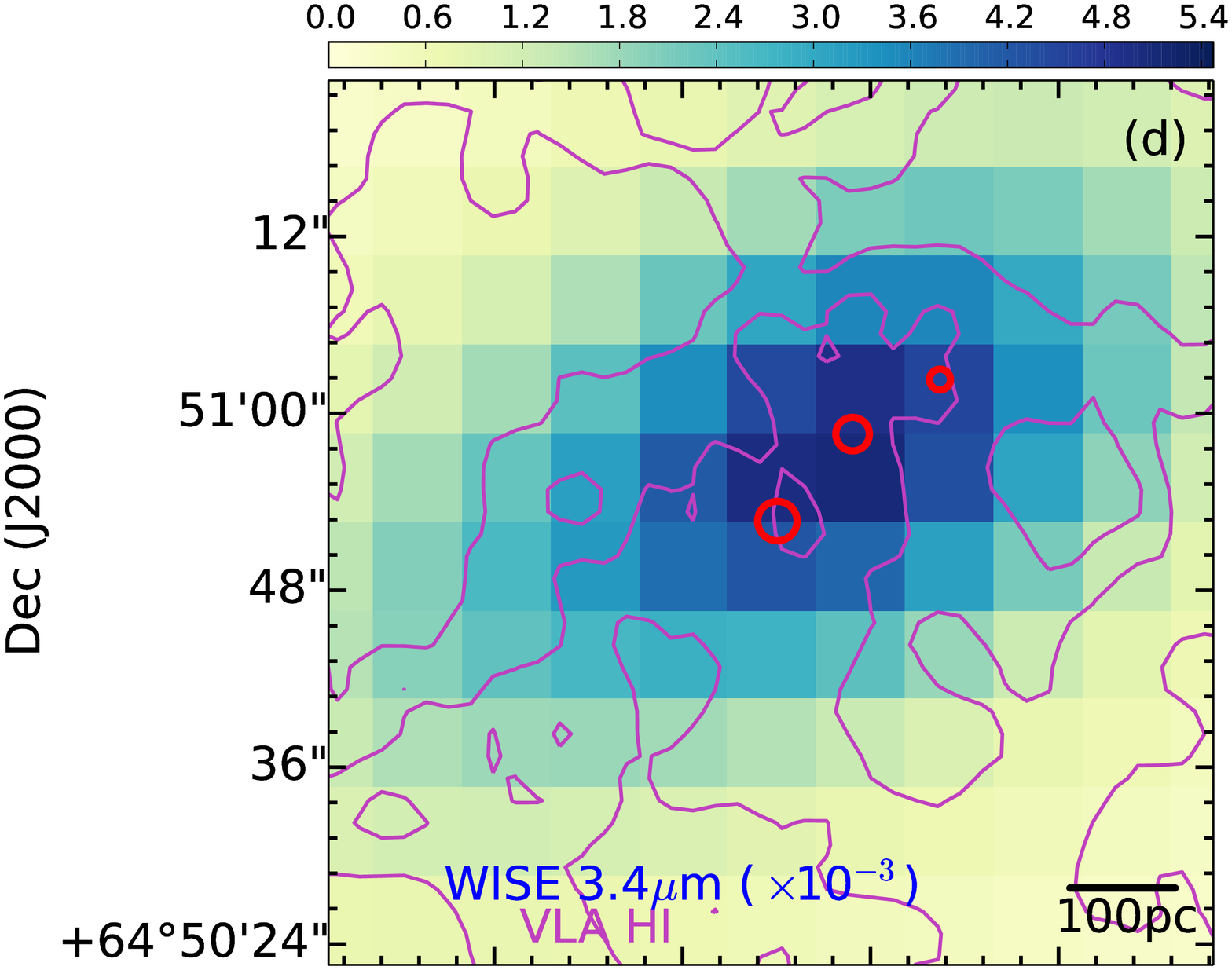}  
      \includegraphics[width=4.8cm,clip]{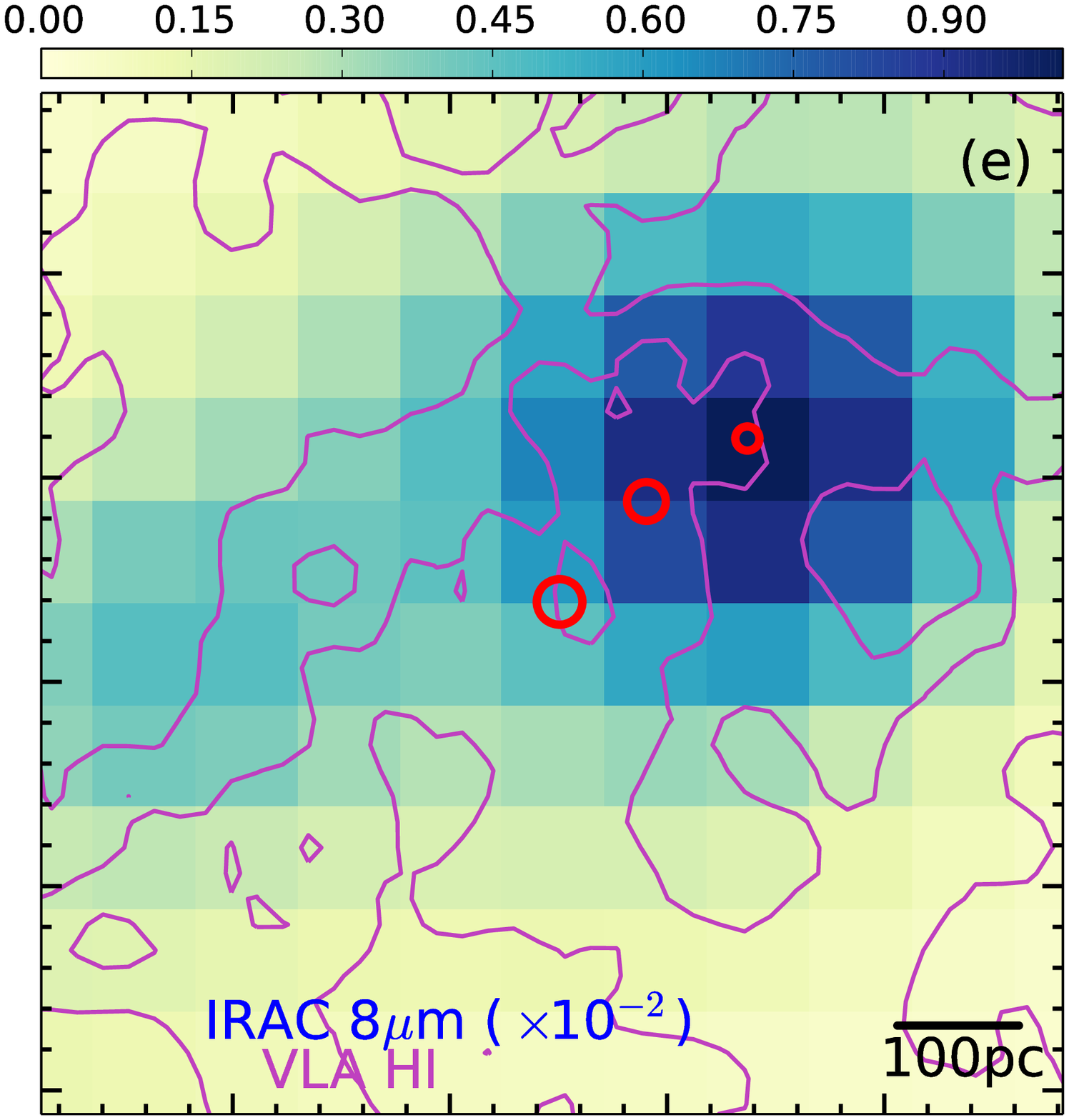} 
      \includegraphics[width=4.8cm,clip]{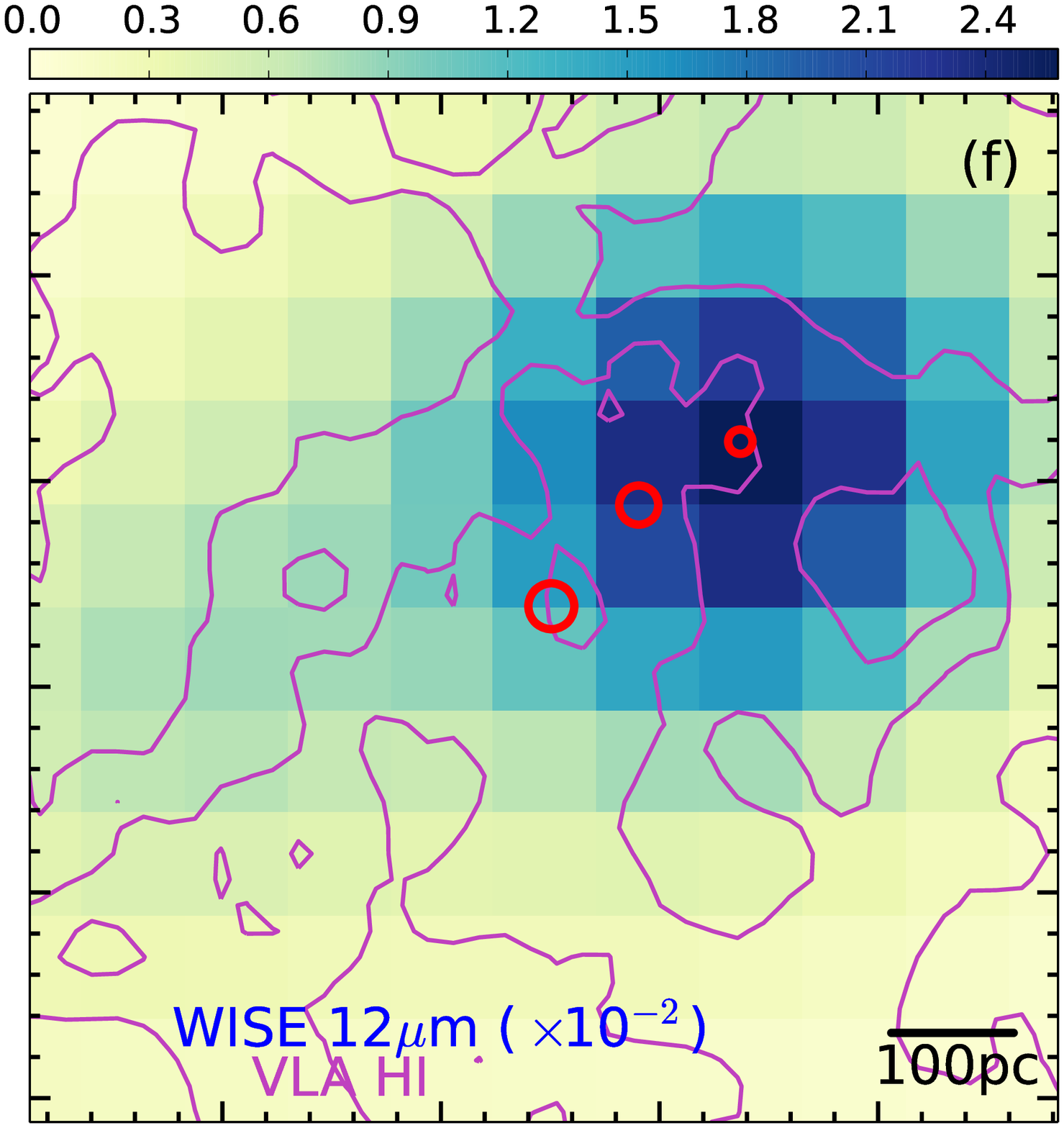}  
      \includegraphics[width=6.3cm,clip]{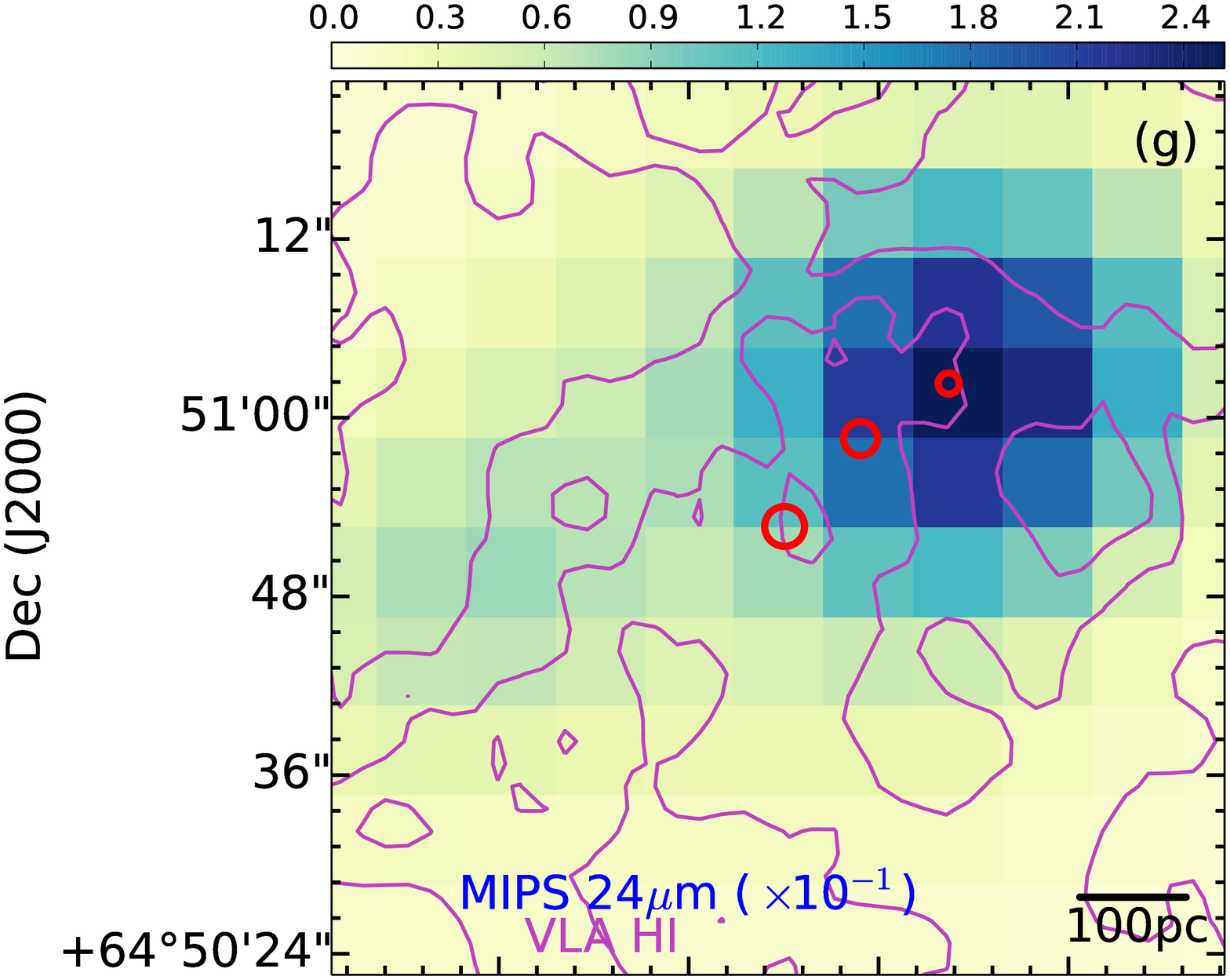}
      \includegraphics[width=5cm,clip]{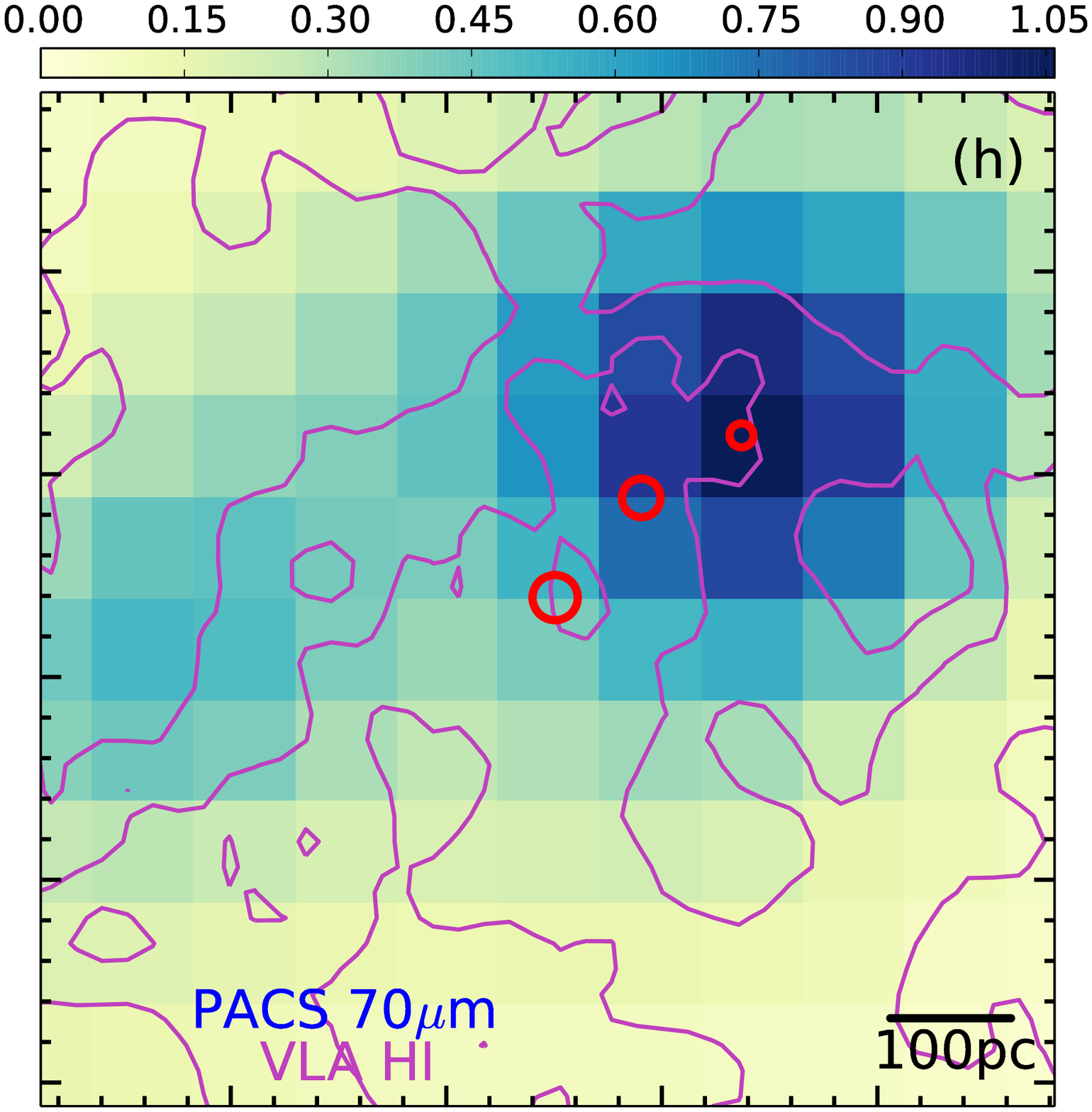} 
      \includegraphics[width=4.9cm,clip]{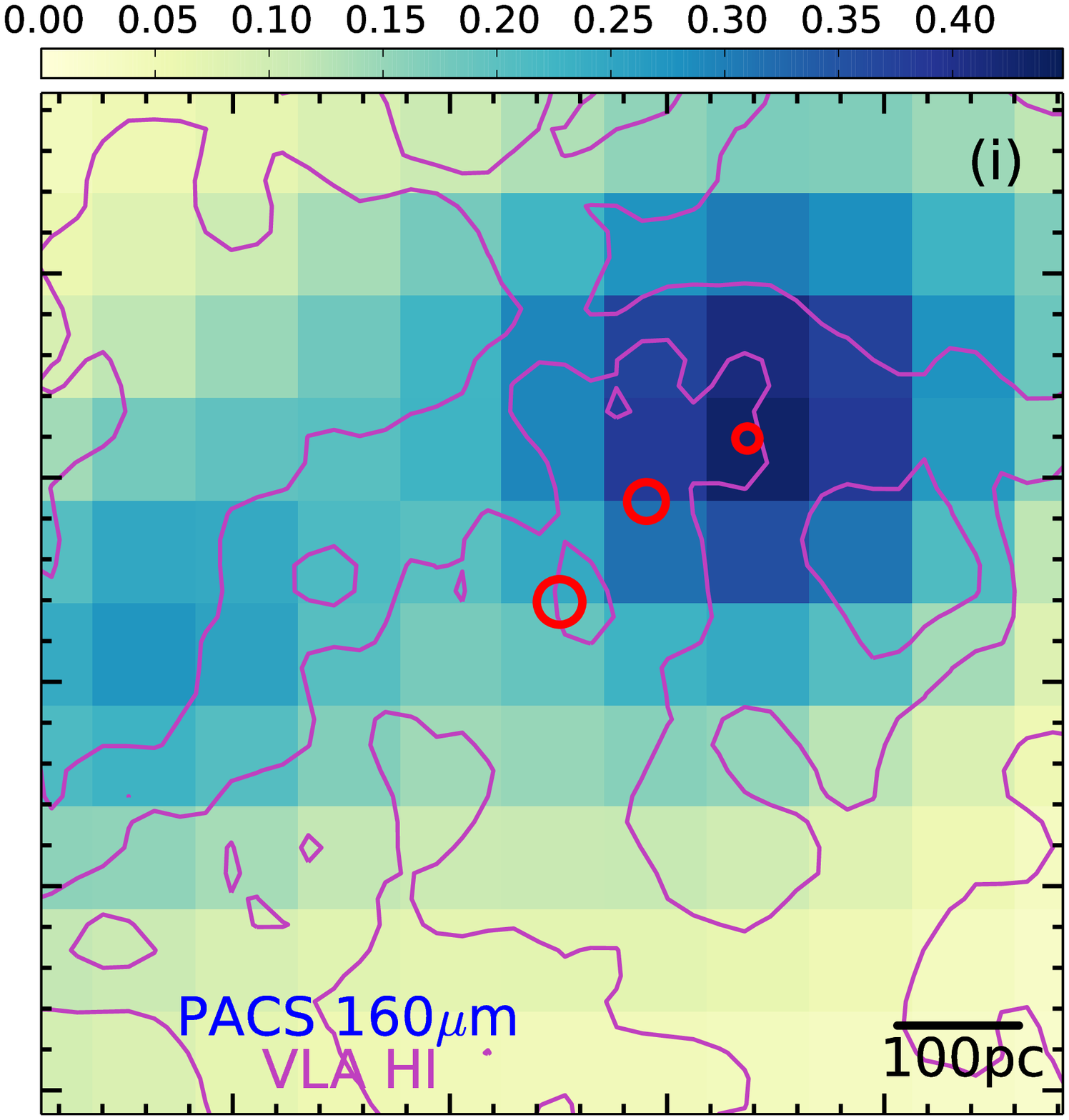}
      \includegraphics[width=6.3cm,clip]{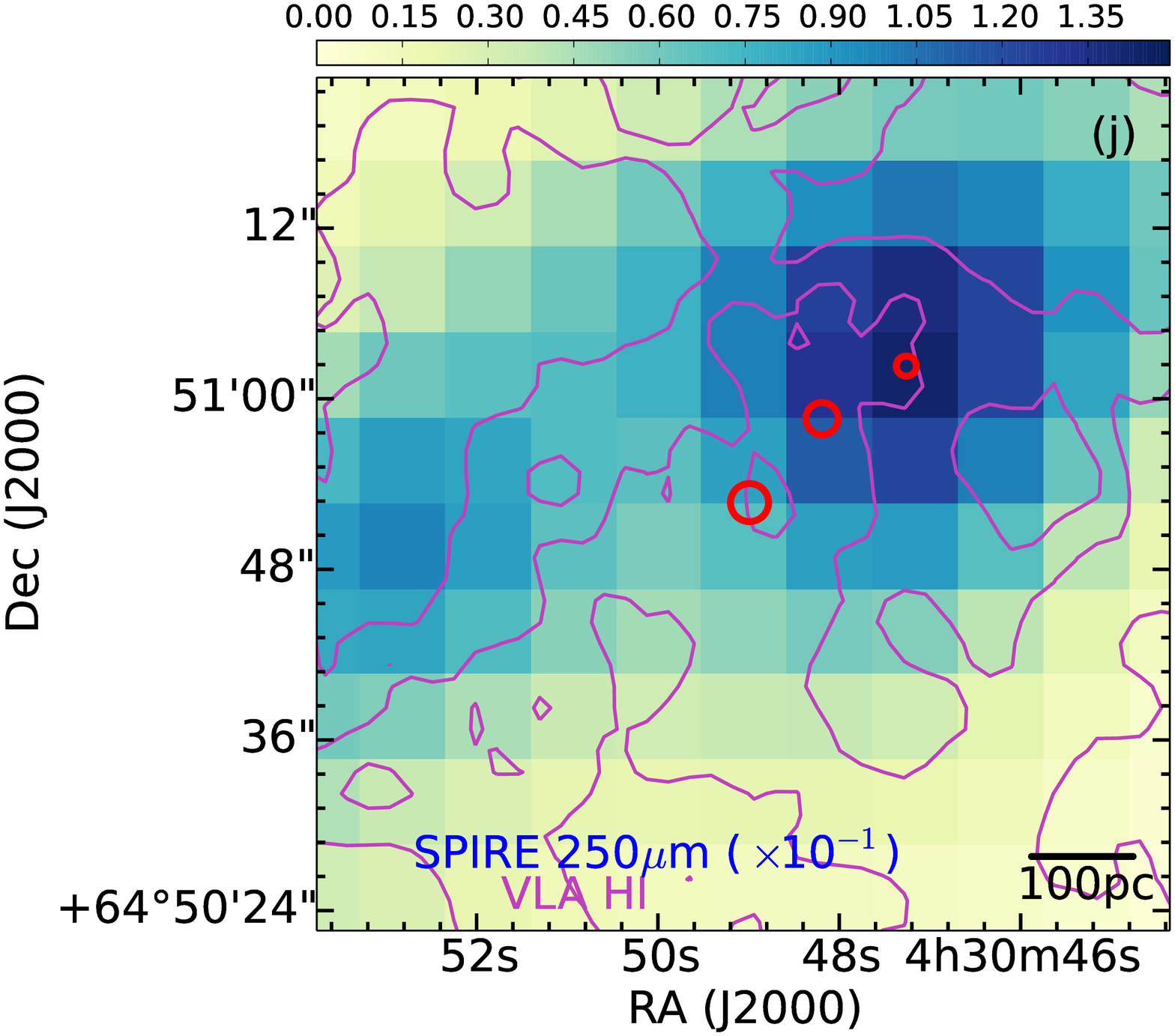}
      \includegraphics[width=4.8cm,clip]{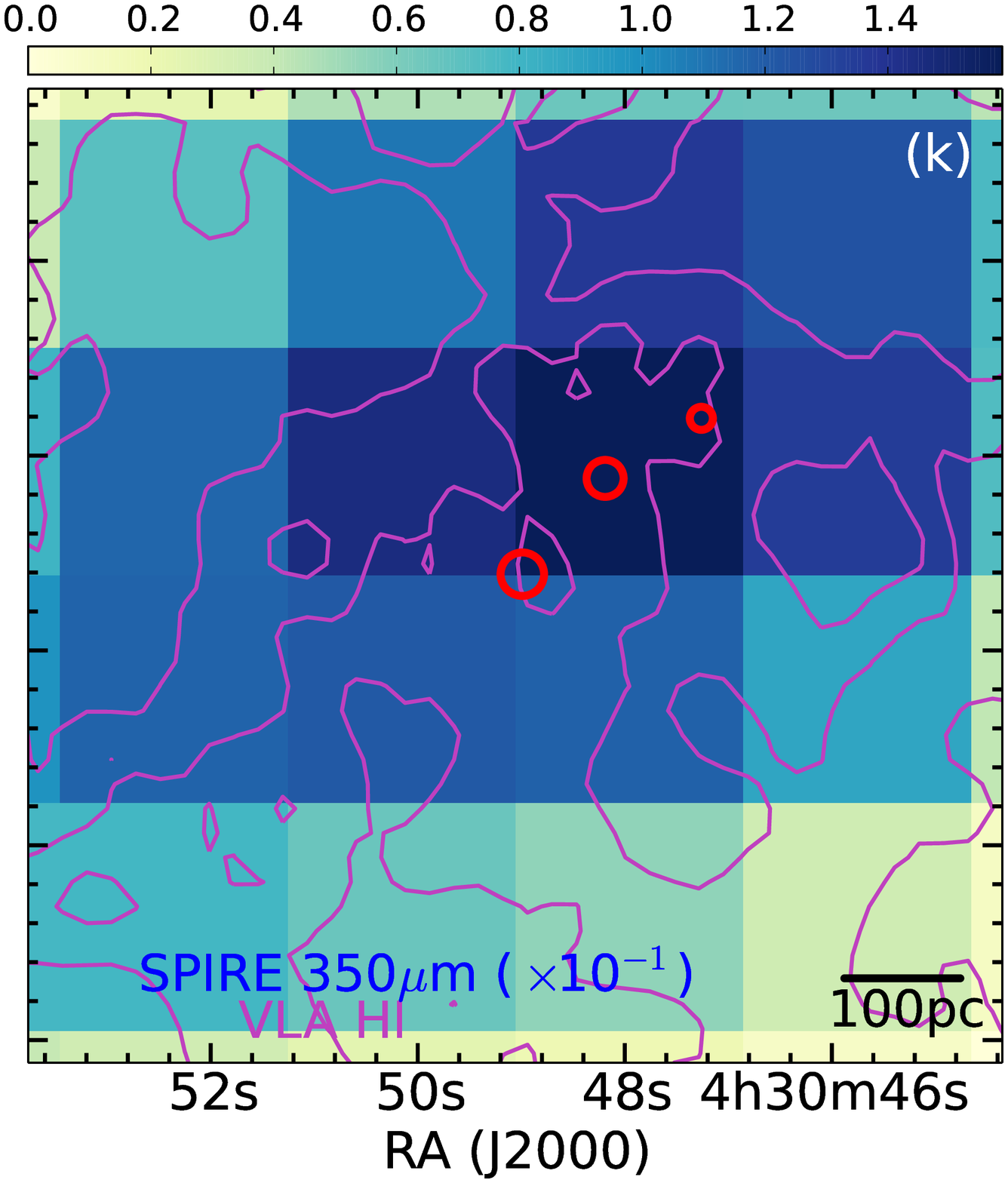}
      \includegraphics[width=4.8cm,clip]{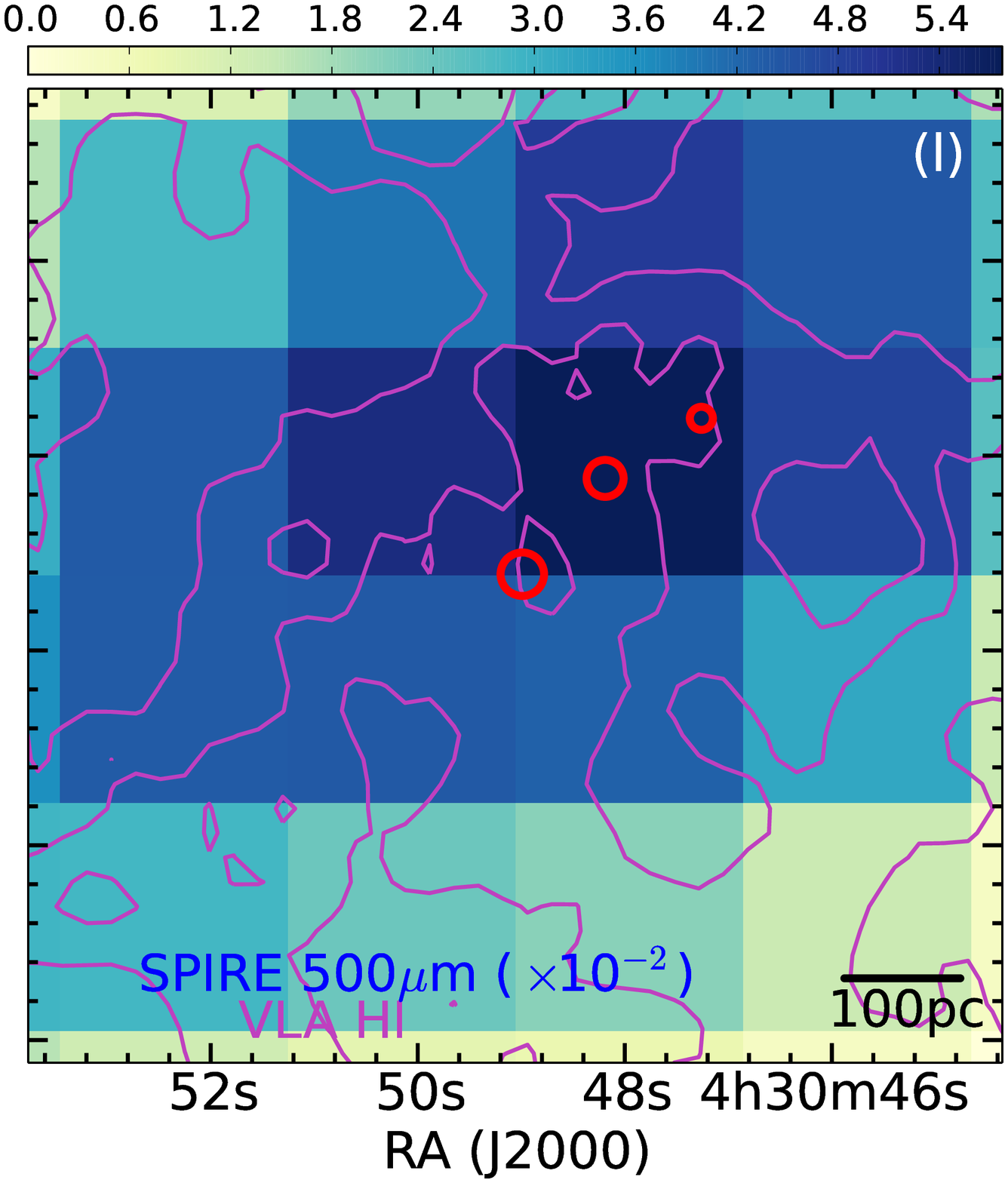}
  \caption{Multi-wavelength flux density maps. The unit of the colorbar is Jy\,/\,pixel. The maps up to SPIRE\,250\,$\mu$m are in the resolution of the SPIRE\,250\,$\mu$m PSF (18.2$\arcsec$ FWHM), while the remaining two maps are in the resolution of the SPIRE\,500\,$\mu$m PSF (36.4$\arcsec$ FWHM). Contours show the \mbox{H\,{\sc i}} emission, with levels ranging from 2\,mJy/pixel$\times$km/sec to 9\,mJy/pixel$\times$km/sec with a step of 3.5\,mJy/pixel$\times$km/sec. The field of view is 60$\arcsec$\,$\times$\,60$\arcsec$. Red solid circles indicate the location of the SSCs and SC\,10.}
  \label{sl_figure2}
 \end{figure*}
%%%%%%%%%%%%%%%%%%%%%%%%%%%%%%%%%%%
%
   Fig.~\ref{sl_figure2} shows the flux density maps of the images at representative filters we use in our SED analysis, so as to illustrate the spatial distribution of their emission. The GALEX FUV emission (panel a) reveals a rather homogeneous and centrally concentrated spatial distribution, which is dominated by the emission of the young stars. The optical B--band (panel b), 2MASS K$_{s}$ (panel c) and WISE 3.4\,$\mu$m (panel d) emission, which trace the young as well as the more evolved stars, show a similar distribution as the FUV emission, albeit more extended. A more confined distribution of the FUV emission as compared to the optical and MIR is also reflected in the corresponding disk scale lengths reported in \citet{Zhang12}. The emission in the IRAC 8\,$\mu$m (panel e) and {\it WISE}\,12\,$\mu$m (panel f) bands is sensitive to the hot dust from polycyclic aromatic hydrocarbons (PAHs), as well as the warm dust continuum. The emission in the MIPS\,24\,$\mu$m and up to the PACS\,160\,$\mu$m traces the warm dust (panels g to i), while the emission longwards to the SPIRE\,250\,$\mu$m (panels j to l) traces the cold dust. At the IRAC 8\,$\mu$m and up to the SPIRE\,250\,$\mu$m, two dust knots are revealed, offset from the central region, as also seen in the warmer dust emitting at 100\,$\mu$m discussed in Fig.~\ref{sl_figure1}.

  The North--Western dust knot is bright at all bands longwards to the IRAC 8\,$\mu$m, while the fainter dust emission knot to the southeast of SSC\,A becomes redder in the MIPS 24\,$\mu$m \citep[see also Fig.~4 in][]{Wu06} and in the PACS\,70\,$\mu$m map. The giant molecular clouds detected by \citet{Taylor99} lie in the brightest North--Western dust knot seen in the warm and cold dust emission. At all MIR to FIR and submm wavelengths, where the angular resolution permits, dust knots coincide with the locations of the ongoing star formation in the periphery of the cavity, the latter indicated with the \mbox{H\,{\sc i}} contours in Fig.~\ref{sl_figure2}. The SPIRE\,350 and 500\,$\mu$m maps show the coldest dust emission to be distributed over the whole body of NGC\,1569. Any substructure in the coldest dust components is not well-resolved in these bands, thus making it challenging to reveal the source of heating.

%
%%%%% FIGURE 3 - Flux density ratio maps %%%%%%%%%% 
 \begin{figure*}
%  \centering
      \includegraphics[width=6.1cm,clip]{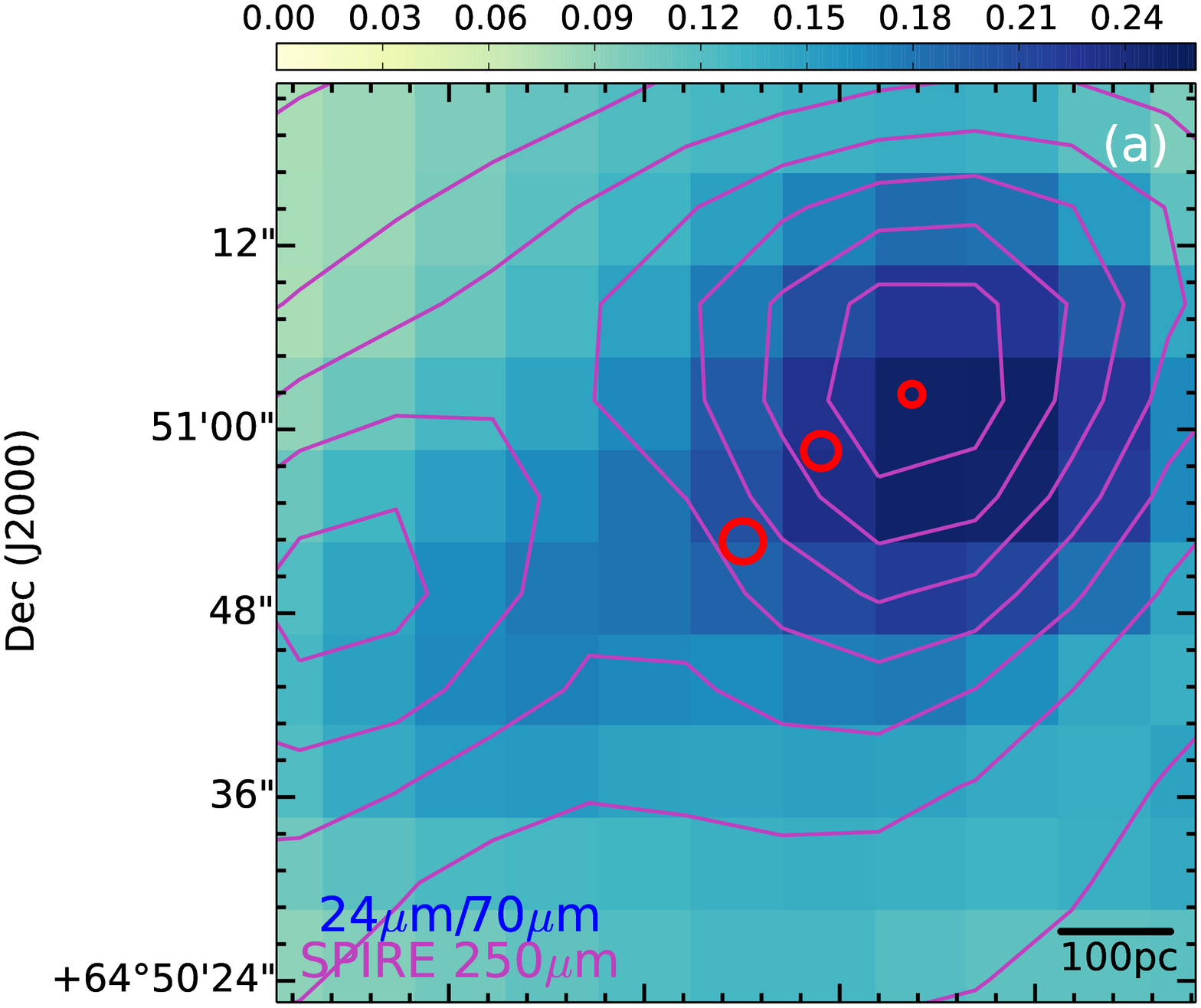}
      \includegraphics[width=4.8cm,clip]{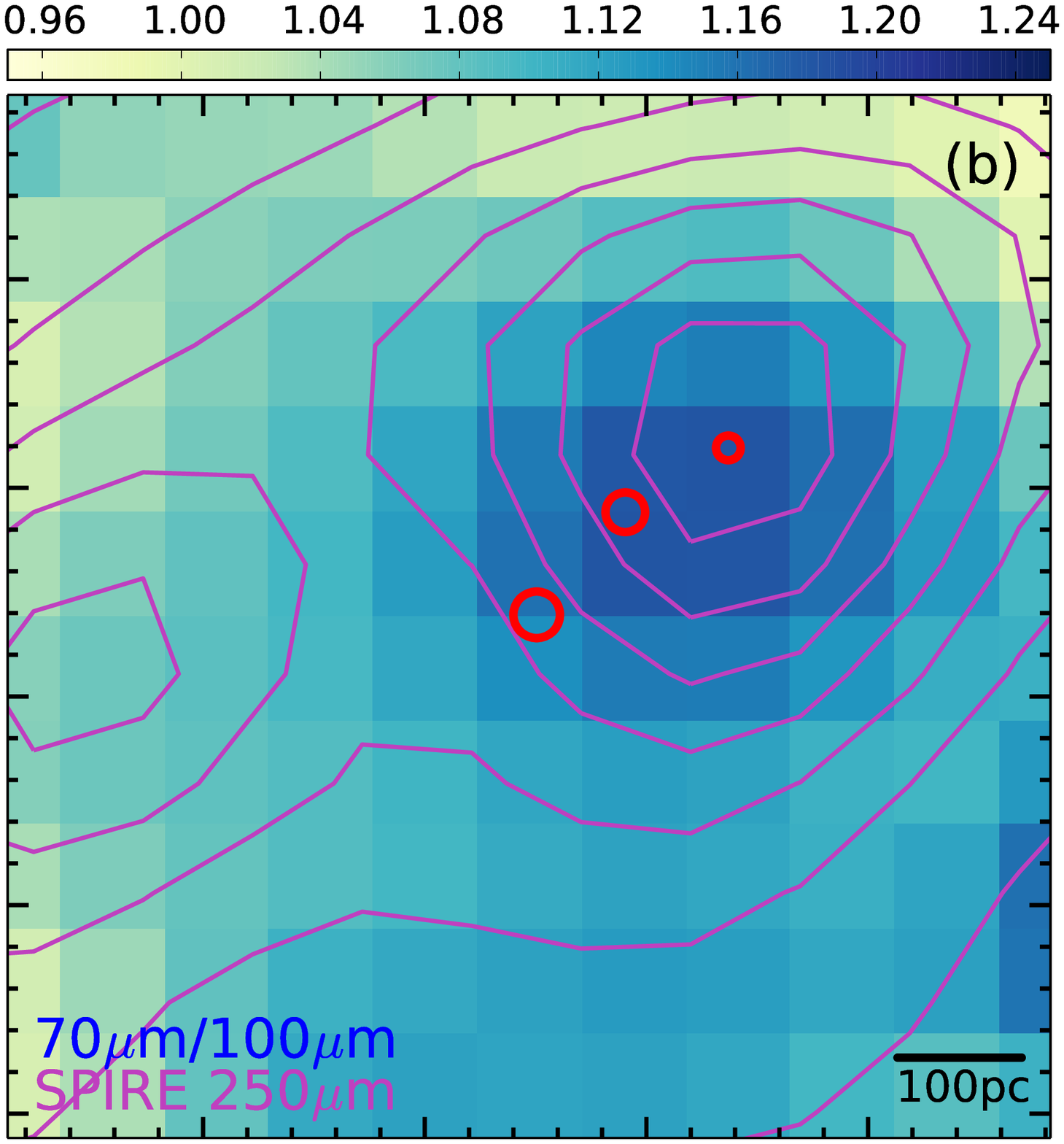}
      \includegraphics[width=4.8cm,clip]{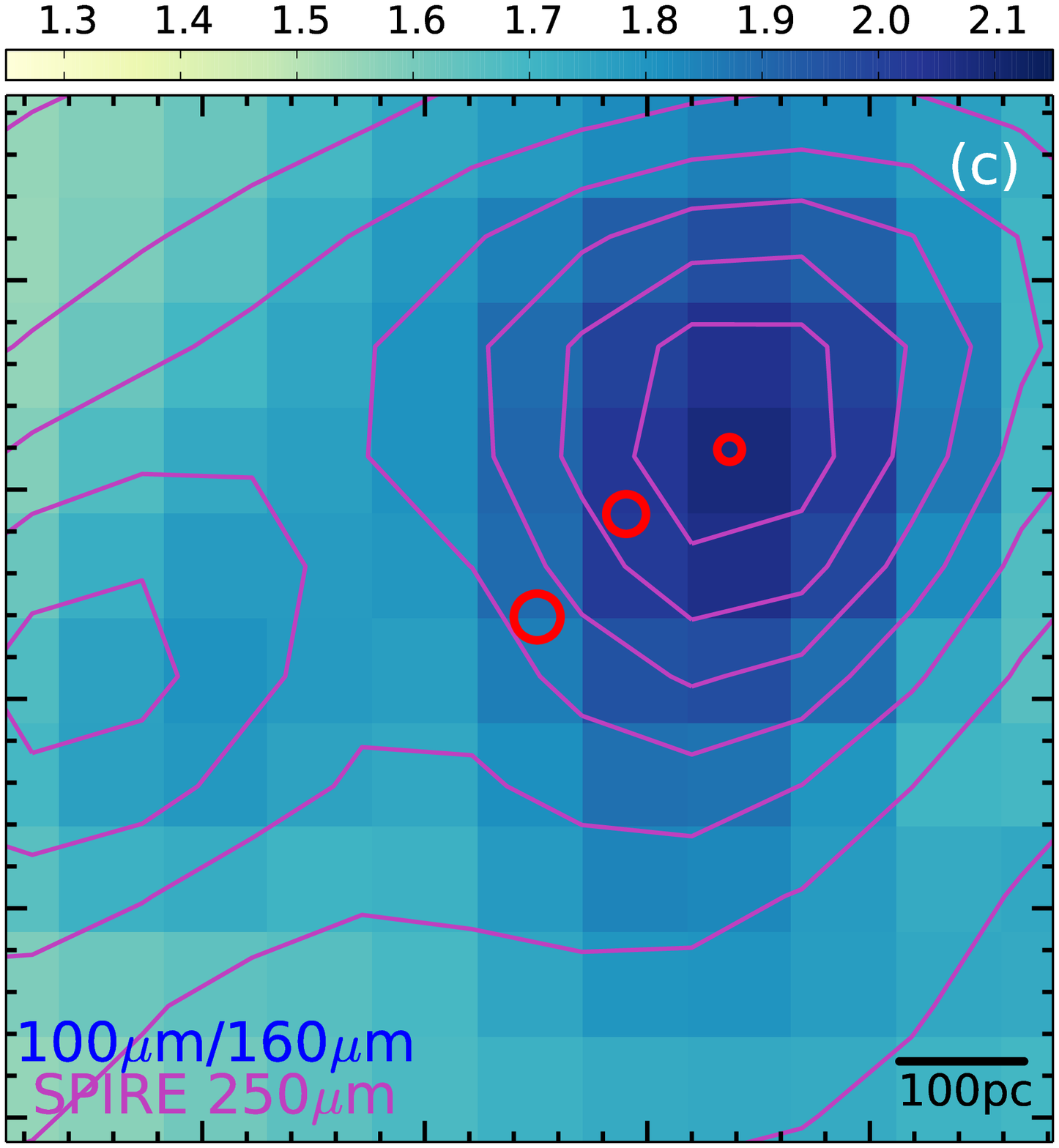}  
      \includegraphics[width=6.1cm,clip]{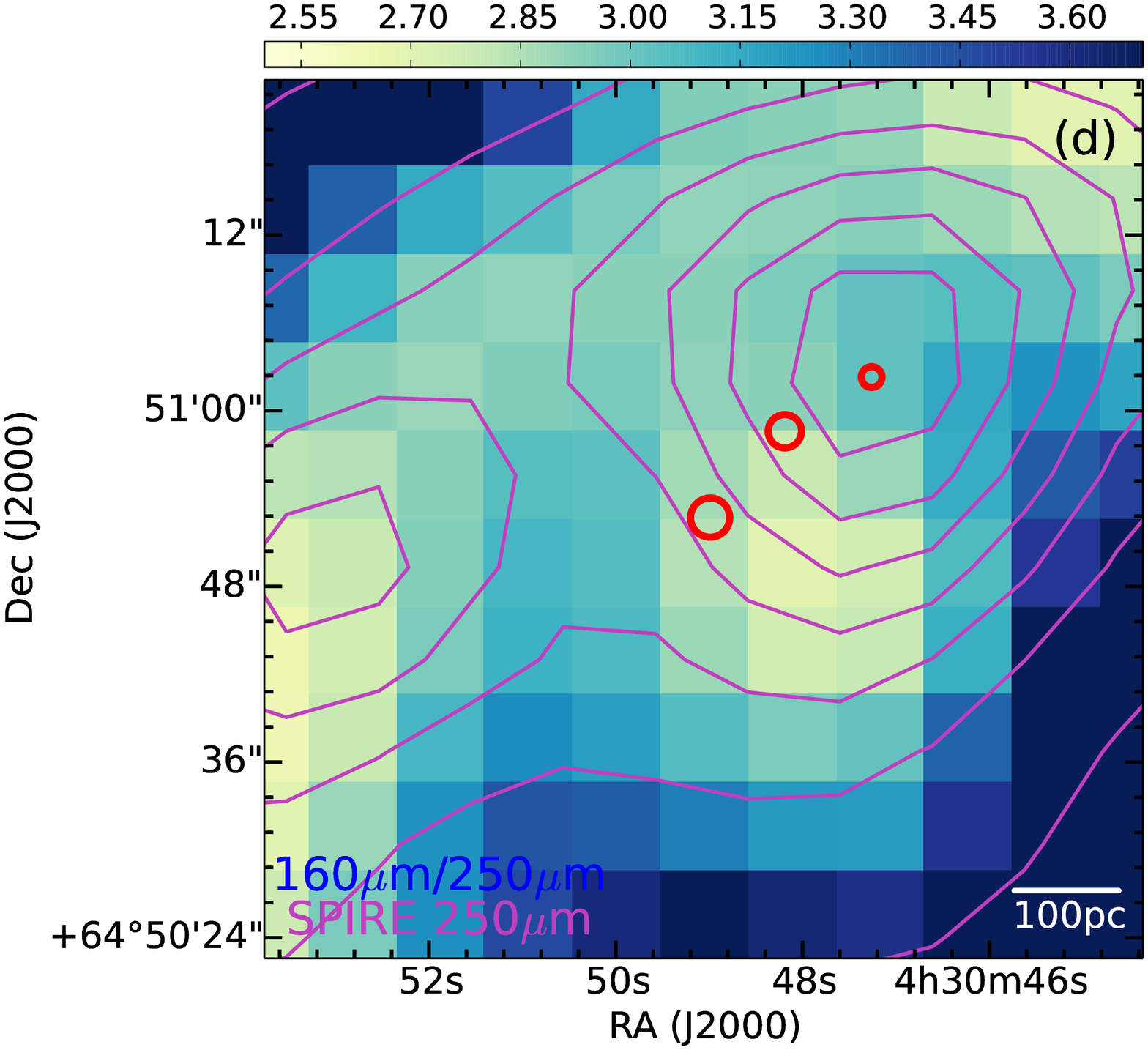}  
      \includegraphics[width=4.9cm,clip]{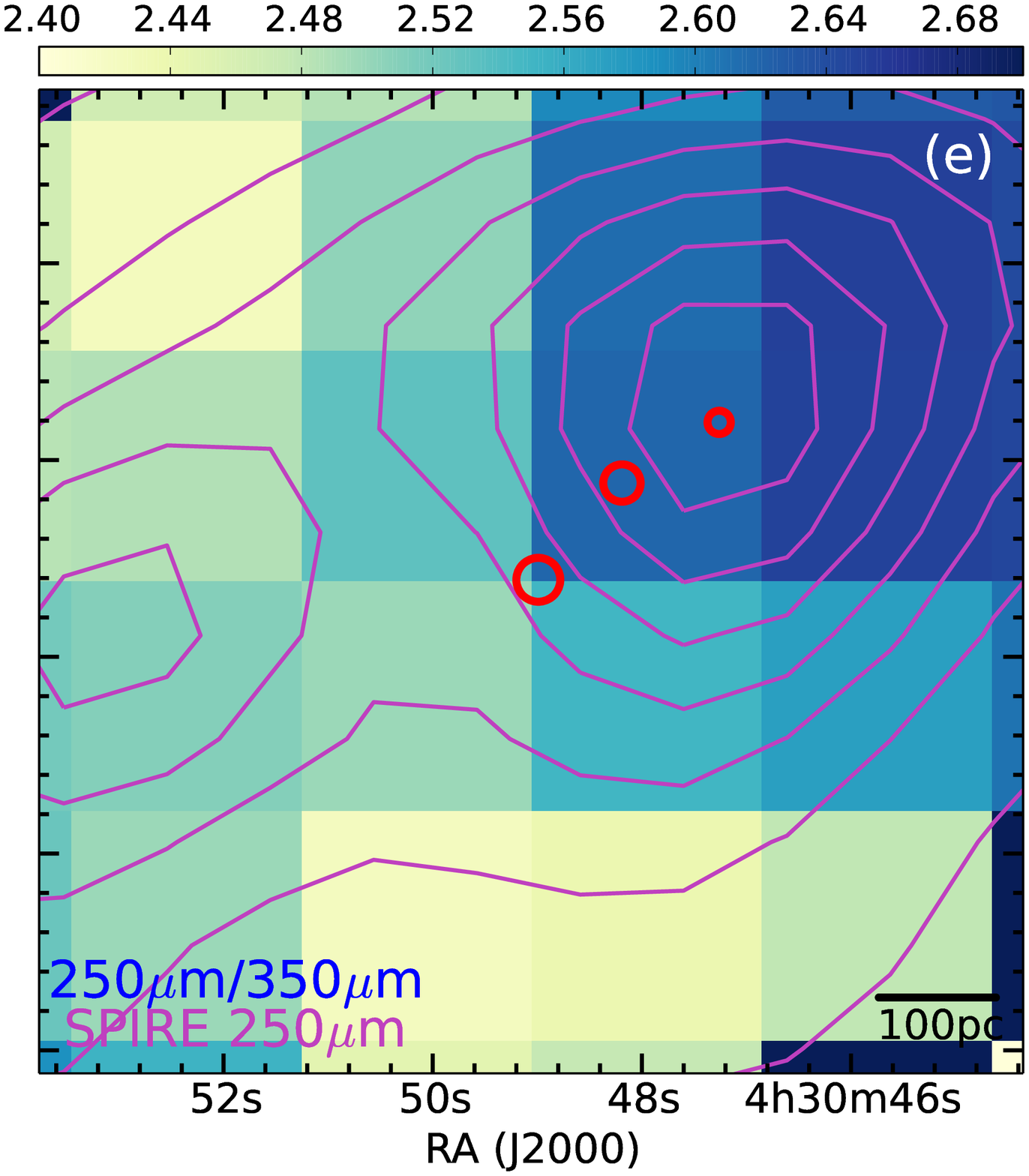} 
      \includegraphics[width=4.8cm,clip]{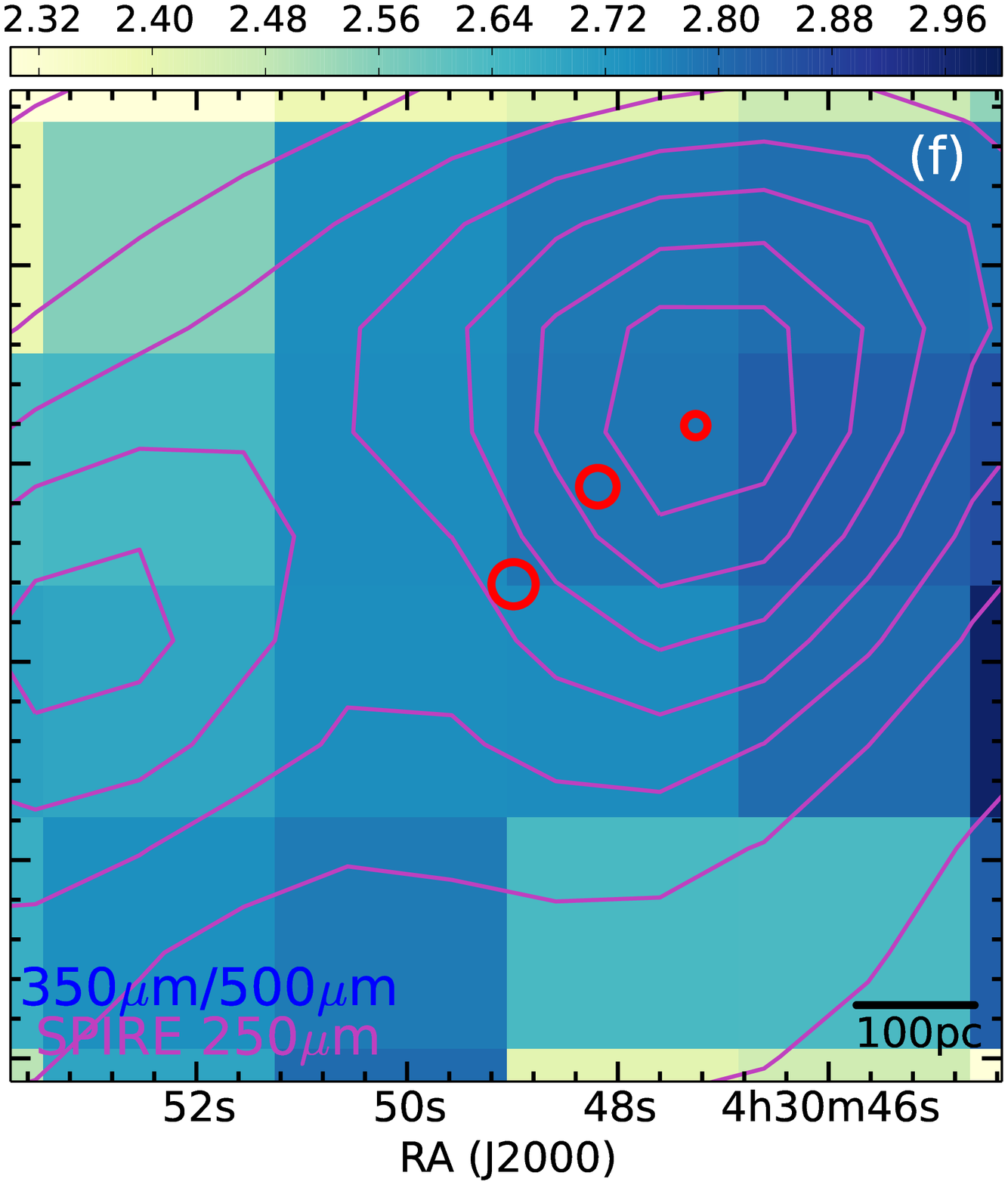}  
  \caption{Flux density ratio maps for the FIR bands. Ratio maps shown up to SPIRE\,250\,$\mu$m are in its angular resolution (18.2$\arcsec$ FWHM), while the remaining maps are shown in the resolution of the SPIRE\,500\,$\mu$m PSF (36.4$\arcsec$ FWHM). Contours show the SPIRE\,250\,$\mu$m emission, with levels ranging from 130 to 10\,mJy with a step of 20\,mJy, per pixel. The field of view is 60$\arcsec$\,$\times$\,60$\arcsec$. Red solid circles indicate the SSCs and SC\,10.}
  \label{sl_figure3}
 \end{figure*}
%%%%%%%%%%%%%%%%%%%%%%%%%%%%%%%%%%%
%
   Fig.~\ref{sl_figure3} shows the flux density ratio maps of the FIR images. These ratio maps reveal overall warmer colours, i.e. having ratios larger than one, in all but the very first map (panel a). The ratio 24\,$\mu$m\,/\,70\,$\mu$m (panel a) reveals a colder colour in the central starburst, with an increasing 24\,$\mu$m emission in the North--Western dust knot.

   Extended emission beyond the main optical body of NGC 1569 is seen at all {\it Herschel} wavelengths up to SPIRE\,250\,$\mu$m. The extended emission in the  
%%%%% FIGURE 4 - Extended emission %%%%%%%%%% 
 \begin{figure*}
%  \centering
      \includegraphics[width=8cm]{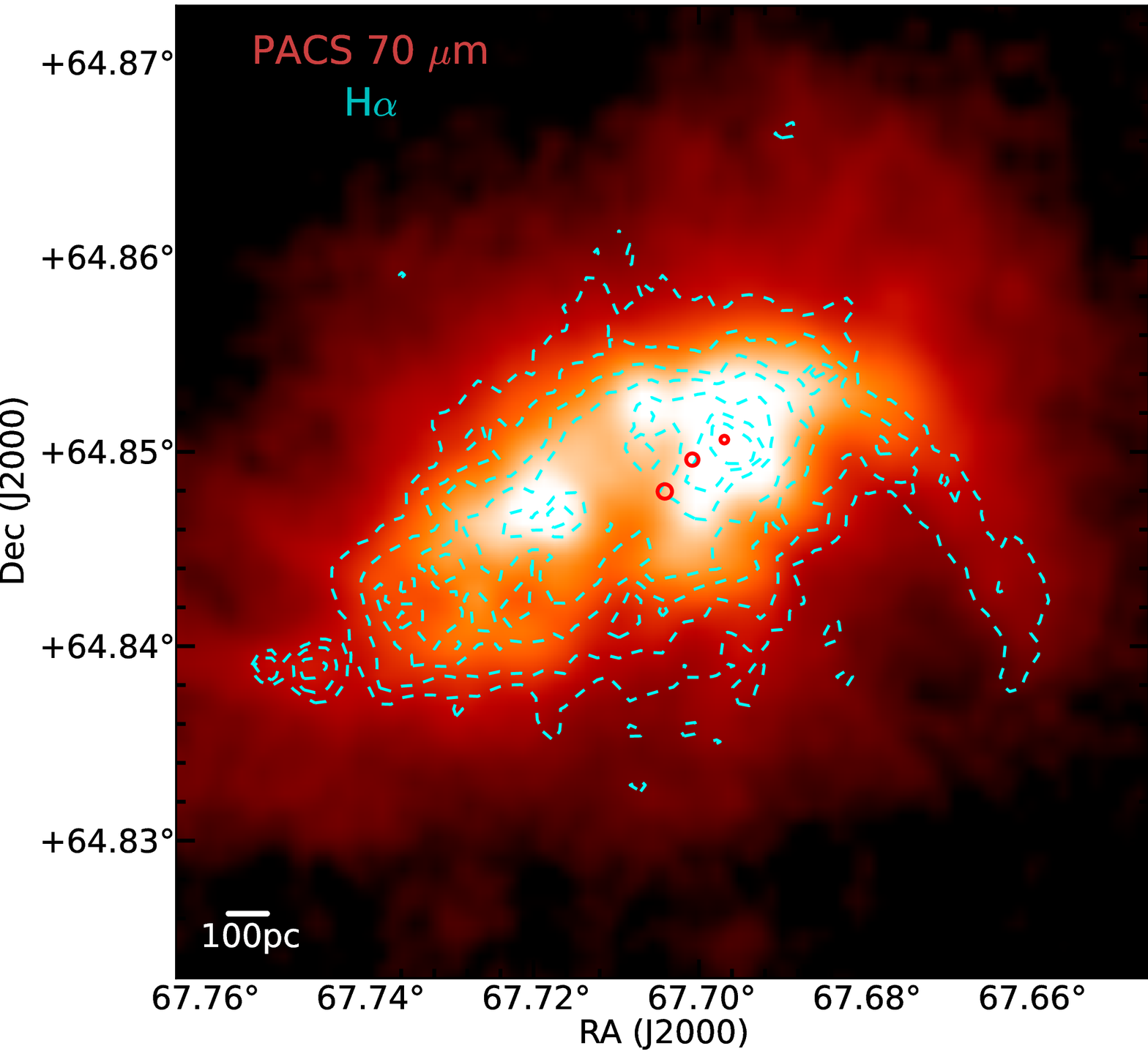}
      \includegraphics[width=8cm]{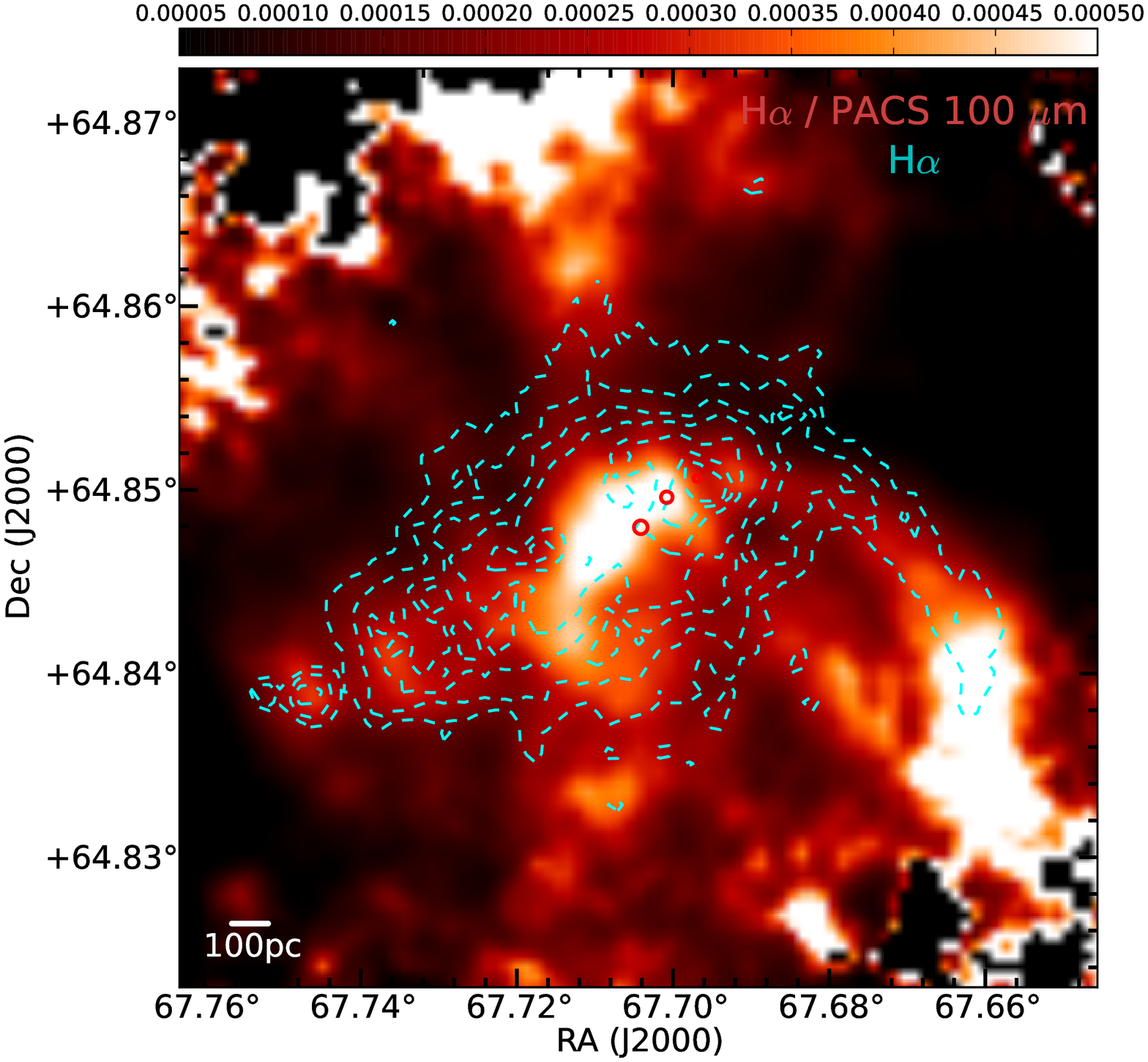}
  \caption{{\it Left panel:} PACS\,70\,$\mu$m map showing the extended emission within 180$\arcsec$\,$\times$\,180$\arcsec$, with contours of H\,$\alpha$ overlaid in dashed cyan. White colour corresponds to higher emission. {\it Right panel:} H\,$\alpha$\,/\,PACS\,100\,$\mu$m ratio map of the same region with the same contours overlaid as in the left panel. White colour corresponds to a higher flux density ratio. The location of the SSCs and SC\,10 is indicated with the red solid circles.}
  \label{sl_figure4}
 \end{figure*}
%%%%%%%%%%%%%%%%%%%%%%%%%%%%%%%%%%%
%
PACS\,70\,$\mu$m map is shown in the left panel of Fig.~\ref{sl_figure4}. The emission is also evident in X-ray and H\,$\alpha$ observations and has been associated with larger scale galactic outflows \citep{Heckman95,Martin98,Westmoquette08}. Such galactic outflows are seen in other nearby starburst galaxies \citep[e.g., NGC\,253, M\,82;][]{Westmoquette11,Heckman90}, and recently shown for NGC\,253 to be accompanied by molecular gas outflows \citep{Bolatto13}, as well as associated with atomic hydrogen plumes \citep{Boomsma05}. In the case of NGC\,1569, its larger scale extended emission is also detected by the warm and cold dust components, from the PACS\,70\,$\mu$m to the SPIRE\,$250$\,$\mu$m. The warm dust traced by the PACS\,70\,$\mu$m emission follows the H\,$\alpha$ morphology in the most extended structures seen, as shown with the contours overlaid of the H\,$\alpha$ emission (left panel of Fig.~\ref{sl_figure4}). The ratio of the warm ionised gas (traced with the H\,$\alpha$ emission) to the warm dust (traced with the 100\,$\mu$m emission) reveals even more prominently the warm colour of the emission (higher ratio) at the sites of the extended emission, alternating by cooler colours (lower ratio) adjacent to the warm ionised gas marking the outflow. 

%%%%% FIGURE 5 - Extended emission radial profile %%%%%%%%%% 
 \begin{figure*}
%  \centering
      \includegraphics[width=5.9cm]{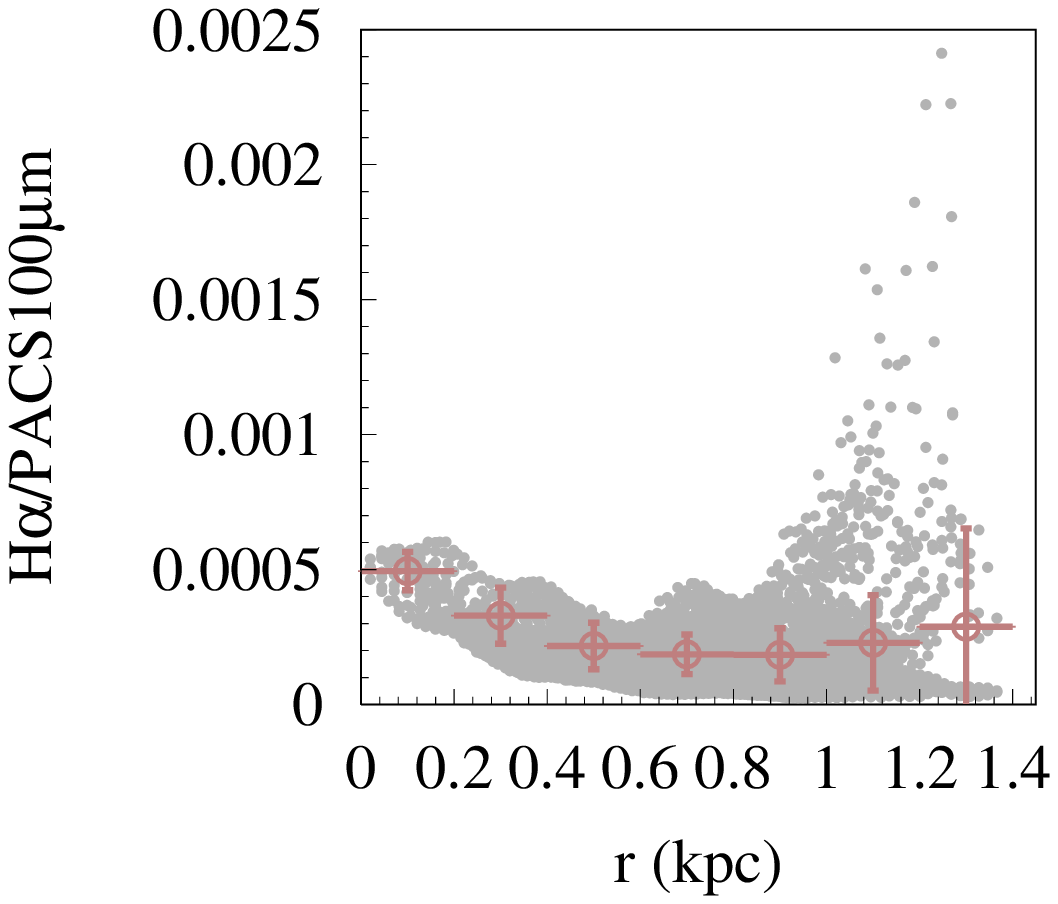}
      \includegraphics[width=5.9cm]{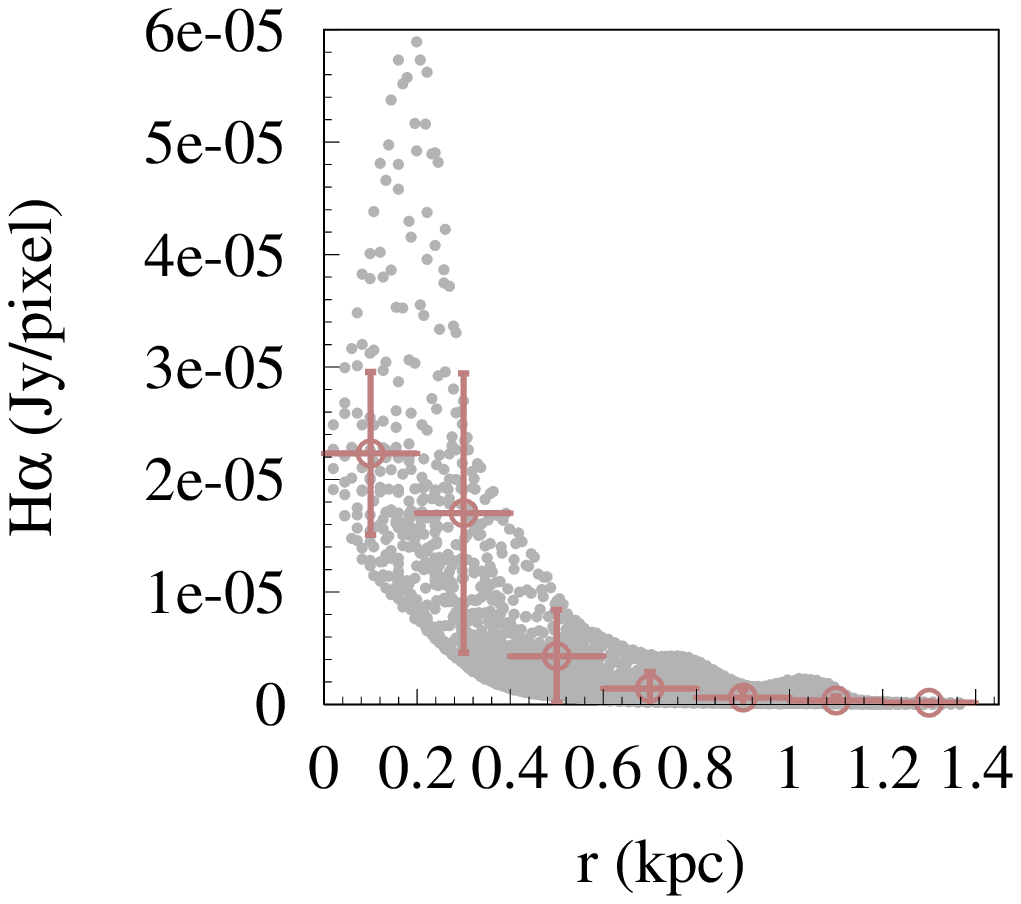}
      \includegraphics[width=5.7cm]{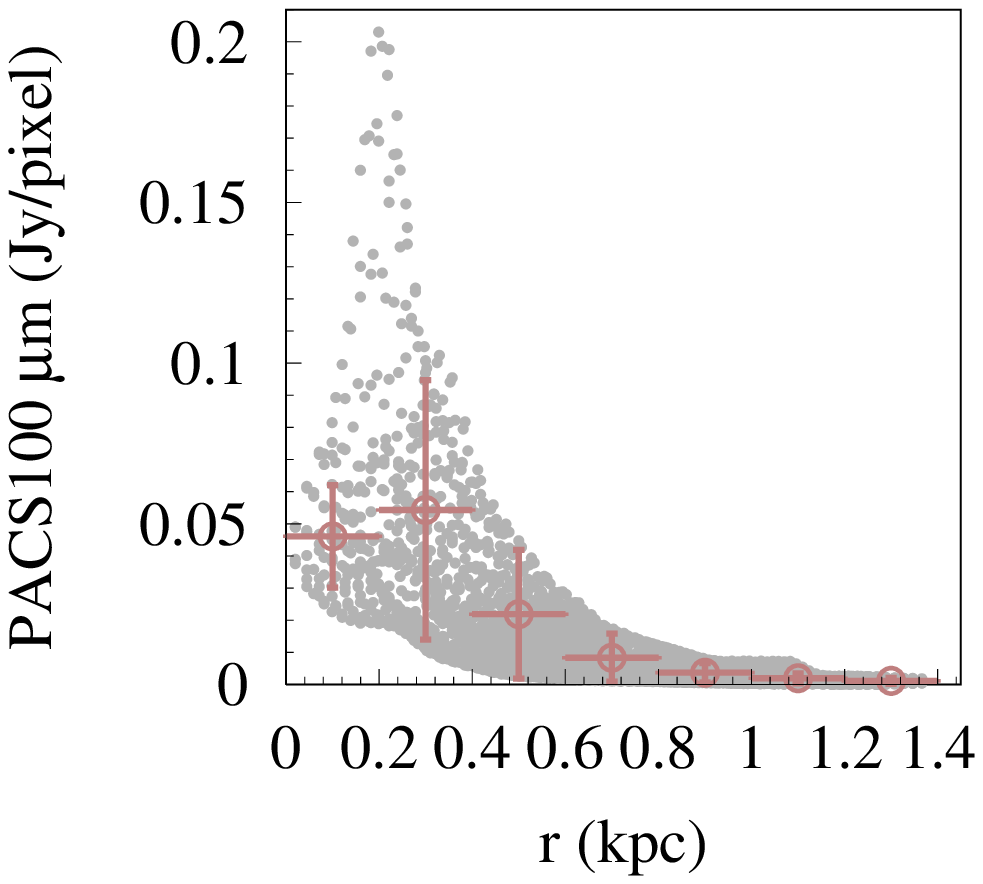}
  \caption {{\it Left panel:} H\,$\alpha$\,/\,PACS\,100\,$\mu$m ratio profile. The red (darker) symbols are data binned within 0.2\,kpc regions. 
{\it Middle panel:} PACS\,100\,$\mu$m flux density profile. The red (darker) symbols are data binned within 0.2\,kpc regions.
{\it Right panel:} H$\alpha$ flux density profile. The red (darker) symbols are data binned within 0.2\,kpc regions. All data have been convolved to the PACS\,100\,$\mu$m PSF. }
  \label{sl_figure5}
 \end{figure*}
%%%%%%%%%%%%%%%%%%%%%%%%%%%%%%%%%%%
%
   The same trends are revealed in the radial profiles shown in Fig.~\ref{sl_figure5}. The left panel shows the H\,$\alpha$\,/\,PACS\,100\,$\mu$m ratio profile. There is the trend of a decreasing ratio with increasing galactocentric radius, with overall cooler colours. The colour, though, becomes warmer at the location of the extended emission, i.e. beyond $\sim$1\,kpc. The middle and right panel of Fig.~\ref{sl_figure5} show the flux density profiles of the H\,$\alpha$ and the 100\,$\mu$m emission, respectively. The flux density enhancements in both profiles occur beyond $\sim$1\,kpc and indicate the extended emission.
%______________________________________________________________

\section{SED modelling with {\it magphys}}
\label{sec:modelling}
   We derive the star formation and dust properties by using {\it magphys} \citep{dacunha08} to model the observed multiwavelength SED of NGC\,1569, initially on a pixel--by--pixel basis, and later on for two individual regions within the galaxy. {\it Magphys} is a simple, physically motivated model to interpret the SED of a galaxy and its basic characteristics are described in \citet{dacunha08}. {\it Magphys} provides a consistent model for the emission of stars and dust to interpret the multiwavelength SEDs of galaxies and extract constraints on their stellar populations and ISM properties. The {\it magphys} model uses the \citet{Bruzual03} stellar population synthesis to model the spectral evolution of a galaxy, together with the model of \citet{Charlot00} to describe the attenuation of stellar emission by dust in the diffuse ISM and in the stellar birth clouds (BCs), which are then coupled to the dust emission model of \citet{dacunha08} via conservation of the energy for the dust absorbed in the BCs and diffuse ISM and re-radiated in the IR. No detailed modelling of the physical properties of the dust grains is performed in {\it magphys}, with the main components of the IR emission briefly described below.

   There are two basic steps involved in the panchromatic SED modelling with {\it magphys}. The first step is the construction of a library of model SEDs ranging in rest--frame wavelength from 912\,\AA\ to 1\,mm, with which the observed SEDs are fitted via $\chi ^{2}$--minimisation. The second step is to construct the marginalised likelihood distribution for each physical parameter of the observed galaxy, through the comparison of the observed and model SEDs. An assumption made in the modelling is that there is contribution of only starlight to the heating of the diffuse ISM, thus ignoring any AGN component. The total dust mass derived in {\it magphys} includes the dust mass contributed by warm dust in the stellar BCs and in the diffuse ISM, as well as the dust mass contributed by cold dust in the diffuse ISM. Components of cold dust clumps not heated by starlight are not included in the modelling with {\it magphys}. Such quiescent cold clumps of dust have been uncovered with radiative transfer modelling of disk galaxies and are proposed in order to reconcile the observed FIR and submm SED \citep[e.g.,][]{deLooze12,Holwerda12,Bianchi11,Baes10}.

   The library of model SEDs covers a wide range of star formation histories (SFHs), metallicities and dust contents and is constructed using two components. The first component of the model library involves the set of stellar population spectra, drawn from the \citet{Bruzual03} stellar population synthesis code and covering the wavelength range of 912\,\AA\ to 160\,$\mu$m. The second component of the model library involves the dust emission spectra, for which dust emission from the stellar BCs and the diffuse ISM is taken into account. These two components of the model library are then linked together via energy conservation, in order to construct the 912\,\AA\, to 1\,mm model SED and fit this to the observed SED.  

   The first component of the model SED library involves the stellar population spectra which are described by:
\begin{itemize}
       \item Parameters related to the SFH assuming the following to hold simultaneously: (1) a continuous SFR, described by the age of the galaxy, t$_{form}$, uniformly distributed between 0.1 to 13.5\,Gyr; (2) a star formation timescale, $\gamma$, uniform in the interval 0 to 0.6\,Gyr$^{-1}$ and exponentially dropping to zero around 1\,Gyr$^{-1}$; (3) bursts of star formation superimposed on the continuous SFR, such that 50 per cent of the stellar population spectra in the model library include a burst in the last 2\,Gyr.
       \item The metallicity, uniformly distributed between 0.02\,Z$_{\odot}$ and 2\,Z$_{\odot}$, where Z\,$_{\odot}$ is the solar metallicity.
       \item The attenuation by dust, described by absorption in the BCs and in the diffuse ISM using the model of \citet{Charlot00}, via the effective absorption optical depth of the dust in the BC, $\hat{\tau}_{\lambda}^{BC}$, and in the diffuse ISM, $\hat{\tau}_{\lambda}^{ISM}$. The effective absorption optical depth is parameterized with the total effective V--band absorption optical depth seen by young stars in the BC, $\hat{\tau}_{V}$, and with the fraction of this contributed by dust in the diffuse ISM, $\mu$ \citep[eqs.\,2--4 in][]{dacunha08}. 
\end{itemize}

   The second component of the model SED library involves the IR dust emission spectra, which are modelled considering the following:
\begin{itemize} 
       \item The stellar BCs contribute to the total dust emission via three components: (1) the PAHs, which are modelled with eq.\,9 in \citet{dacunha08}, assuming the fixed template spectrum of M17SW \citep{Cesarsky96,Madden06}, while the NIR continuum emission associated with PAHs is modelled with a modified blackbody assuming an emissivity index $\beta$\,=\,1; (2) a MIR continuum to characterise the emission from hot grains at temperatures between 130--250\,K, which is modelled with the sum of two modified blackbodies each with temperature 130\,K and 250\,K; (3) and warm dust grains in thermal equilibrium, modelled with a modified blackbody assuming an emissivity index $\beta$\,=\,1.5, and with temperatures T$_{W}^{BC}$ allowed to vary between 30--60\,K \citep[eq.\,13 in][]{dacunha08}. 
        \item The diffuse ISM contributes to the dust emission with the same components as described above for the BCs, but with ratios of the components fixed to reproduce the spectral shape of the diffuse cirrus emission of the Milky Way \citep[see eq.\,17 in][]{dacunha08}, and with an additional fourth component to describe the cold dust from grains in thermal equilibrium, modelled with a modified blackbody assuming an emissivity index $\beta$\,=\,2, and with temperature T$_{C}^{ISM}$ allowed to range between 15\,K to 25\,K. 
\end{itemize}
   
    The second step involved in SED modelling with {\it magphys} is the Bayesian approach to interpret the multiwavelength SED of a galaxy. The motivation to the Bayesian approach is to quantify the inherent degeneracy tied to the modelling technique, where for the same set of input parameters several models can be accommodated. With the Bayesian approach, model SEDs, drawn from the model library, are fit via $\chi ^{2}$--minimisation to the observations and a probability, based on the $\chi ^{2}$, is assigned to each of the fitted models and the accompanied physical parameters. Thus, a probability density function is constructed for these parameters, and the median of the resulting probability density function is adopted as the best estimate of that parameter. The 16$^{\rm th}$--84$^{\rm th}$ percentile range of the probability density function is then used to reflect the associated confidence interval. 

    Models similar to {\it magphys}, which fit the panchromatic SEDs of galaxies, are available in the literature, such as, for example, CIGALE \citep{Noll09,Serra11}. Model fits to the full multiwavelength data set can provide a consistent constraint to interpret the physical properties of a galaxy, as discussed in \citet{Hayward14}. The choice of {\it magphys} for this work is based on its wide range of application to reproduce the stellar and dust properties of several galaxy types in several environments, ranging from nearby galaxies and dwarf galaxies to galaxies in a group environment to star--forming galaxies \citep{dacunha08,dacunha10a,dacunha10b,Bitsakis11,Bitsakis14,Viaene14}. In all these cases, the Bayesian approach in {\it magphys} allows the authors to derive statistical constraints (median likelihood estimates) for a set of physical parameters, such as the star formation rate (SFR) or the mass of the dust (M$_{D}$). Likewise, and even though {\it magphys} additionally gives the best--fit model, we adopt for our study the Bayesian approach and consider the median likelihood estimates of the physical parameters of our interest, rather than the physical parameters inherent to the best--fit SED model. The physical parameters we consider in our study are: 
total dust mass, M$_{D}$; 
fraction of M$_{D}$ included in the diffuse ISM, f$\mu$; 
equilibrium temperature of cold dust in the diffuse ISM, T$_{C}^{ISM}$; 
equilibrium temperature of warm dust in the birth clouds, T$_{W}^{BC}$; 
star formation rate averaged over the last 100\,Myr, SFR;
specific SFR, sSFR, which is defined as the ratio of the SFR (averaged over the last 100\,Myr) to M$_{\star}$, where M$_{\star}$ is the stellar mass. 

   We use {\it magphys} initially to perform an investigation of whether the properties we derive when we drop the 350\,$\mu$m and 500\,$\mu$m SPIRE bands, give consistent results against the properties derived when we use these. Keeping only up to the SPIRE\,250\,$\mu$m has the meaning of gaining in spatial resolution, while the investigation has the meaning to explore possible differences due to keeping only up to SPIRE\,250\,$\mu$m. For this investigation, we re-sample the 2 sets of images described in Sec.~2 to a pixel size of 42$\arcsec$, i.e., larger than the SPIRE\,500\,$\mu$m PSF. Subsequently, we model the SEDs on a pixel--by--pixel basis, and we compare the derived physical parameters of interest. For this comparison, we use pixels with a signal--to--noise ratio (S\,/\,N) larger than 2 in the SPIRE\,500\,$\mu$m band. This results to a total of 6\,pixels with an effective area of about 126$\arcsec$\,$\times$\,126$\arcsec$, or 1764\,pc\,$\times$\,1764\,pc. This effective area covers 43 per cent the area defined by its optical B--band D$_{25}$ diameter \citep[216$\arcsec$;][]{Hunter06,Johnson12}. 

%%%%% FIGURE 6 - SEDs at Spire 500 PSF %%%%%%%%%% 
 \begin{figure*}
%  \centering
      \includegraphics[width=8cm,clip]{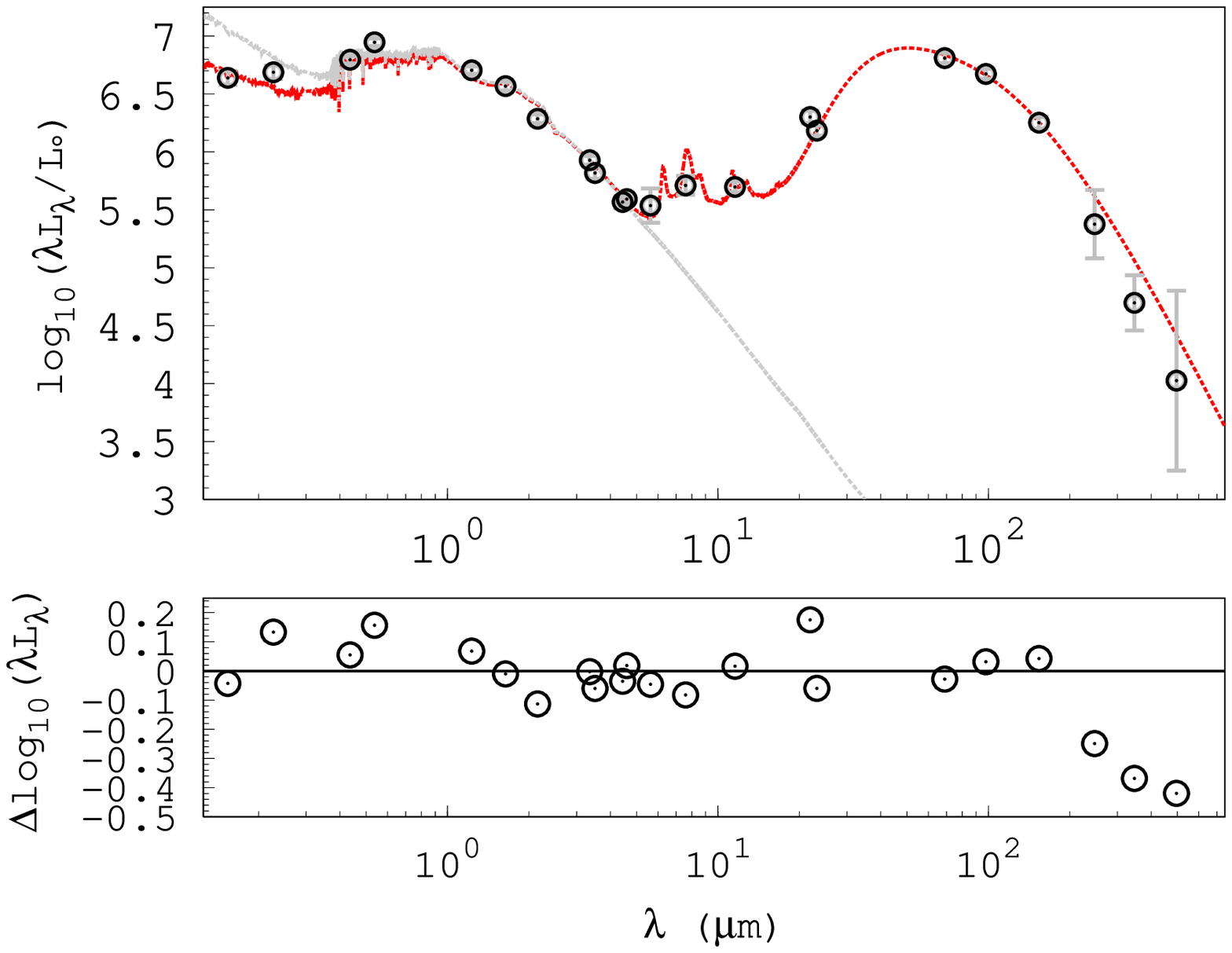}
      \includegraphics[width=8cm,clip]{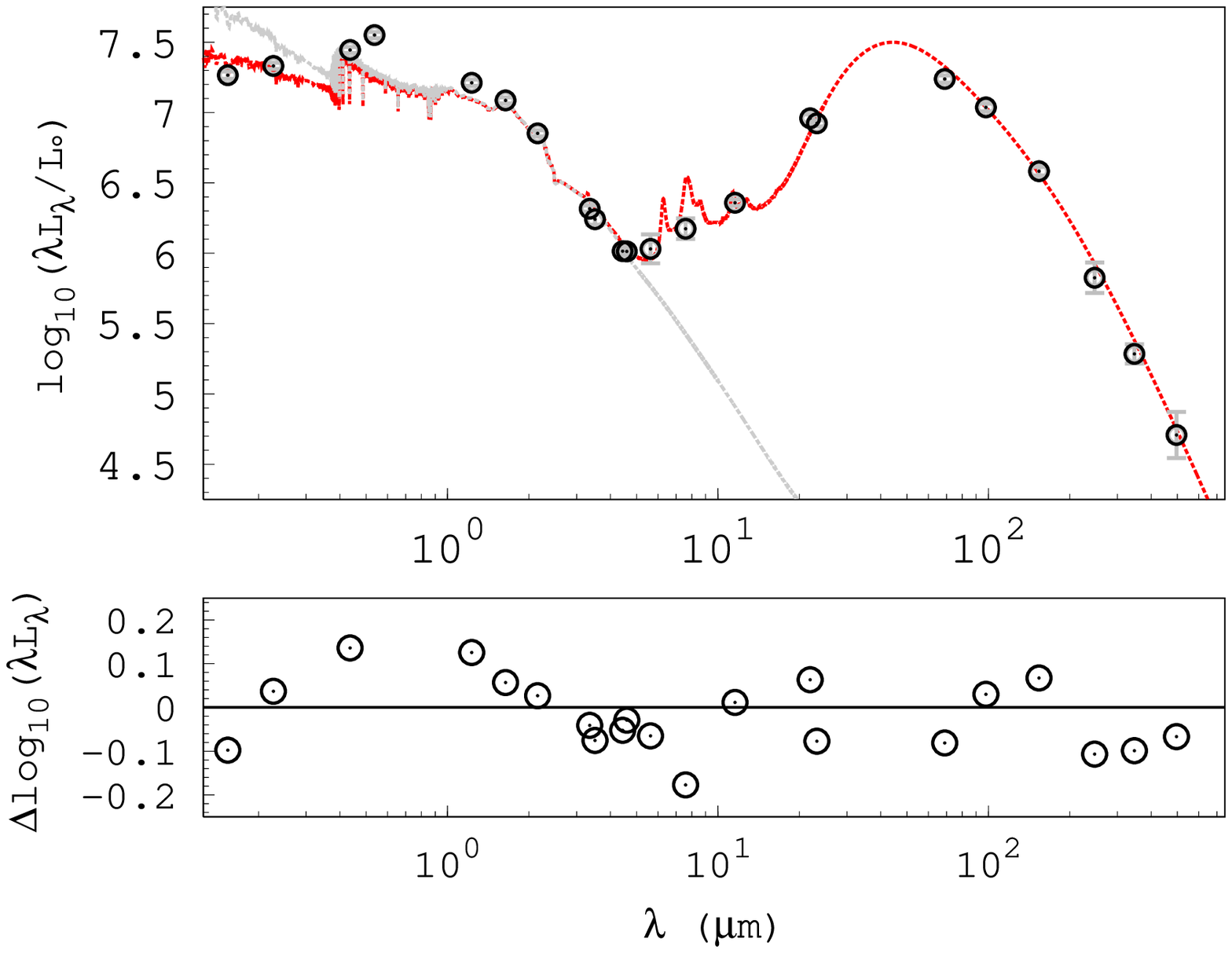}
      \includegraphics[width=8cm,clip]{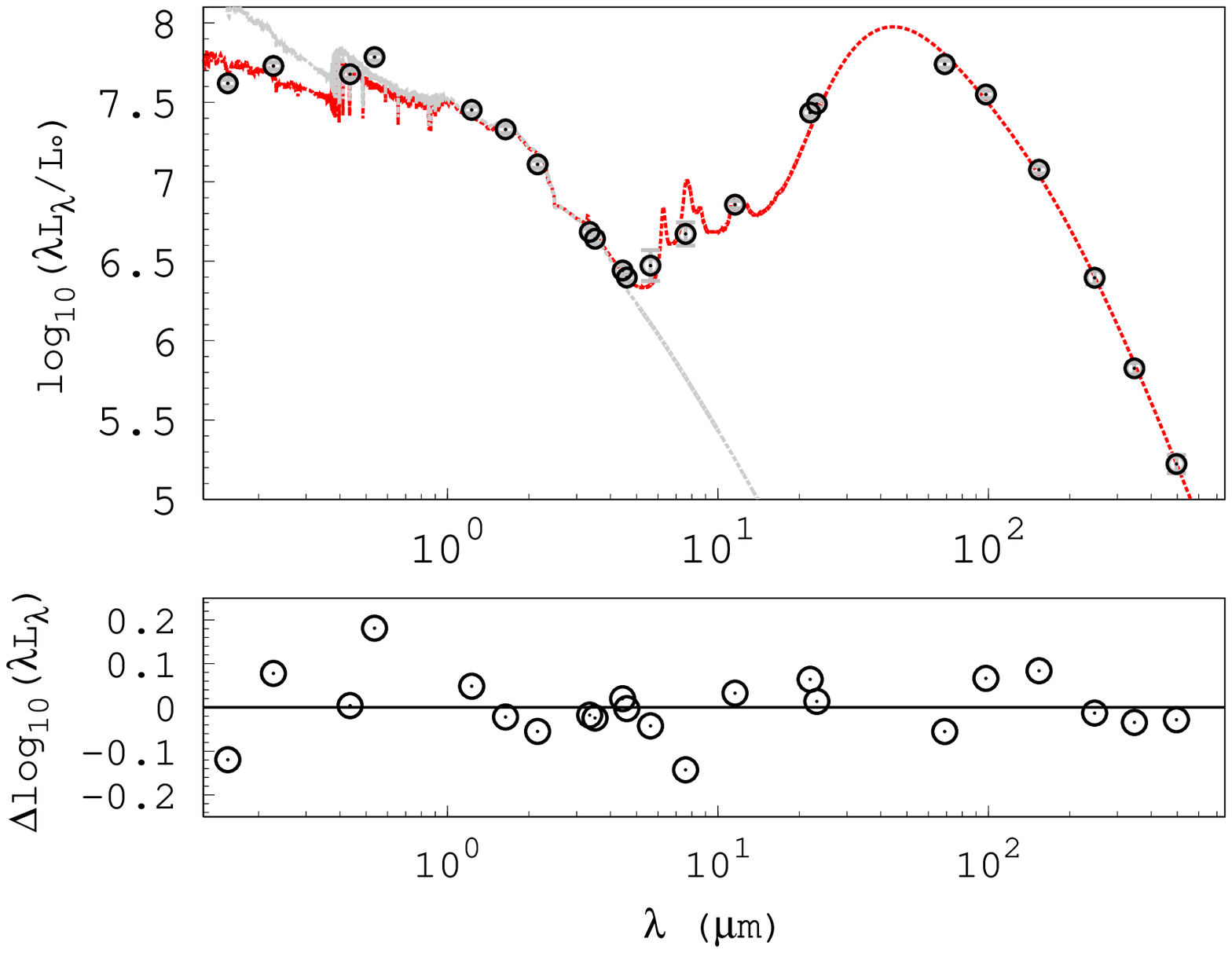}
      \includegraphics[width=8cm,clip]{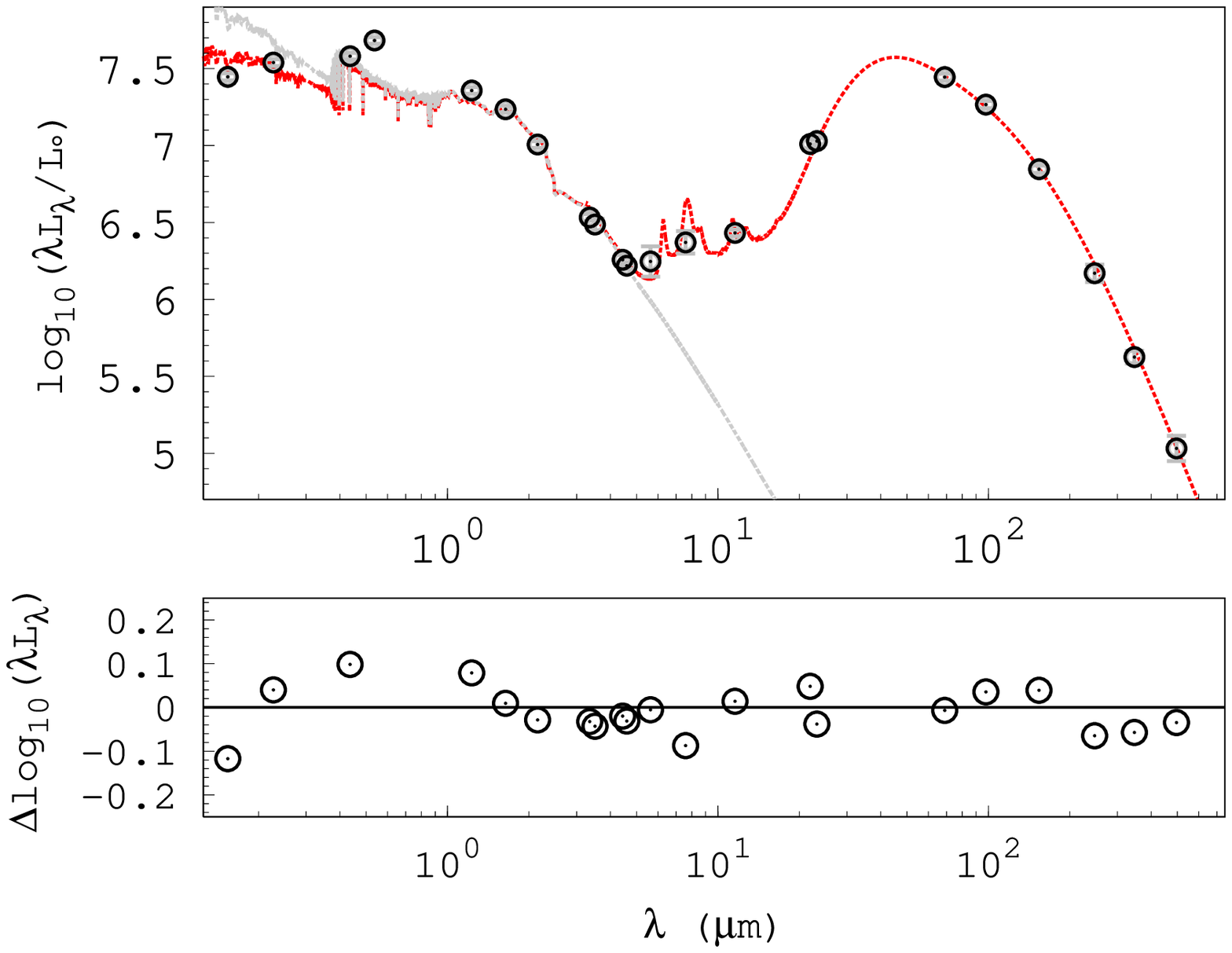}
      \includegraphics[width=8cm,clip]{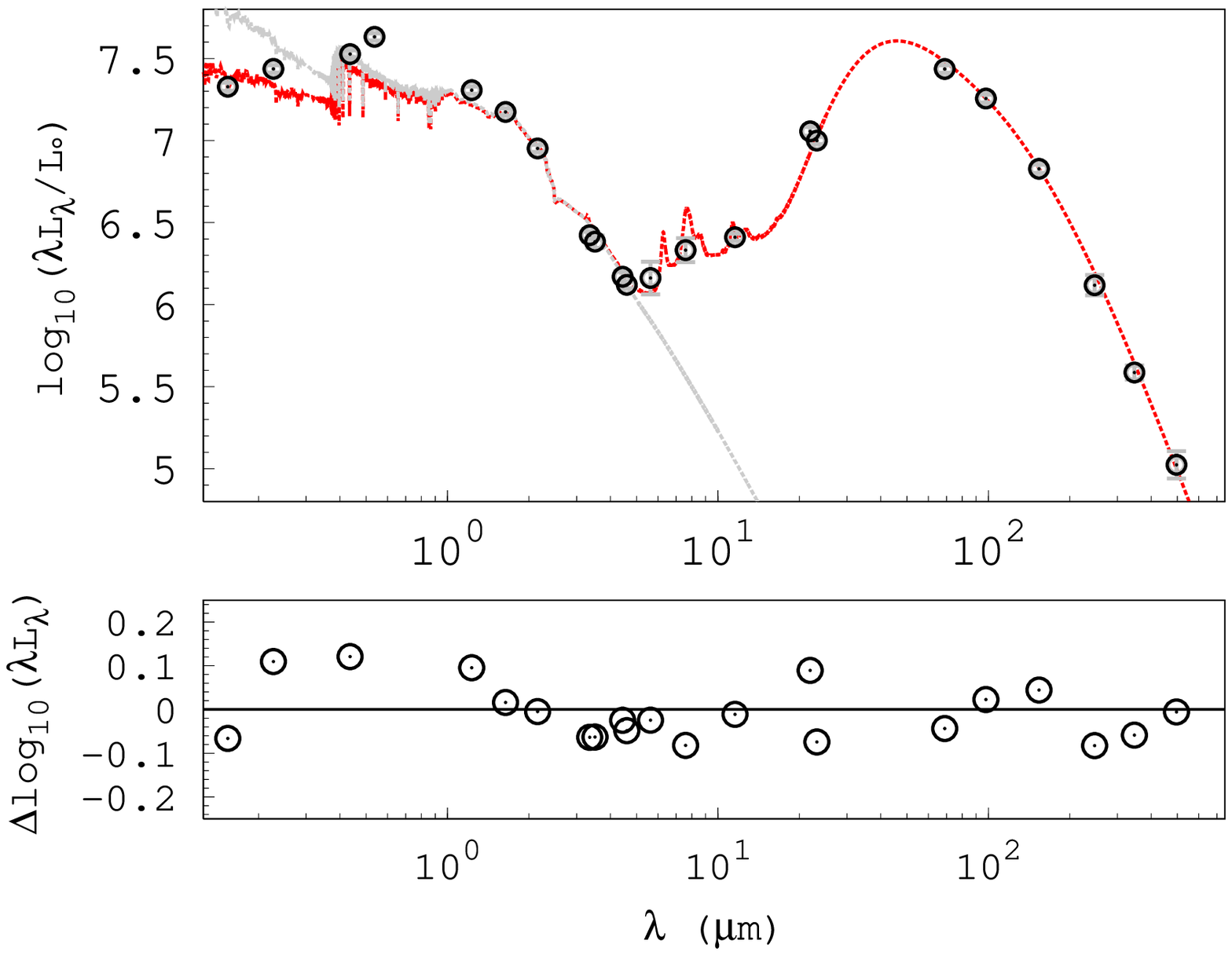}
      \includegraphics[width=8cm,clip]{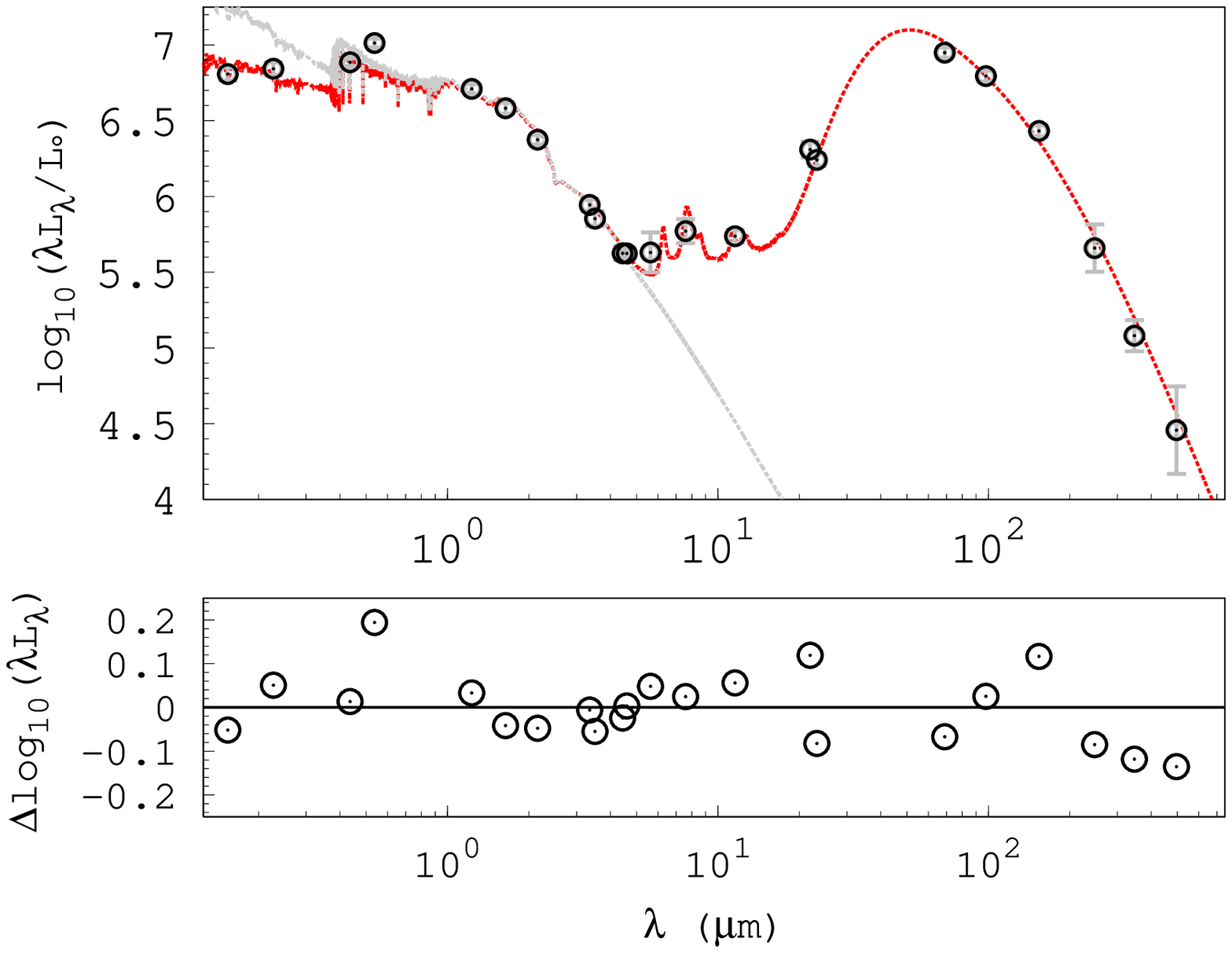}
  \caption{Best--fit model SEDs for the six pixels used to compare the two sets of convolved images. Here shown are the models using images convolved at the SPIRE\,500\,$\mu$m resolution (36.4$\arcsec$ FWHM), while the lower sub--panels show the residuals between the model and the observations. The middle left SED contains the cavity region and the SC\,10, while North is to the left and East is to the bottom.}
  \label{sl_figure6}
 \end{figure*}
%%%%%%%%%%%%%%%%%%%%%%%%%%%%%%%%%%%
%
    Fig.~\ref{sl_figure6} shows the resulting best--fit model SEDs for the six pixels used in this comparison, at the SPIRE\,500\,$\mu$m resolution. As the pixels are large (42$\arcsec$), the cavity region with the SSCS, SC\,10 and the bright north--western dust knot seen in Fig.~\ref{sl_figure1} are all contained in one pixel. The corresponding SED to this pixel is shown in the middle left panel of Fig.~\ref{sl_figure6}, and the MIR and FIR emission there is brighter compared to the rest SEDs. The MIR emission drops in regions further from the SSCs.
%%%%% FIGURE 7 - Sanity comparisons I %%%%%%%%%% 
 \begin{figure*}
%  \centering
      \includegraphics[width=8cm,clip]{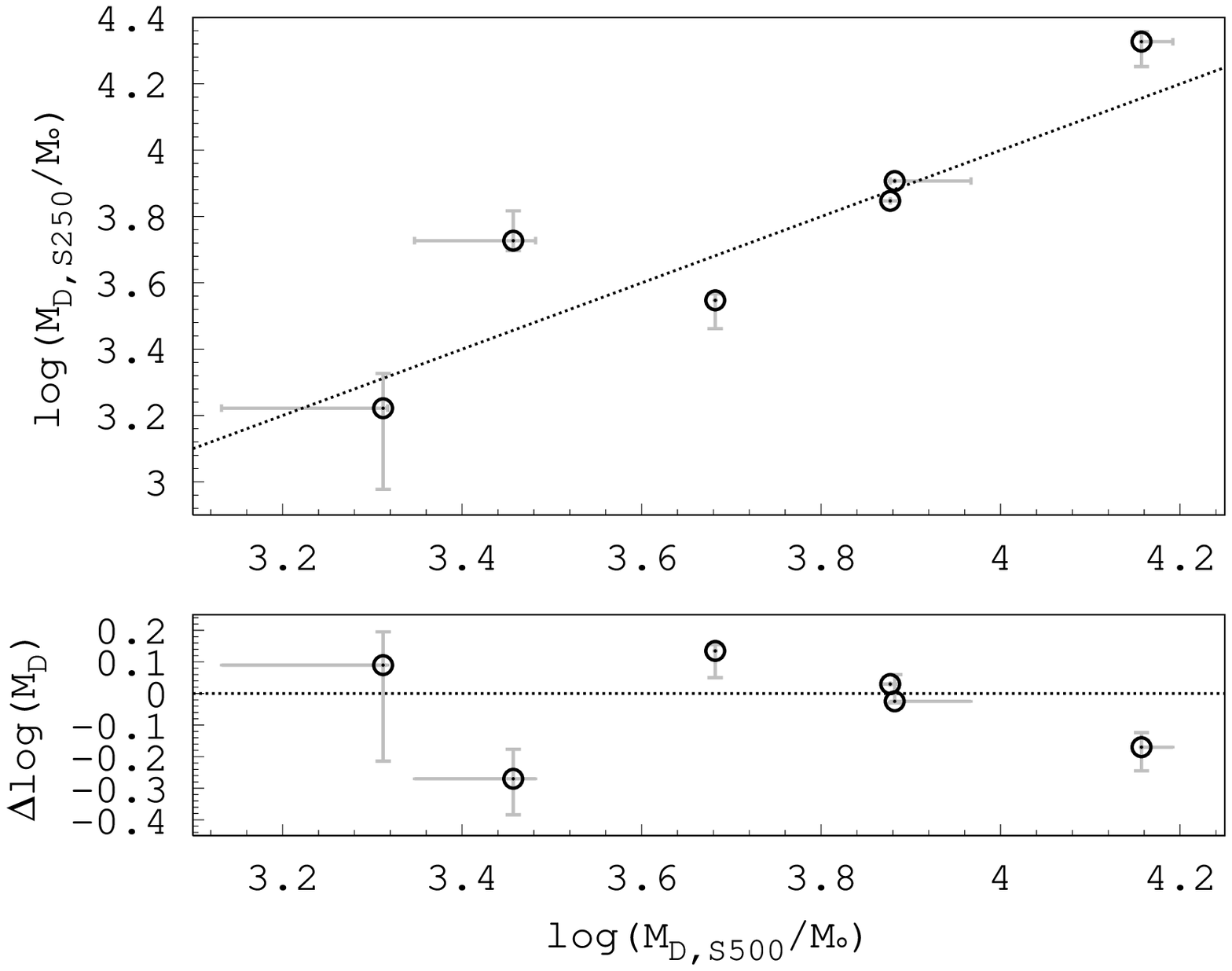}
      \includegraphics[width=8cm,clip]{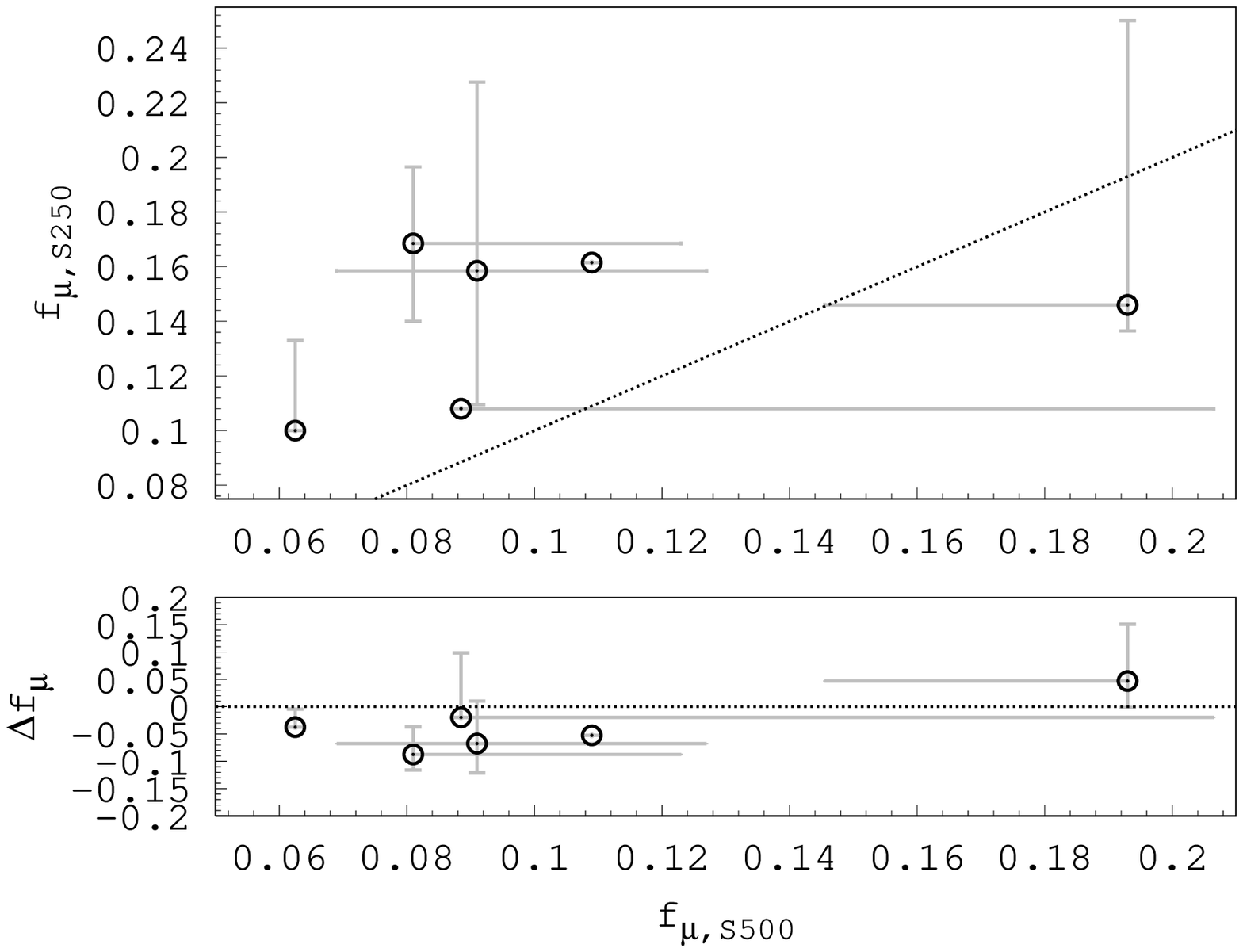}
      \includegraphics[width=8cm,clip]{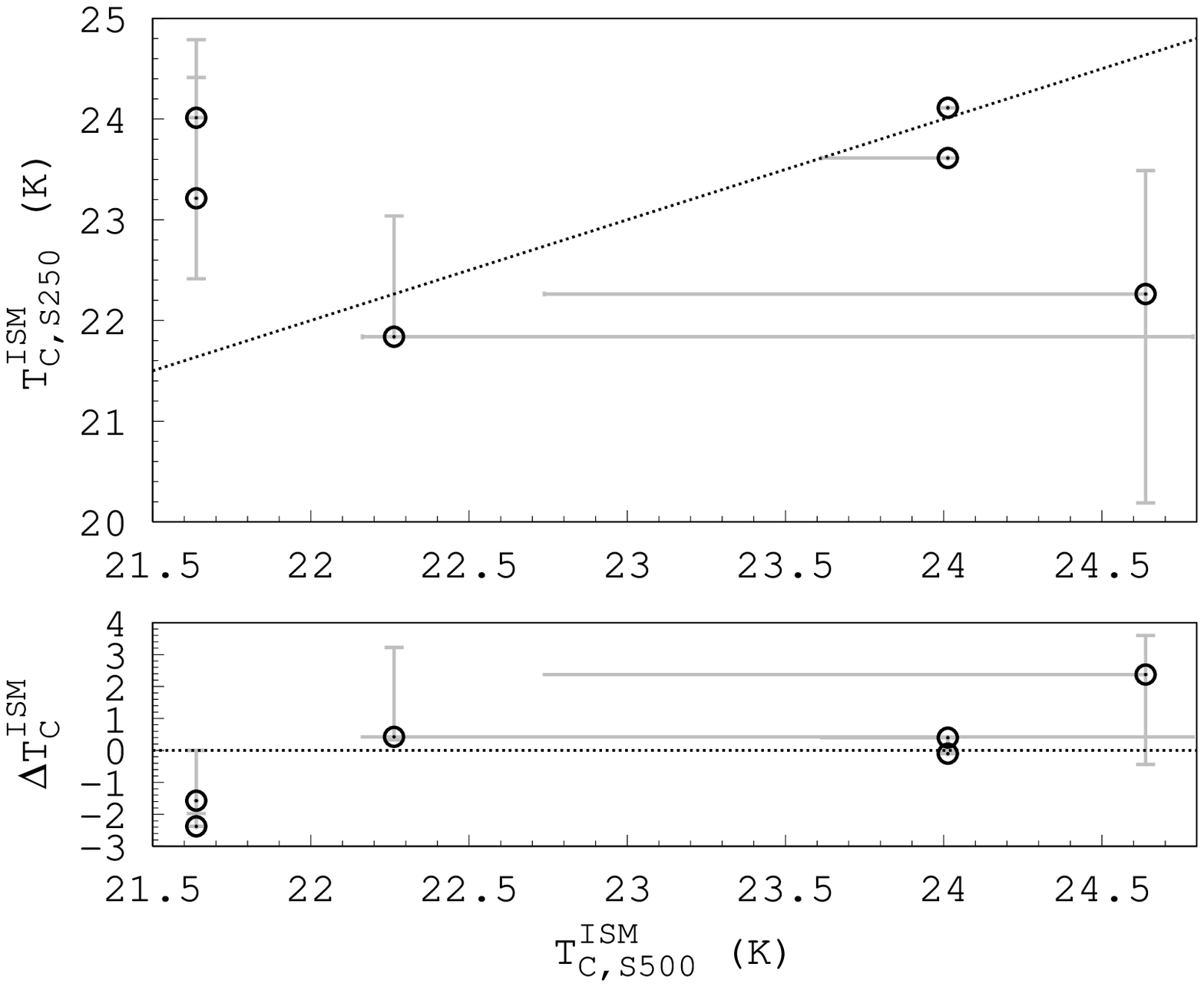}
      \includegraphics[width=8cm,clip]{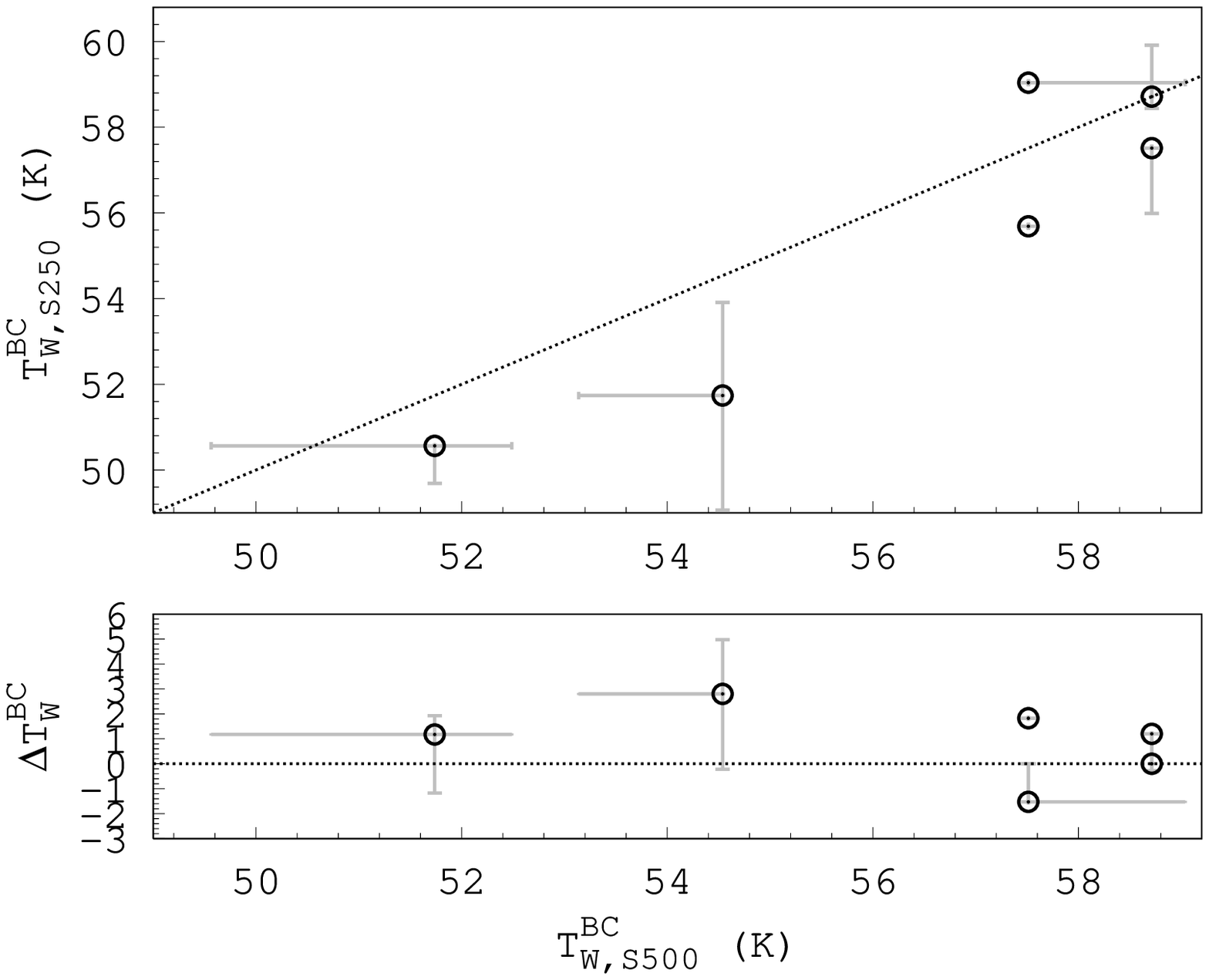} 
      \includegraphics[width=8cm,clip]{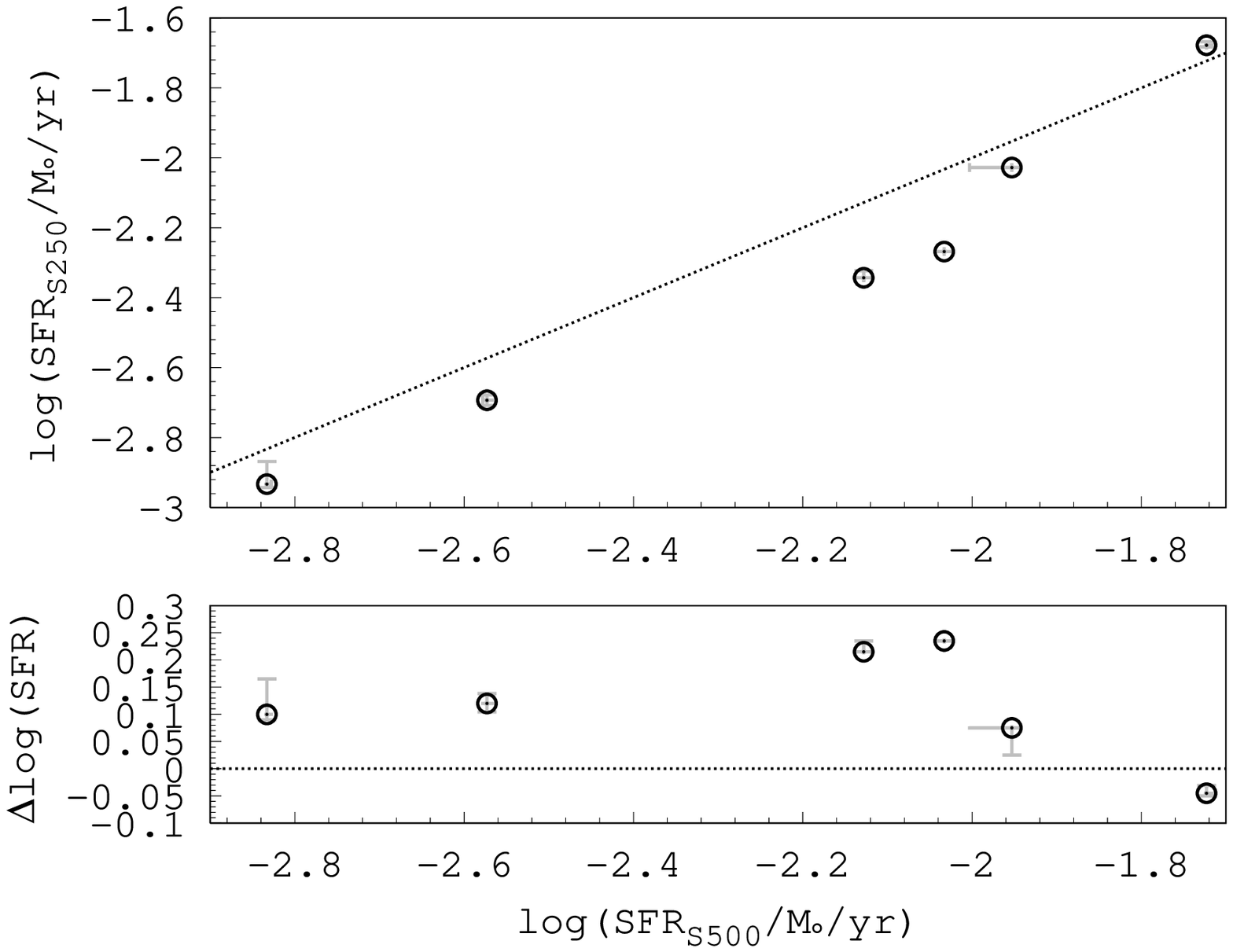}
      \includegraphics[width=8cm,clip]{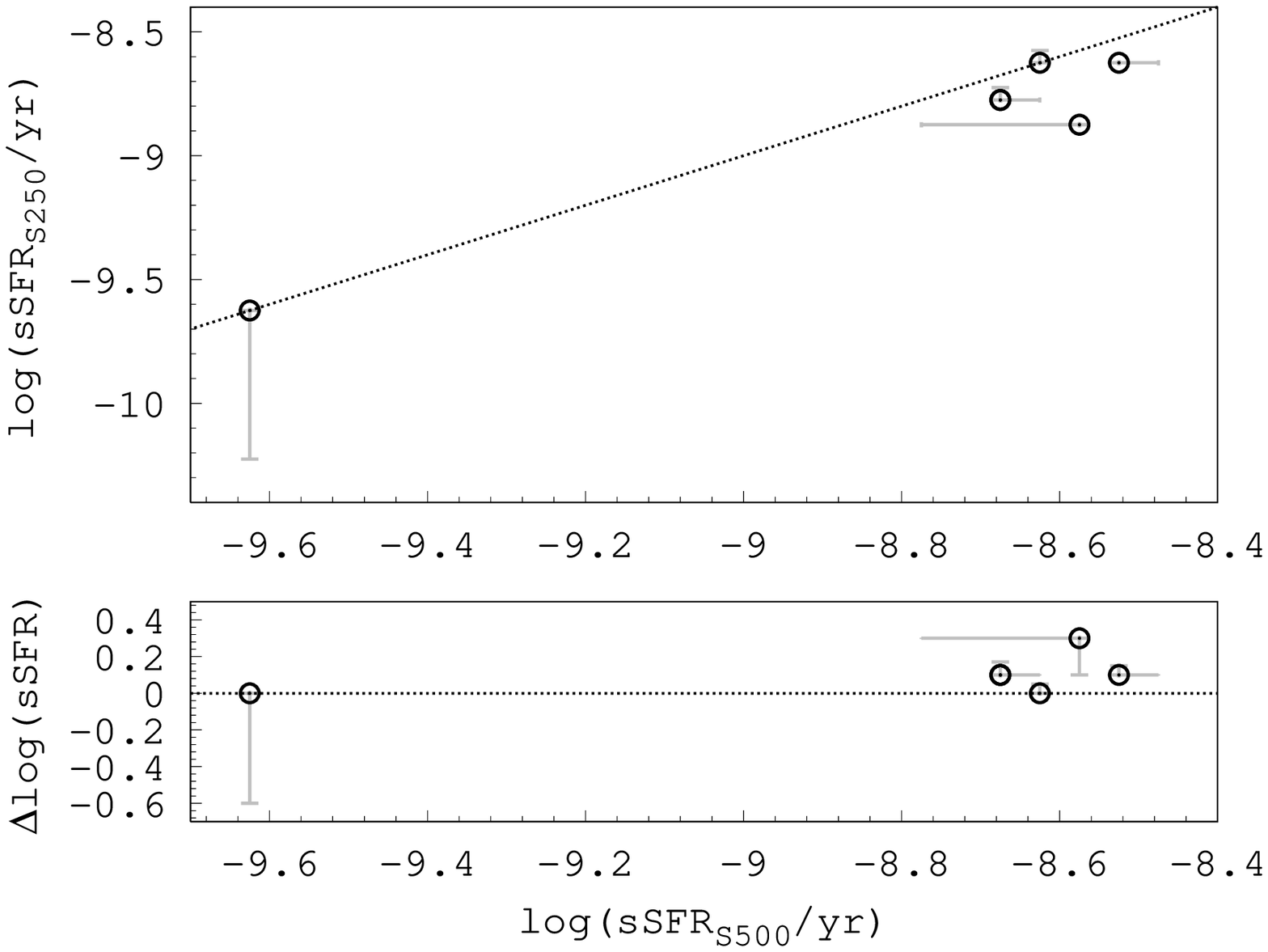}
  \caption{Physical parameters derived for the set of images convolved to the SPIRE\,500\,$\mu$m PSF (x--axis) and SPIRE\,250\,$\mu$m PSF (y--axis) and compared on a pixel--by--pixel basis with a pixel size of 42$\arcsec$. The physical parameters are the total dust mass, M$_{D}$, the fraction of the dust in the diffuse ISM versus the total dust mass, f$\mu$, the cold dust temperature in the diffuse ISM, T$_{C}^{ISM}$, the warm dust temperature in the birth clouds, T$_{W}^{BC}$, the star formation rate averaged over the last 100\,Myr, SFR, and the specific SFR, sSFR. The lower panels show the difference of the physical parameter in the y--axis (i.e., at the SPIRE\,250\,$\mu$m resolution) and in the x--axis (i.e., at the SPIRE\,500\,$\mu$m resolution), $\Delta$(Physical Parameter), as a function of the one at the x--axis (i.e., at the SPIRE\,500\,$\mu$m resolution). The dotted line in the upper sub--panels indicate unity.}
  \label{sl_figure7}
 \end{figure*}
%%%%%%%%%%%%%%%%%%%%%%%%%%%%%%%%%%%
%
  The results of the comparison between the median likelihood estimates of the physical parameters for the 2 sets of images are shown in Fig.~\ref{sl_figure7}. The error bars reflect the percentile ranges of the median likelihood estimates of the physical parameters, while the difference between these, $\Delta$(Physical Parameter), are shown in the lower sub-panels. 

   Fig.~\ref{sl_figure7} suggests overall that the properties determined with the longest wavelength constraint being SPIRE\,500\,$\mu$m versus SPIRE\,250\,$\mu$m are consistent with each other, taking into account the uncertainties. The two parameters that seem to deviate the most are the f$\mu$ and T$^{ISM}_{C}$. Decomposing the ISM into the different phases requires to take into account information from specific gas tracers, and this approach is lacking in the modelling technique used here. The standard deviation of the differences for the properties quoted above are: $\sigma_{\rm \Delta log(M_{D})}$=\,0.16\,dex; $\sigma_{\rm \Delta f_{\mu}}$=\,0.05; $\sigma_{\rm \Delta T_{C}^{ISM}}$=\,1.7\,K; $\sigma_{\rm \Delta T_{W}^{BC}}$=\,1.5\,K; $\sigma_{\rm \Delta log(SFR)}$\,=\,0.1\,dex; $\sigma_{\rm \Delta log(sSFR)}$\,=\,0.1\,dex. The consistency between results with and without the SPIRE\,350 \& 500\,$\mu$m data means that it is reasonable to use the latter for our quantitative analysis and thus gain in spatial resolution of 21$\arcsec$. The assumption of quiescent cold dust clumps not being taken into account in this modelling technique also points to data longwards to SPIRE\,250\,$\mu$m not needed. Moreover, a pixel size of 21$\arcsec$ points to the fact that within our pixels there is averaging of several different components of the ISM. For example, with a size for a giant molecular cloud of 100\,pc, any cold ISM component will be well mixed with warmer regions. This means that our results may indicate warmer temperatures and smaller dust masses, as discussed in Sec.~\ref{sec:g2d} and Sec.~\ref{sec:regions}.
%______________________________________________________________

\section{Properties on a pixel--by--pixel basis}
\label{sec:pixels}

   In the pixel--by--pixel analysis, we focus on the parameters of M$_{D}$, f$\mu$, T$_{C}^{ISM}$, T$_{W}^{BC}$, SFR, and sSFR to see how these vary across the central starburst of NGC\,1569. With a pixel size of 21\,$\arcsec$ in the maps, the spatial scales probed are 294\,pc on a side, i.e. larger than the size of the cavity (198\,pc in diameter). The results obtained in this section are not used in subsequent sections to derive any new properties, and they are only used for comparison reasons. 
%%%%% FIGURE 8 - Pixel--by--pixel maps at SPIRE 250 resolution %%%%%%%%%% 
 \begin{figure*}
%  \centering
      \includegraphics[width=8cm,clip]{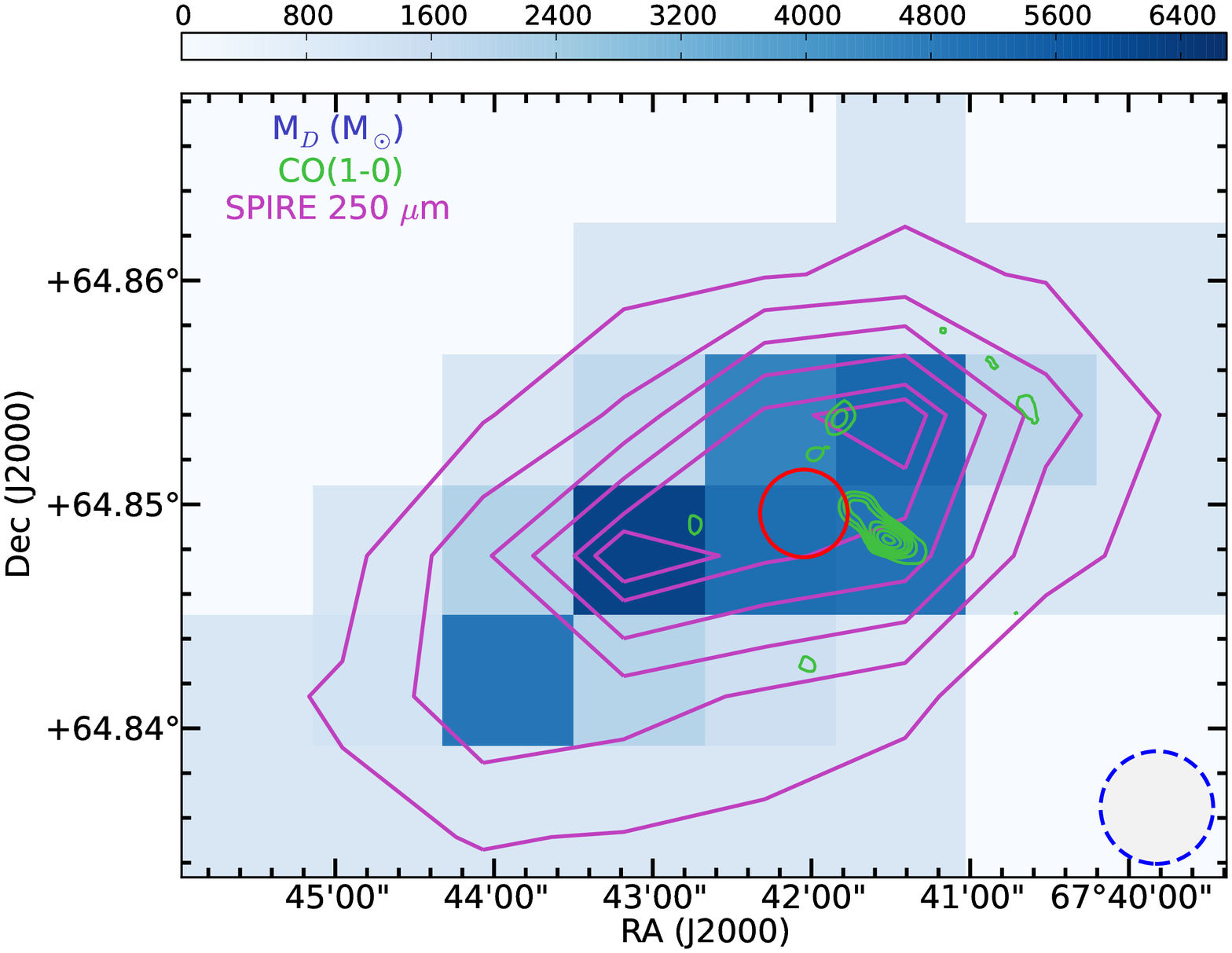}
      \includegraphics[width=8cm,clip]{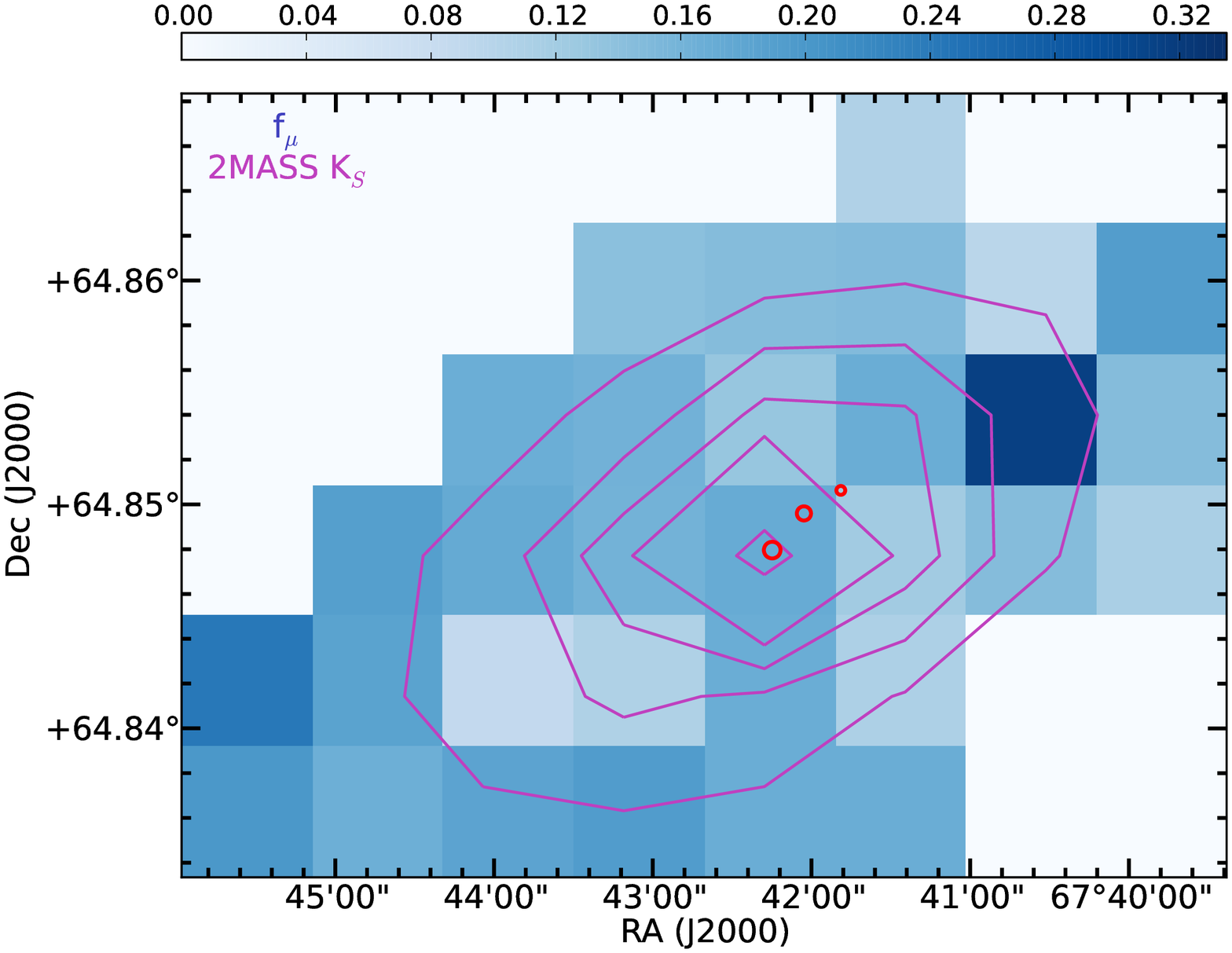}
      \includegraphics[width=8cm,clip]{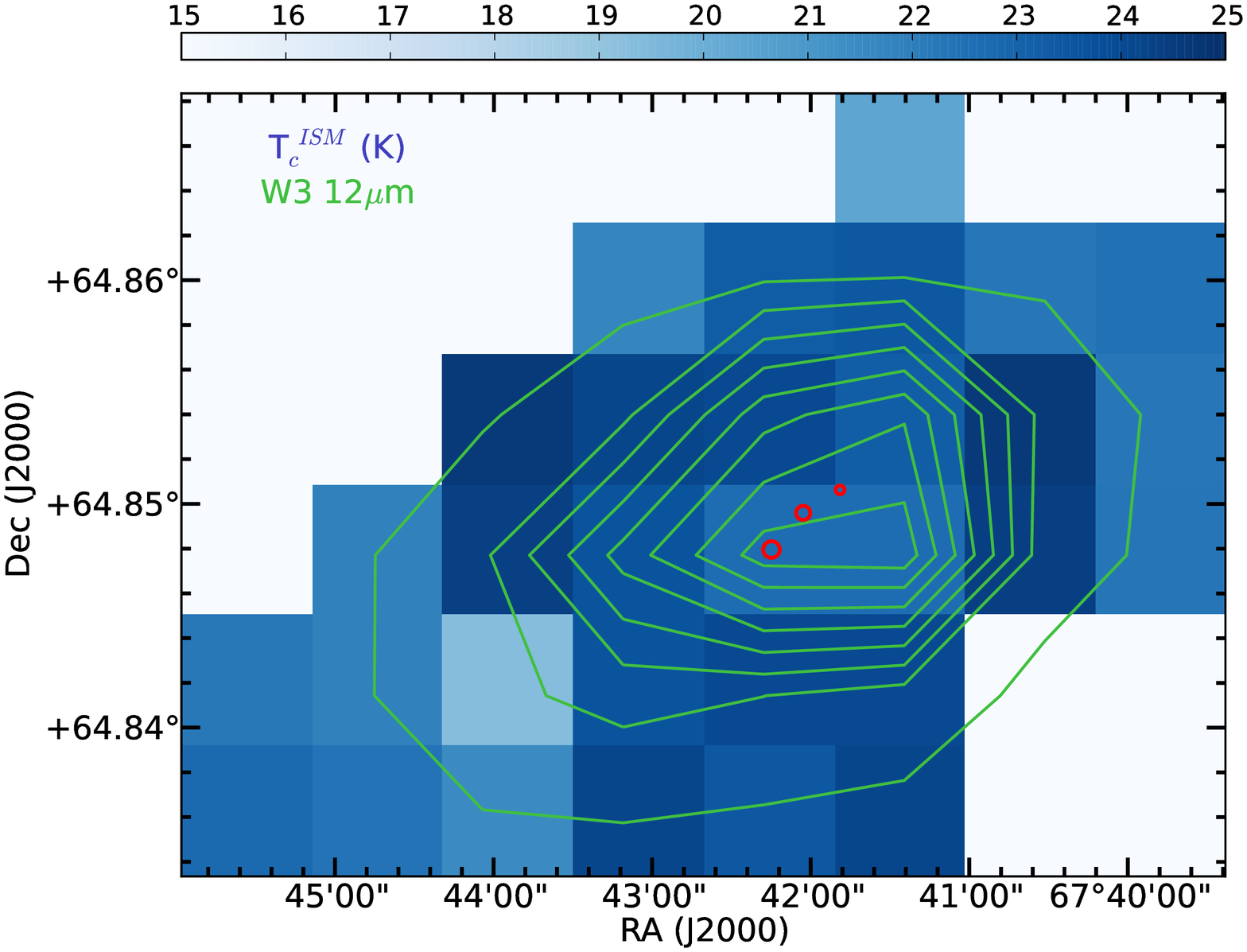}
      \includegraphics[width=8cm,clip]{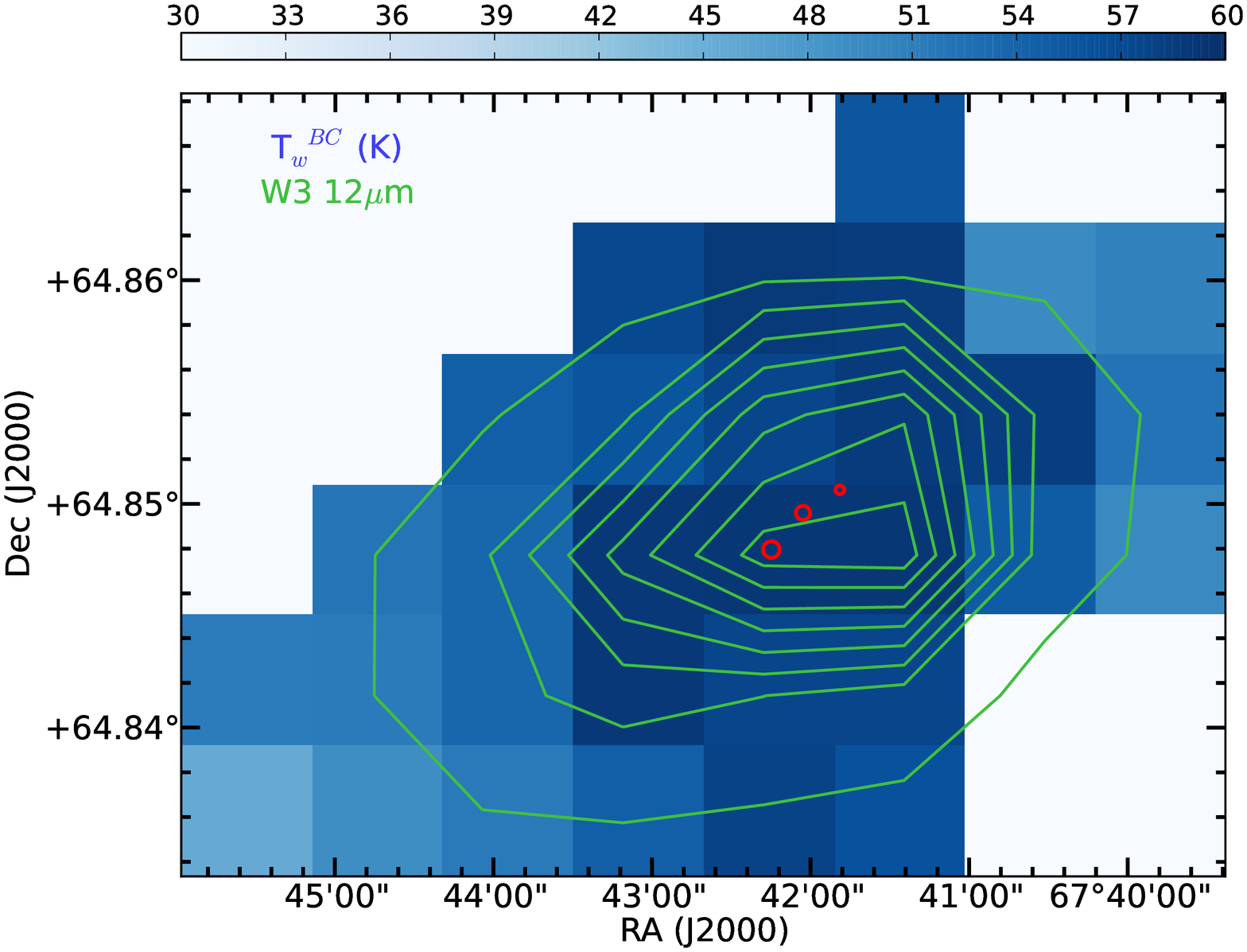}
      \includegraphics[width=8cm,clip]{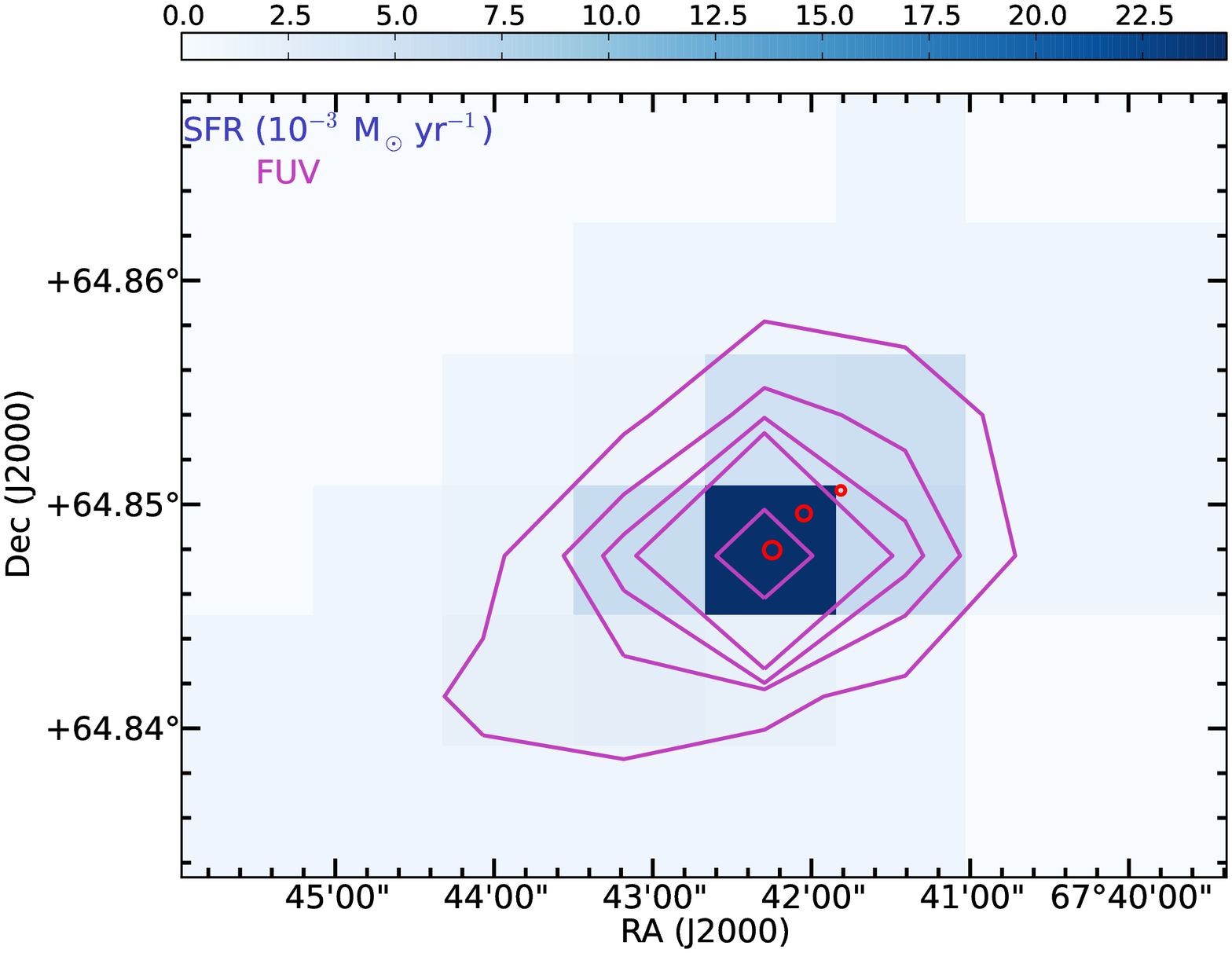}
      \includegraphics[width=8cm,clip]{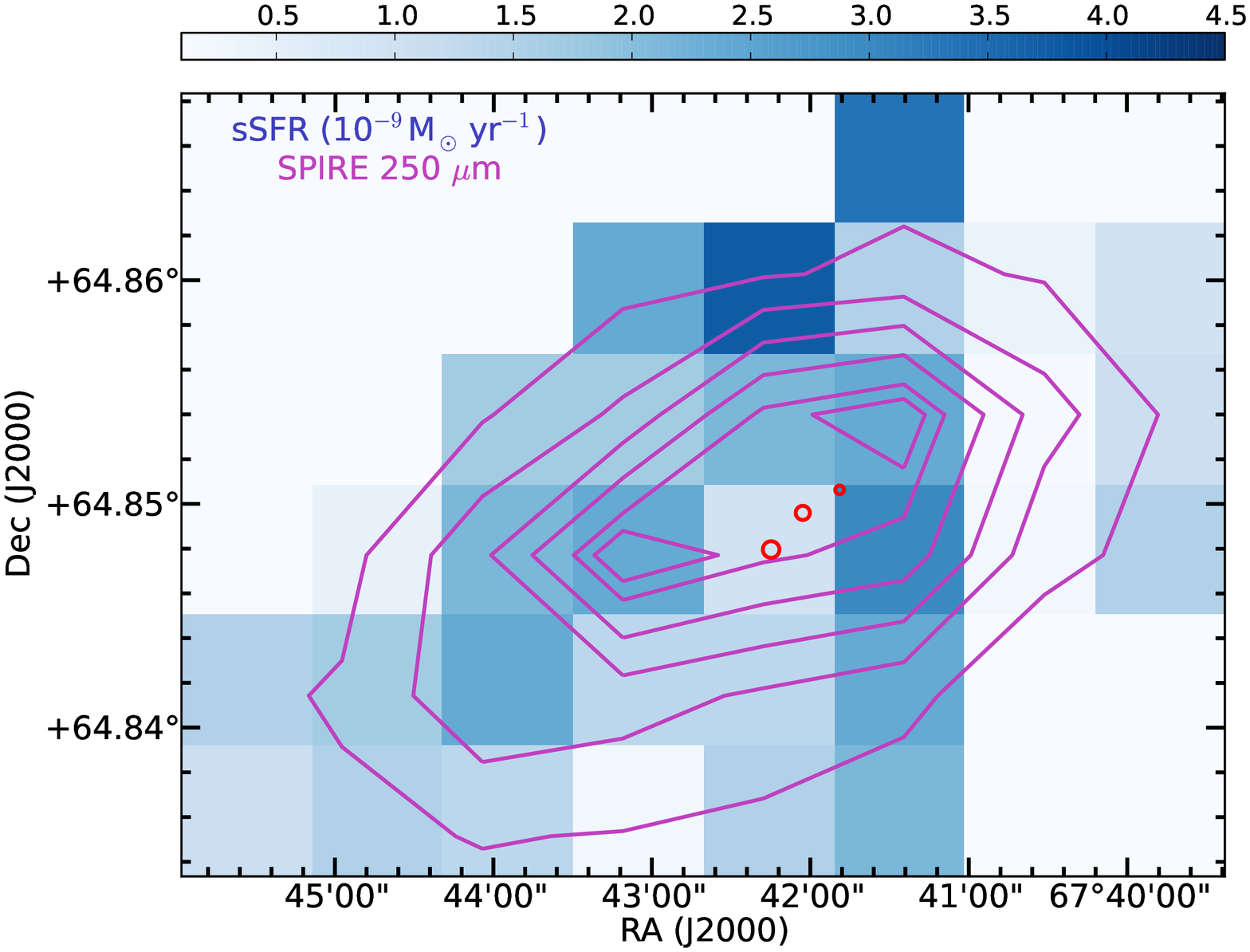}
  \caption{M$_{D}$, f$\mu$, T$_{C}^{ISM}$, T$_{W}^{BC}$, SFR, and sSFR maps at the SPIRE 250\,$\mu$m resolution (18.2$\arcsec$ FWHM) and pixel scale 21$\arcsec$. Pixels with zero values have S\,/\,N ratio lower than the imposed cutoff. The contour levels are 1\,Jy to 0.1\,Jy with a step of 0.2\,Jy for SPIRE\,250$\mu$m, [100, 55, 40, 25, 10]\,mJy for K$_{S}$, and [90, 70, 35, 12.5, 8, 6, 3]\,mJy for FUV. The red solid circle in the M$_{D}$ map shows the cavity, while in the remaining maps the small red circles show the location of the SSCs and SC\,10. The blue dashed circle in the M$_{D}$ map shows the size of the SPIRE 250\,$\mu$m PSF.}
  \label{sl_figure8}
 \end{figure*}
%%%%%%%%%%%%%%%%%%%%%%%%%%%%%%%%%%%
%
The results on these parameters are shown in the maps of Fig.~\ref{sl_figure8} and correspond to the median likelihood estimates for each parameter. The area of NGC\,1569 covered in these maps is roughly 168$\arcsec \times$126$\arcsec$, while the true area is exactly 1.4$\times$10$^{4}$\,sq.\,arcsec (or 2.7\,Mpc$^{2}$; a total of 31\,pixels), and is larger than the one shown in Fig.~\ref{sl_figure2}. The area of NGC\,1569 covered in Fig.~\ref{sl_figure8} corresponds to 37 per cent of its total area, which is 3.7$\times$10$^{4}$\,sq.\,arcsec (or 7.3\,Mpc$^{2}$) assuming a D$_{25}$ diameter of 216$\arcsec$. We are again restricted to pixels with S\,/\,N higher than 2 in the SPIRE\,250\,$\mu$m band. We discard pixels with S\,/\,N lower than that, and we flag these pixels' properties with zero. 

   The upper left panel of Fig.~\ref{sl_figure8} shows M$_{D}$, which ranges from 6.2$^{+0.6}_{-1.2}$ to 1.0$^{+0.3}_{-0.3}$\,$\times$10\,$^{3}$\,M$_{\odot}$. The lower value, 1.0$^{+0.3}_{-0.3}$\,$\times$10\,$^{3}$\,M$_{\odot}$, in the outskirts corresponds to an upper limit, due to binning effects in the recorded probability density function of this parameter in {\it magphys}. The seemingly homogeneous distribution across the outer part of the galaxy's mapped area hides the true variation in this parameter. At the central part of the galaxy, M$_{D}$ fluctuates to higher values, with high amounts found in the periphery of the cavity, to the North--West and East from SSC\,A, and coinciding with the dust knots seen in the PACS and SPIRE bands. Given the large angular size of the pixels, the pixel west and northwest to the SSC\,A also includes the molecular clouds that are located in the periphery of the cavity \citep[see upper left panel of Fig.~\ref{sl_figure8}; CO contours adapted from][]{Taylor99}. On the other hand, the pixels to the East of SSC\,A contain several of the \mbox{H\,{\sc ii}} regions studied in \citet[][see for example their Fig.~3b]{Waller91}. Thus, the higher values of the M$_{D}$ there reflect the embedded and ongoing star formation occurring in the molecular clouds and \mbox{H\,{\sc ii}} regions \citep{Waller91,Taylor99,Tokura06}. The pixel that contains SSC\,A and SSC\,B has a statistically lower value (M$_{D}$\,=\,5.1$^{+0.3}_{-0.6}$\,$\times$10\,$^{3}$\,M$_{\odot}$) than its star-forming vicinity, as already noted in the earlier qualitative analysis of Sec.~\ref{sec:morphology}. This pixel probes the part of the cavity seen in both warm ionised and atomic hydrogen observations, but several other cavities in the ISM of smaller scale have been uncovered in the central part of the galaxy \citep[e.g.,][]{Pasquali11}. These smaller scale cavities are smoothed out due to the relatively large angular resolution. The dust content around the SSCs and in the cavity is the subject of Sec.~\ref{sec:regions}.

   The total integrated M$_{D}$ obtained from the dust mass map is 60.4$^{+1.2}_{-3.5}$\,$\times$\,10$^{3}$\,M$_{\odot}$. The M$_{D}$ value is about an order of magnitude less than derived in \citet[][M$_{D}$=\,2.8$^{+1.5}_{-1.4} \times$10$^{5}$\,M$_{\odot}$]{Remy13}, who consider modified blackbody fits to the dust SED in a circular aperture radius of 150$\arcsec$, and than derived in \citet[][M$_{D}$=\,(1.6--3.4)$\times$10$^{5}$\,M$_{\odot}$]{Galliano03}, who consider a detailed dust model for the galaxy within a radius of 0.65\,kpc. The area of the galaxy used to model the SEDs between this work and the works of \citet{Remy13} and \citet{Galliano03} is different, thus the derived M$_{D}$ refer to different regions of the galaxy, to begin with. Dividing our dust mass by the area probed, i.e. 1.4$\times$10$^{4}$\,sq.\,arcsec, we derive 2.3$^{+1.2}_{-0.7}$\,$\times$\,10$^{4}$\,M$_{\odot}$\,kpc$^{-2}$. Normalising the dust mass found in \citet{Remy13} by their considered area, i.e. 7.1$\times$10$^{4}$\,sq.\,arcsec, we derive 2.0$^{+1.1}_{-1.0}$\,$\times$\,10$^{4}$\,M$_{\odot}$\,kpc$^{-2}$. The two dust mass density values are consistent within the uncertainties. As NGC\,1569 has extended emission in the warm and cold ISM components, averaging the dust mass over the area probed is meaningful out to distances where such an extended emission is seen. Fig.~\ref{sl_figure4} implies that the extended emission goes beyond 180$\arcsec$. 

   Comparing with the results of \citet{Galliano03}, we normalise their dust mass by their considered area, i.e. 1.5\,kpc$^{2}$, to derive a dust mass density of (1.1--2.3)\,$\times$\,10$^{5}$\,M$_{\odot}$\,kpc$^{-2}$. This is a larger dust mass per unit area than what we derive here, larger by a factor of $\sim$7. The difference in the two dust mass density values can be further understood as due to the differences in the modelling techniques between these works. Indeed, \citet{Galliano03} develop a detailed dust model to derive the dust grain properties, such as grain size distribution, and dust masses. Their modelling assumptions and approach are different from the ones in {\it magphys}, which give rise to differences in the derived dust masses. \citet{Galliano03} include a very cold grain component in their model, in order to fit the excess emission in the millimetre wavelength range. They model this very cold grain component with a modified blackbody with fixed $\beta$ of 1, and they find that 40 per cent to 70 per cent of the cold dust mass resides in this very cold grain component.

   The source of heating for the diffuse ISM is the old and intermediate age stars with ages larger than 10\,Myr. The upper right panel of Fig.~\ref{sl_figure8} shows the distribution of f$_{\mu}$ in the central part of NGC\,1569, which ranges from 0.31$\pm$0.01 to 0.09$^{+0.02}_{-0.01}$. The spatial distribution of f$_{\mu}$ is consistent with a homogeneous distribution. We would expect the diffuse ISM in the very central starburst to be negligible as compared to the denser phases in the same region. The pixel that contains the two SSCs has f$_{\mu}$ of 0.17$^{+0.01}_{-0.01}$. The overlaid K$_{S}$ contours show a homogeneous distribution of the stellar mass in the galaxy, as also found for the distribution of the intermediate--age (50\,Myr to 1\,Gyr) and old resolved stellar populations \citep[with ages larger than 1\,Gyr;][]{Aloisi01}. 
%%%%% FIGURE 9 - Modelled versus observed correlations at SPIRE 250 resolution %%%%%%%%%% 
 \begin{figure}
%  \centering
      \includegraphics[width=8.5cm,clip]{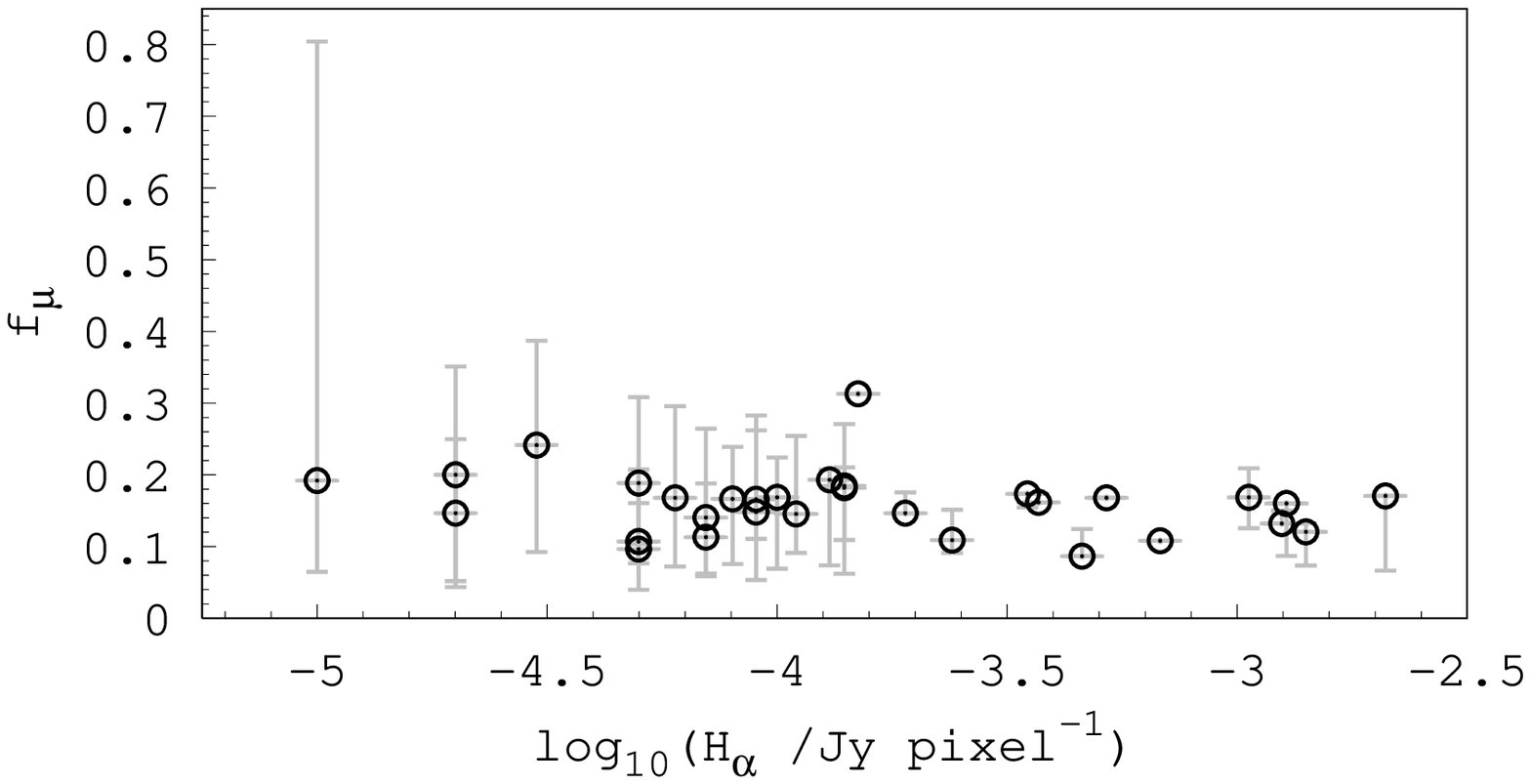}
      \includegraphics[width=8.5cm,clip]{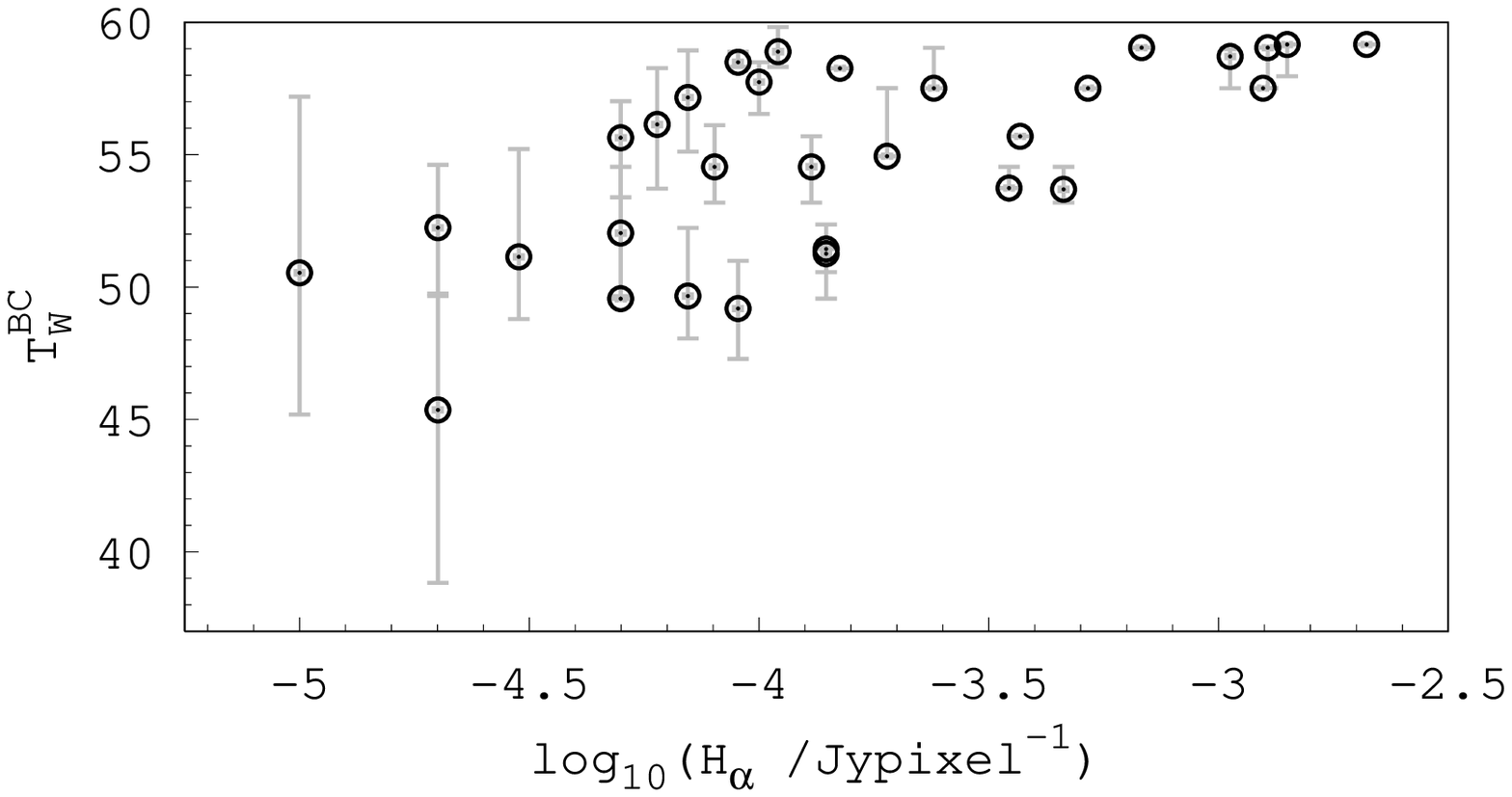}
      \includegraphics[width=8.5cm,clip]{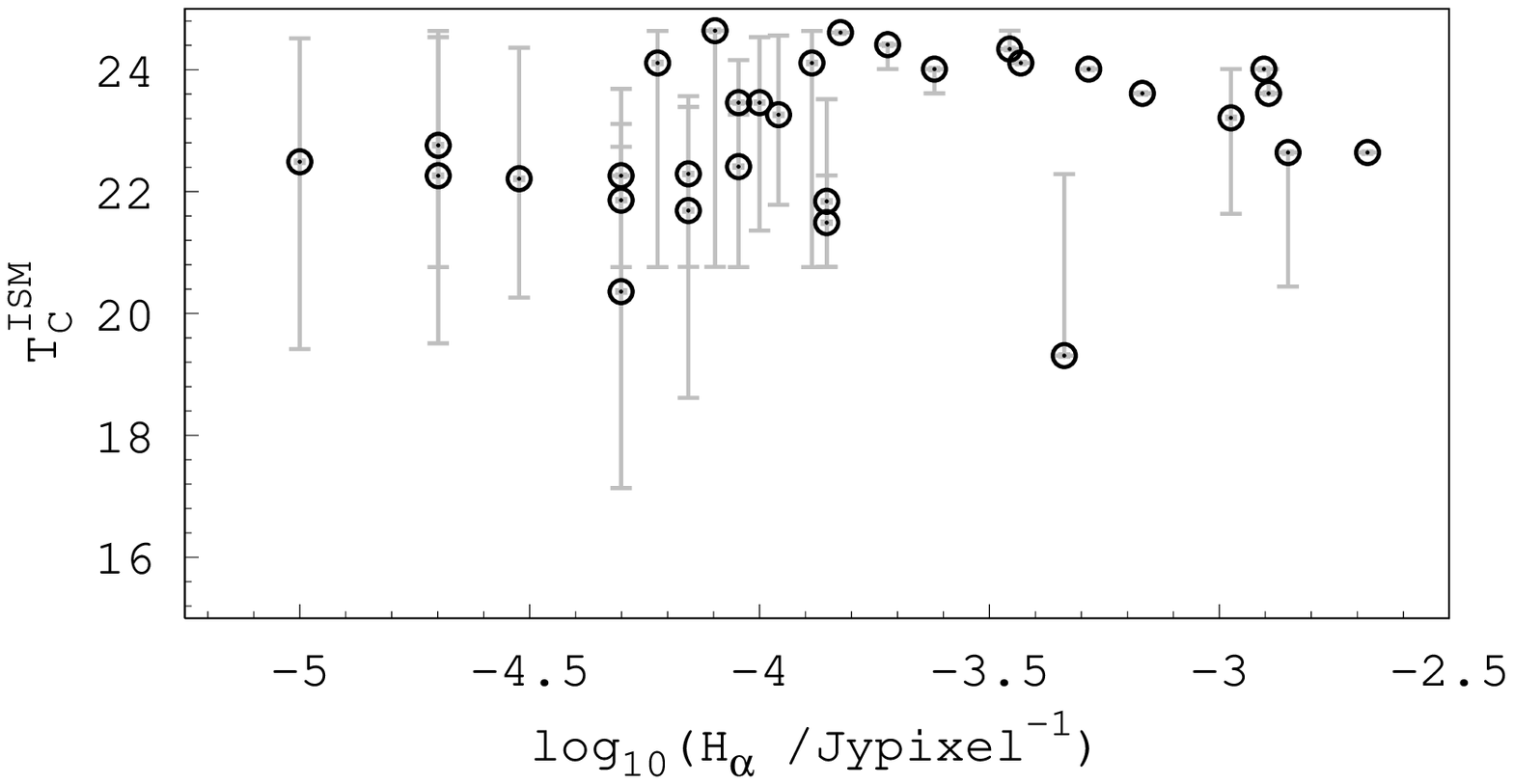}
  \caption{Derived properties with {\it magphys} versus observed properties.}
  \label{sl_figure9}
 \end{figure}
%%%%%%%%%%%%%%%%%%%%%%%%%%%%%%%%%%%
%
Fig.~\ref{sl_figure9} shows f$_{\mu}$ as a function of the H\,$\alpha$ flux density, per pixel. The two quantities show no correlation. As stars younger than 10\,Myr contribute to the H\,$\alpha$ emission, the lack of correlation may indicate that the diffuse ISM is heated by stars with ages larger than those traced with H\,$\alpha$. Alternatively, the spatial distribution of stars with ages younger than 10\,Myr is different than those with older ages, consistent with the findings of \citet{Aloisi01}. 

    \citet{dacunha08} consider in their model warm and cold grains in thermal equilibrium. The warm grains in thermal equilibrium reside in the diffuse ISM, characterised by a temperature T$_{W}^{ISM}$, and in the birth clouds, characterised by a temperature T$_{W}^{BC}$. The cold grains in thermal equilibrium reside in the diffuse ISM and are characterised by a temperature T$_{C}^{ISM}$. \citet{dacunha08} fix T$_{W}^{ISM}$ to 45\,K, and adopt an emissivity index $\beta$\,=\,1.5 for the warm grains in the birth clouds, and $\beta$\,=\,2 for the cold dust in the diffuse ISM. The thermal equilibrium temperatures are shown in the middle left and right panels of Fig.~\ref{sl_figure8}, for the cold dust in the diffuse ISM and for the warm dust temperature in the birth clouds, respectively. These parameters are allowed to vary between 15\,K to 25\,K for the cold dust, and between 30\,K to 60\,K for the warm dust in the birth clouds. The range mapped in T$_{C}$ is from 19.3$^{+3.0}_{-1.0}$\,K to 24.6$^{+0.4}_{-3.9}$\,K, while the T$_{W}$ map ranges from 45.4$^{+4.4}_{-6.5}$\,K to 59.2$^{+0.6}_{-1.2}$\,K. The lowest T$_{C}$ is located roughly southeast of the cavity, where dense \mbox{H\,{\sc i}} clouds are also found \citep[][their Fig.~18]{Johnson12}. The PAH--sensitive emission, indicated with the W3 12$\mu$m contours, shows that brighter emission is associated with colder equilibrium temperature in the diffuse ISM and warmer equilibrium temperature in the BCs. Fig.~\ref{sl_figure9} shows T$_{W}^{BC}$ and T$_{C}^{ISM}$ as a function of the H\,$\alpha$ flux density per pixel. The pixels with warmer dust in the birth clouds have higher H\,$\alpha$ emission, while the pixels with colder dust show no correlation with stars younger than 10\,Myr.

   The SFR in {\it magphys} is derived from the stellar population synthesis component of the SED, and is averaged over the past 100\,Myr. Thus, it is tied to the stellar emission directly, rather than the dust emission. The SFR, shown in the lower left panel of Fig.~\ref{sl_figure8} ranges from 1.0$^{+0.2}_{-0.1}$\,$\times$10\,$^{-3}$ to 24.4$^{+1.0}_{-11.0}$\,$\times$10\,$^{-3}$\,M$_{\odot}$\,yr$^{-1}$. 
The lower value, 1.0$^{+0.2}_{-0.1}$\,$\times$10\,$^{-3}$\,M$_{\odot}$\,yr$^{-1}$, in the outskirts corresponds to an upper limit, due to binning effects in the recorded probability density function of this parameter in {\it magphys}. The highest SFR is found in the pixel that contains SSC\,A, which reflects the massive star formation that occurs in the form of the SSCs. The integrated SFR in the central part of the galaxy is 76.8$^{+0.4}_{-0.1}$\,$\times$10\,$^{-3}$\,M$_{\odot}$\,yr$^{-1}$. \citet{Hunter10} find a SFR$_{FUV}$ of 13.8\,($\pm$0.0)\,$\times$10\,$^{-2}$\,M$_{\odot}$\,yr$^{-1}$, and a SFR$_{H\alpha}$ of 12.6\,($\pm$0.0)\,$\times$10\,$^{-2}$\,M$_{\odot}$\,yr$^{-1}$, while \citet[][their Table 1]{McQuinn12} find a SFR$_{SFH}$ of 76$\pm$2\,$\times$10\,$^{-3}$\,M$_{\odot}$\,yr$^{-1}$. All these results compare well with our derived value, which is limited to the very central region of the galaxy.

%%%%% FIGURE 10 - SFR comparison at SPIRE 250 resolution %%%%%%%%%% 
 \begin{figure}
      \includegraphics[width=8.5cm,clip]{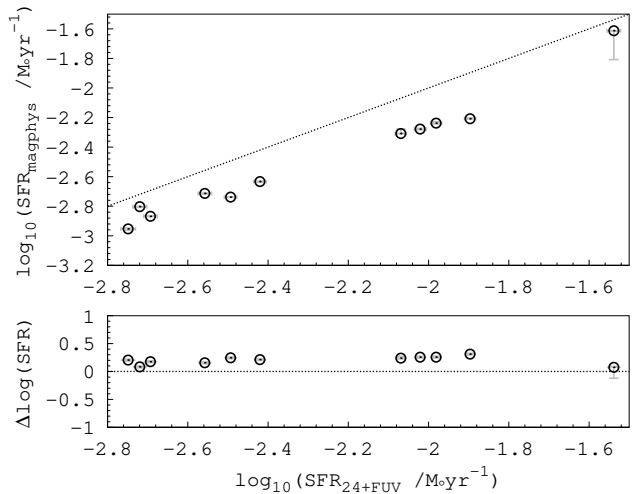}
  \caption{SFR (median likelihood) derived with {\it magphys} versus SFR$_{24+FUV}$. The lower panel shows their difference in the sense SFR$_{24+FUV}-$SFR$_{magphys}$.}
  \label{sl_figure10}
 \end{figure}
%%%%%%%%%%%%%%%%%%%%%%%%%%%%%%%%%%%
%
   Fig.~\ref{sl_figure10} compares the SFR derived with {\it magphys} and the one derived from the combination of FUV and 24\,$\mu$m \citep{Leroy08}. We do not correct for effects of the old stellar population \citep{Ford13}, as the central region studied here is dominated by the starburst \citep{Aloisi01}. When we compute the SFR from the combination of FUV and 24\,$\mu$m, we correct the FUV for internal extinction by a factor of 2.63, corresponding to the average optical extinction found by \citet[][$\langle$A$_{V}\rangle$\,=\,0.40\,mag]{Pasquali11}. The pixel--by--pixel comparison between the two tracers in Fig.~\ref{sl_figure10} shows that the SFR derived with {\it magphys} is lower than the one derived with the combination of FUV and 24\,$\mu$m, with a mean difference of 0.2\,$\pm$\,0.1\,dex. \citet{Viaene14} also report a similar offset in their comparison of the same two SFRs, and attribute the difference to the different way these SFR tracers are derived. The SFR based on the combination of FUV and 24\,$\mu$m is calibrated for much more massive galaxies and on global scales, while the SFR derived with {\it magphys} here refers to local scales within a low-mass dwarf galaxy.

   The sSFR map is shown in Fig.~\ref{sl_figure8} with values ranging between 0.2$^{+0.1}_{-0.1}$\,$\times$10\,$^{-9}$\,yr$^{-1}$
and 3.8$^{+1.3}_{-3.9}$\,$\times$10\,$^{-9}$\,yr$^{-1}$. The sSFR is higher in the periphery of the cavity. The sSFR map shows that this physical property is more sensitive to the details of the local star formation in the periphery of the cavity. The sSFR variations is driven by the stellar mass variations, which are highest in the central region of the cavity and SSCs.
%______________________________________________________________

\section{Gas--to--dust mass ratio}
\label{sec:g2d}

   In order to derive the gas--to--dust (G/D) mass ratio on a pixel--by--pixel basis, we use the dust masses from Sec.~\ref{sec:pixels} (the pixel--by--pixel analysis) and the total gas mass, which we derive in what follows. There are three gas components present in NGC\,1569: the atomic hydrogen; the warm ionised hydrogen; and the molecular hydrogen. Literature values give an atomic hydrogen mass of 2.5$\times$10$^{8}$\,M$_{\odot}$ \citep{Hunter12}, a warm ionised gas of 1.3$\times$10$^{5}$\,M$_{\odot}$ \citep{Hunter10}, and a CO--traced molecular gas mass of 2$\times$10$^{6}$\,M$_{\odot}$ \citep{Greve96,Taylor99}.  

   Both the \mbox{H\,{\sc i}} and H\,$\alpha$ morphology of NGC\,1569 shows extended emission beyond the optical body of the galaxy, as well as cavities within. On the other hand, the CO morphology shows emission concentrated primarily in the west and to a lesser degree in the southeast to the cavity with no detectable contributions elsewhere \citep{Greve96}, and most of this emission is in the form of giant molecular clouds located in the west and northwest of the cavity \citep{Taylor99}. The interferometric observations of \citet{Taylor99} may have missed any diffuse CO emission present. Thus, the warm ionised and atomic hydrogen gas are expected to have equally low contributions within the cavity, while elsewhere the atomic hydrogen should dominate by mass. The CO--derived molecular hydrogen is expected to contribute significantly in the northwest and west of the cavity, while elsewhere the atomic hydrogen mass should dominate. 

%%%%% FIGURE 11 - HI map %%%%%%%%%% 
 \begin{figure}
%  \centering
      \includegraphics[width=8cm,clip]{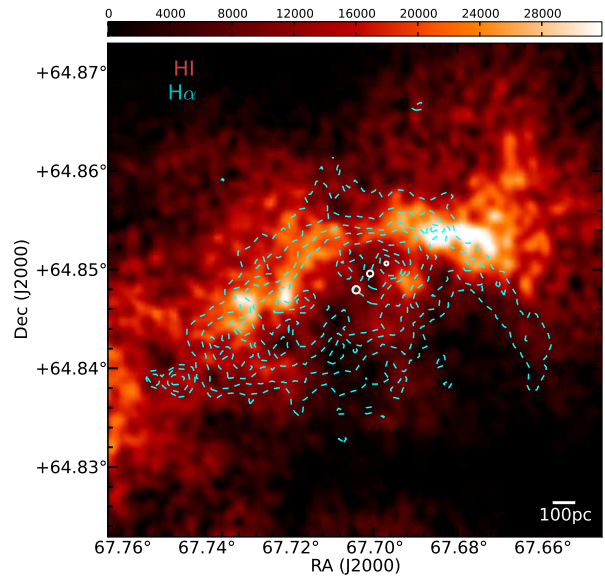}
  \caption{\mbox{H\,{\sc i}} map within 180$\arcsec \times$180$\arcsec$, with contours of H\,$\alpha$ overlaid in dashed cyan, as in Fig.~\ref{sl_figure4}. The white solid circles indicate the location of the SSCs and SC\,10. The units of the map are mJy\,/\,pixel\,$\times$\,km\,/\,sec.}
  \label{sl_figure11}
 \end{figure}
%%%%%%%%%%%%%%%%%%%%%%%%%%%%%%%%%%%
%
   To obtain the atomic hydrogen mass on a pixel--by--pixel basis, we use the Very Large Array (VLA) data set presented in LITTLE THINGS by \citet{Hunter12} and retrieved from NRAO\footnote{\url{https://science.nrao.edu/science/surveys/littlethings/data/n1569.html}}. We use their primary--beam--corrected moment 0 map, shown in Fig.~\ref{sl_figure11}, in conjunction with the equation: 
\begin{equation}
M_{\mbox{H\,{\sc i}}}\,({\rm M_{\odot}}) = 235.6\,D^{2} \sum (S\, \Delta V)
\end{equation}
where M$_{\mbox{H\,{\sc i}}}$ is the atomic hydrogen mass, $D$ is the distance of NGC\,1569 in Mpc, $S$ is the \mbox{H\,{\sc i}} flux density in mJy, integrated along the velocity axis, $\Delta V$ (2.6 km sec$^{-1}$).

   To obtain the warm ionised hydrogen mass on a pixel--by--pixel basis, we use the H\,$\alpha$ map (stellar continuum and background corrected) from \citet{Hunter06}, in conjunction with eq.\,10 in \citet[][see also \citealt{Goudfrooij94}]{Kulkarni14}: 
\begin{equation}
 M_{\rm H \alpha}\,({\rm M_{\odot}}) = 2.33 \times 10^{3} \left(\frac{L_{\rm H \alpha}}{10^{39}}\right) \left(\frac{10^{3}}{n_{e}}\right)
\end{equation}
where  M$_{\rm H \alpha}$ is the warm ionised gas mass, L$_{\rm H \alpha}$ is the H$\alpha$ luminosity (erg/sec), and n$_{e}$ the electron density, assumed equal to 10$^{3}$\,cm$^{-3}$. 

    The molecular hydrogen mass is usually traced through CO observations, assuming a conversion factor between them, which depends on the metallicity and is thus largely uncertain \citep{Wilson95,Remy14}. Additionally, \citet{Madden13} and \citet{Cormier14} discuss that CO is a poor tracer of molecular hydrogen gas in low--metallicity dwarf galaxies, while the \mbox{[C\,{\sc ii}]}\,158\,$\mu$m fine-structure line becomes a better tracer and provides better insights on the presence of the molecular gas component. NGC\,1569 is more luminous in \mbox{[C\,{\sc ii}]} than CO, with luminosity ratio L$_{\mbox{[C\,{\sc ii}]}}$\,/\,L$_{CO}$ of 3.7$\times$10$^{4}$ \citep{Hunter01}, and further modelling of the line strengths of forbidden lines suggests a molecular gas mass of 3.2$\times$10$^{7}$\,M$_{\odot}$ \citep[][their Table 6]{Hunter01}, i.e. 13 per cent of the atomic hydrogen mass. 

   Here, we use the CO observations from \citet{Taylor99} to include the molecular gas mass in our analysis, as follows. The contribution from the CO-traced molecular gas is in the 5 giant molecular clouds, each of which with masses listed in Table 3 of \citet{Taylor99}. The giant molecular clouds designated with numbers 1,2, \& 3 in \citet{Taylor99} are located in one pixel, west of SSC\,A (see Fig.~\ref{sl_figure8}), and have a total molecular gas mass of (multiplied by a factor of 1.8 to correct for the distance adopted here) 26.1($\pm$4)$\times$10$^{5}$\,M$_{\odot}$. We add this mass to the total gas mass in the pixel west of SSC\,A. The giant molecular clouds designated with numbers 4 \& 5 in \citet{Taylor99} are located in two pixels, north and northwest of SSC\,A (see Fig.~\ref{sl_figure8}), and have a total molecular gas mass of 3.9($\pm$0.7)$\times$10$^{5}$\,M$_{\odot}$ (corrected for our adopted distance). We divide this amount of mass to the total gas mass in the two pixels, north and northwest to the SSC\,A. 

   The conversion factor between CO and H$_{2}$ used in \citet[][their Table 3]{Taylor99} is modified from the Galactic value using the virial theorem, with an average value of 6.6$\pm$1.5 times that of the galactic value. As discussed in \citet{Schruba12}, using the virial theorem to derive the conversion factor may lead to underestimated values for the molecular gas mass. These authors assume a constant star formation efficiency to derive a functional form for the metallicity dependence of the conversion factor for a sample of dwarf galaxies. Using the relation between the metallicity and the conversion factor derived for their whole HERACLES sample \citep[see Table 7 in][]{Schruba12}, and assuming a gas phase metallicity of 8.02$\pm$0.02 \citep{Madden13}, the conversion factor is 183$^{+137}_{-77}$\,M$_{\odot}$\,pc$^{-2}($K\,km\,s$^{-1})^{-1}$, and the ratio of this with the Galactic conversion factor is 42$^{+31}_{-18}$. Comparing this ratio with the average ratio 6.6$\pm$1.5 of \citet[][]{Taylor99} indicates that the total molecular gas mass in the giant molecular clouds, which we use in this study, may have been underestimated by a factor of $\sim$6, leading to an even higher total gas mass, as well as a higher G\,/\,D mass ratio in those pixels that contain the giant molecular clouds considered here. 

   The total gas mass is the sum of all gas components. We also consider the helium gas and the gaseous metals, thus the total gas mass is expressed as \citep[see eq.\,3 in][]{Remy14}: 
\begin{equation}
M_{gas} = 1.38 \times (M_{\mbox{H\,{\sc i}}} + M_{\rm H_{2}} + M_{\rm H \alpha})
\end{equation}
where M$_{\mbox{H\,{\sc i}}}$ is the atomic hydrogen mass, M$_{\rm H_{2}}$ is the molecular hydrogen mass, and M$_{\rm H\, \alpha}$ is the warm ionised hydrogen mass. The 1.38 factor includes the contribution from the helium gas and the gaseous metals. Before summing up the individual gas components, we process the maps in the same way as described in Sec.~\ref{sec:observations}, i.e. convolving these to the SPIRE\,250\,$\mu$m PSF and re-sampling to a 21$\arcsec$ pixel size, in accord with the dust mass maps shown in Fig.~\ref{sl_figure8}.

%%%%% FIGURE 12 - Total gas mass map and G/D mass ratio at SPIRE 250 resolution %%%%%%%%%% 
 \begin{figure}
%  \centering
      \includegraphics[width=8cm,clip]{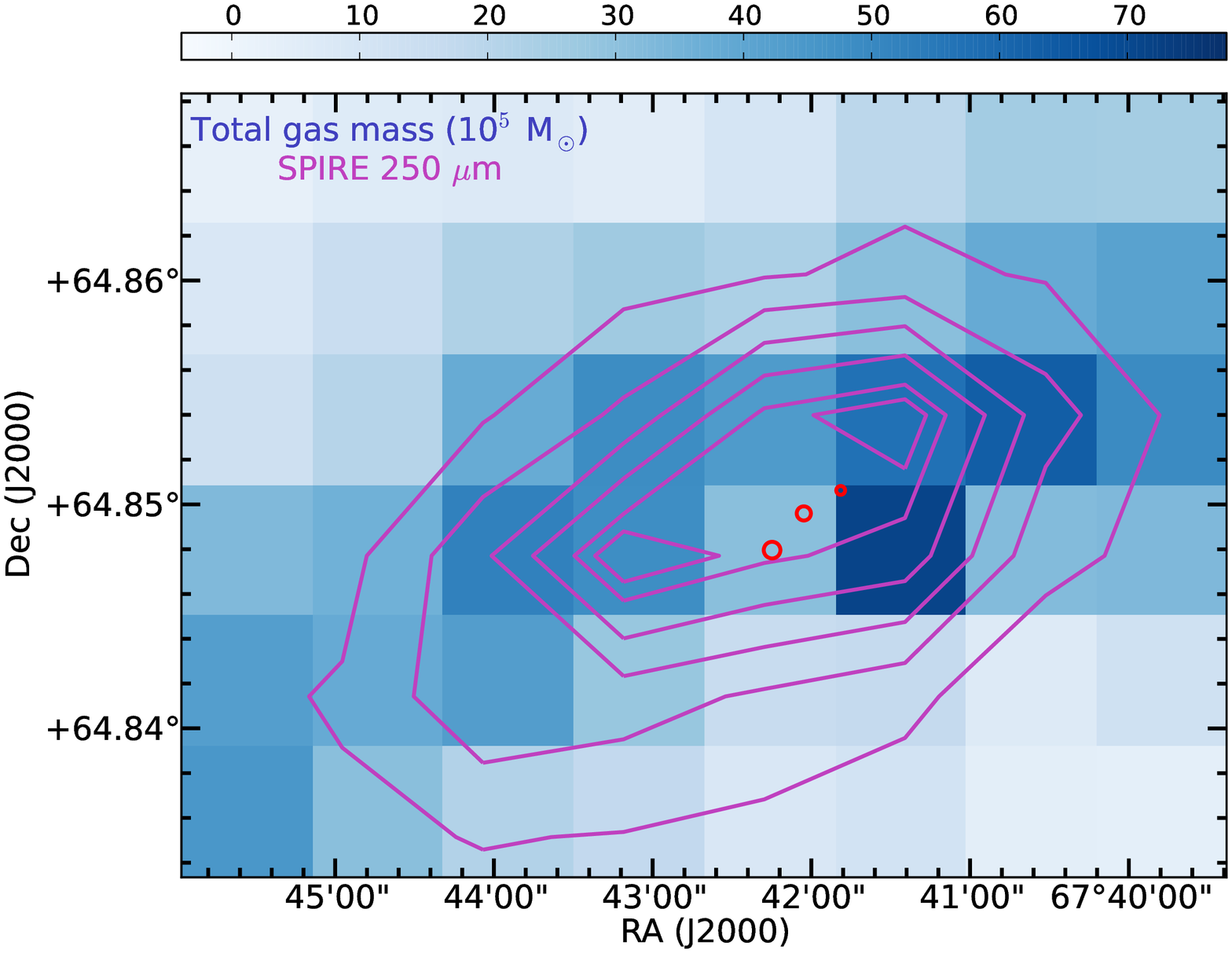}
      \includegraphics[width=8cm,clip]{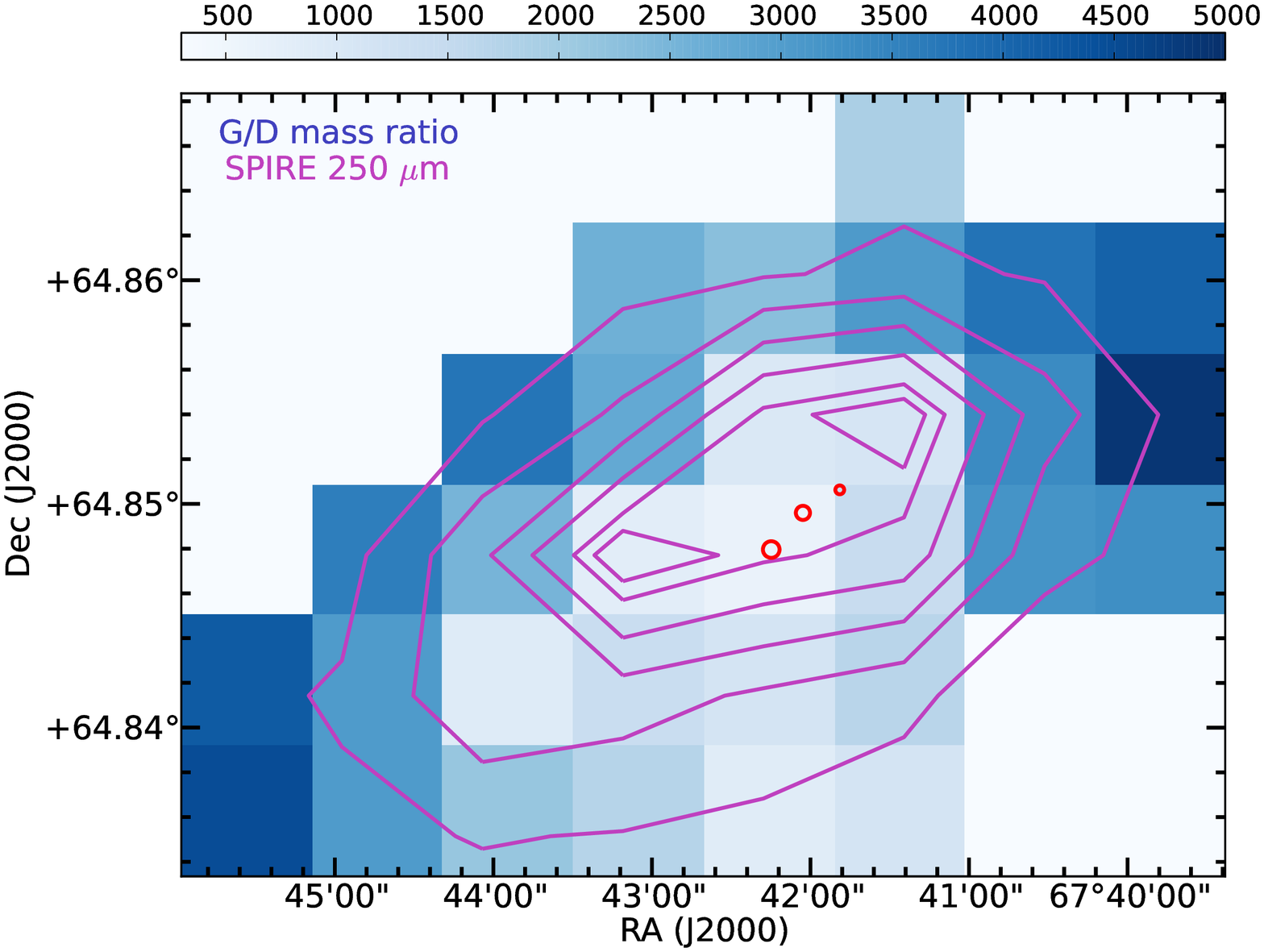}
      \includegraphics[width=8cm,clip]{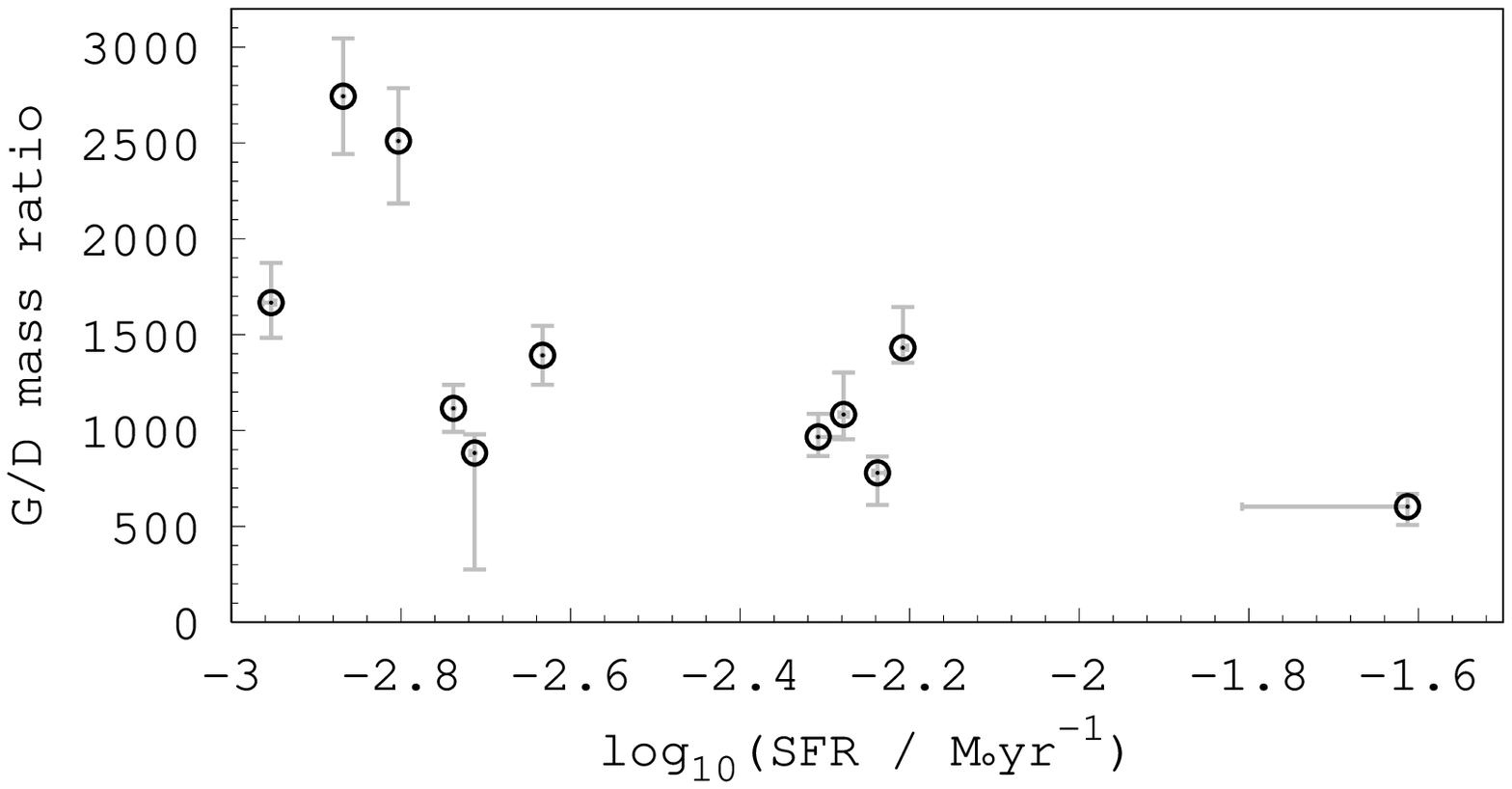}
  \caption{Total gas mass map (upper panel), G/D mass ratio map (middle panel), and G/D mass ratio versus SFR based on the {\it magphys} modelling (lower panel).}
  \label{sl_figure12}
 \end{figure}
%%%%%%%%%%%%%%%%%%%%%%%%%%%%%%%%%%%
%
   The resulting total gas mass map for the same central region as the dust mass map (shown in Fig.~\ref{sl_figure8}) is shown in the upper panel of Fig.~\ref{sl_figure12}. We obtain ranges for the total gas mass between 0.27($\pm$0.03)\,$\times$\,10$^{6}$\,M$_{\odot}$ to 7.10($\pm$0.39)\,$\times$\,10$^{6}$\,M$_{\odot}$. The total gas mass increases towards the periphery of the cavity, with the maximum mass observed west of SSC\,A. The cavity and SSCs region has a total gas mass of 3.1($\pm$0.3)$\times$10$^{6}$\,M$_{\odot}$, primarily driven by the atomic gas component. The size of the pixel (21$\arcsec$) is large enough to include atomic hydrogen emission from the vicinity of the cavity (see Fig.~\ref{sl_figure11}).

   We use the dust mass on a pixel--by--pixel basis derived in Sec.~\ref{sec:pixels} to get the G/D mass ratio variation in the central starburst region of NGC\,1569, shown in the middle panel of Fig.~\ref{sl_figure12}. We obtain G/D mass ratio ranging between 0.60$^{+0.07}_{-0.10}$\,$\times$\,10$^{3}$ to 4.88$^{+0.54}_{-0.54}$\,$\times$\,10$^{3}$. The G/D mass ratio becomes larger beyond the periphery of the cavity, where the total gas mass dominates over the dust mass, showing the gas--rich nature of the galaxy as seen in the \mbox{H\,{\sc i}} map in Fig.~\ref{sl_figure11}. The variation of the G/D mass ratio in the very central starburst is driven by the variation of the dust mass there, which is larger than the outer parts (see also the dust mass map in Fig.~\ref{sl_figure8}). The  G/D mass ratio has the smallest value in the cavity and SSCs region, primarily driven by the total gas mass there. At the location of the cavity, the total gas decreases due to the decrease of the atomic hydrogen gas, as seen in the \mbox{H\,{\sc i}} map in Fig.~\ref{sl_figure11} and in the total gas map in Fig.~\ref{sl_figure12} (upper panel). 

   \citet{Kobulnicky97} report that NGC\,1569 doesn't show any significant oxygen abundance gradient, which they limit to less than 0.05\,dex per kpc. The G/D mass ratio shows a dependence on metallicity \citep[][and references therein]{Remy14,Sandstrom13}, and, due to the flat abundance gradient in NGC\,1569, the G/D mass ratio would be expected to remain constant. The preceding paragraph discusses how the variation of the total gas and of the dust mass drives the variation of the G/D mass ratio in NGC\,1569 across the central starburst. Moreover, an additional component of quiescent cold dust clumps, which is not included in the modelling here, could drive the dust masses in the outer parts to larger values, and thus reconcile the G/D mass ratio to a constant value across the dwarf galaxy. Assuming an average total gas mass per pixel of $\sim$3$\times$10$^{6}$\,M$_{\odot}$ and an average G/D mass ratio per pixel of $\sim$1000 yields an average dust mass of $\sim$3$\times$10$^{3}$\,M$_{\odot}$ per pixel. Thus, about 2$\times$10$^{3}$\,M$_{\odot}$ of dust mass larger than what inferred with the dust modelling, per pixel. Assuming an average size of 300\,pc per pixel, this translates to $\sim$7\,M$_{\odot}$ per pc of dust mass needed in quiescent cold dust clumps. For comparison, \citet[their Table 5]{Rathborne10} derive dust masses of $\sim$50\,M$_{\odot}$ per pc for quiescent clump cores in Galactic infrared-dark clouds, with the latter harbouring the earliest phases of star cluster formation \citep{Rathborne06}. Whether such quiescent cold dust clumps exist in NGC\,1569 remains to be uncovered.

   The integrated G/D mass ratio in the central starburst region results to 75.5($^{+1.8}_{-1.9})$\,$\times$\,10$^{3}$. Such large G/D mass ratios are also observed in other low--metallicity dwarf galaxies \citep{Remy14}. The lower panel of Fig.~\ref{sl_figure12} shows the G/D mass ratio as a function of the SFR, per pixel, derived from the {\it magphys} modelling in Sec.~\ref{sec:pixels}. There is a trend of regions with higher SFRs to have smaller G/D mass ratio, consistent with what is expected in terms of gas fuelling star formation. The highest SFR is observed in the cavity region, associated with the SSCs, and there is where the lowest G/D mass ratio is also observed. In the periphery of the cavity there is embedded and ongoing star formation, and there the G/D mass ratio is lower compared to the outer parts. In the outer parts, the SFR is low, while the G/D mass ratio is large. 
%______________________________________________________________

\section{Cavity and SSCs}
\label{sec:regions}
   
   The winds from the SSCs have impacted the ISM around them by creating the cavity seen in warm ionised and atomic hydrogen observations. While the angular resolution of our images does not allow us to zoom into the properties of the SSCs, the set of convolved images at the SPIRE\,250\,$\mu$m PSF allows us to study the cavity region. It is thus interesting to compare the star and dust properties of the cavity versus the central starburst, as a way to understand the impact of the SSCs on the surrounding ISM. For this analysis, we model the SED anew for the cavity and the central starburst region. 

%%%%% FIGURE 13 - Apertures for individual regions + Cavity versus whole, at S250 and 21'' pixelsize %%%%%%%%%% 
 \begin{figure*}
%  \centering
      \includegraphics[width=7.3cm,clip]{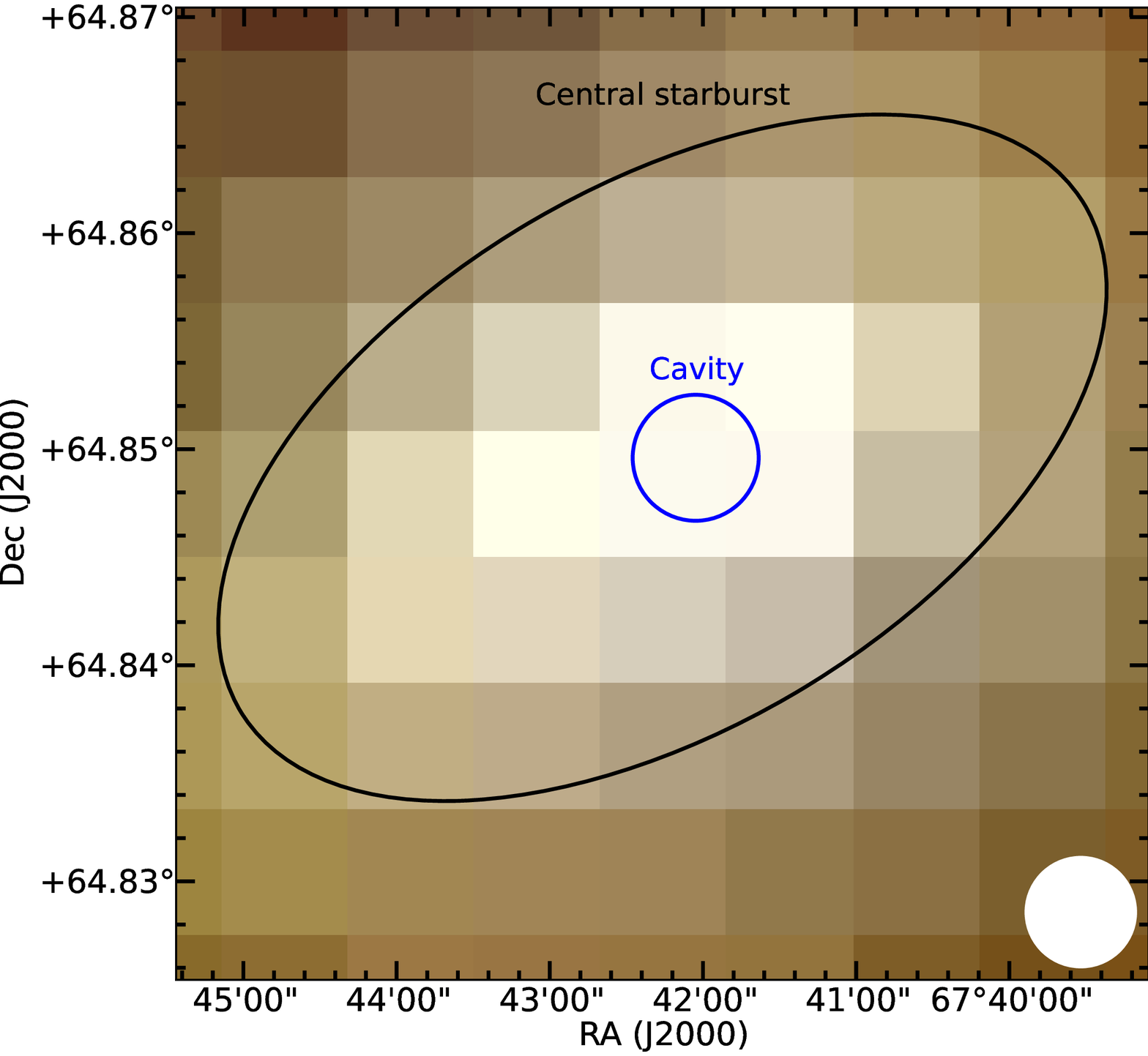}
      \includegraphics[width=8.7cm]{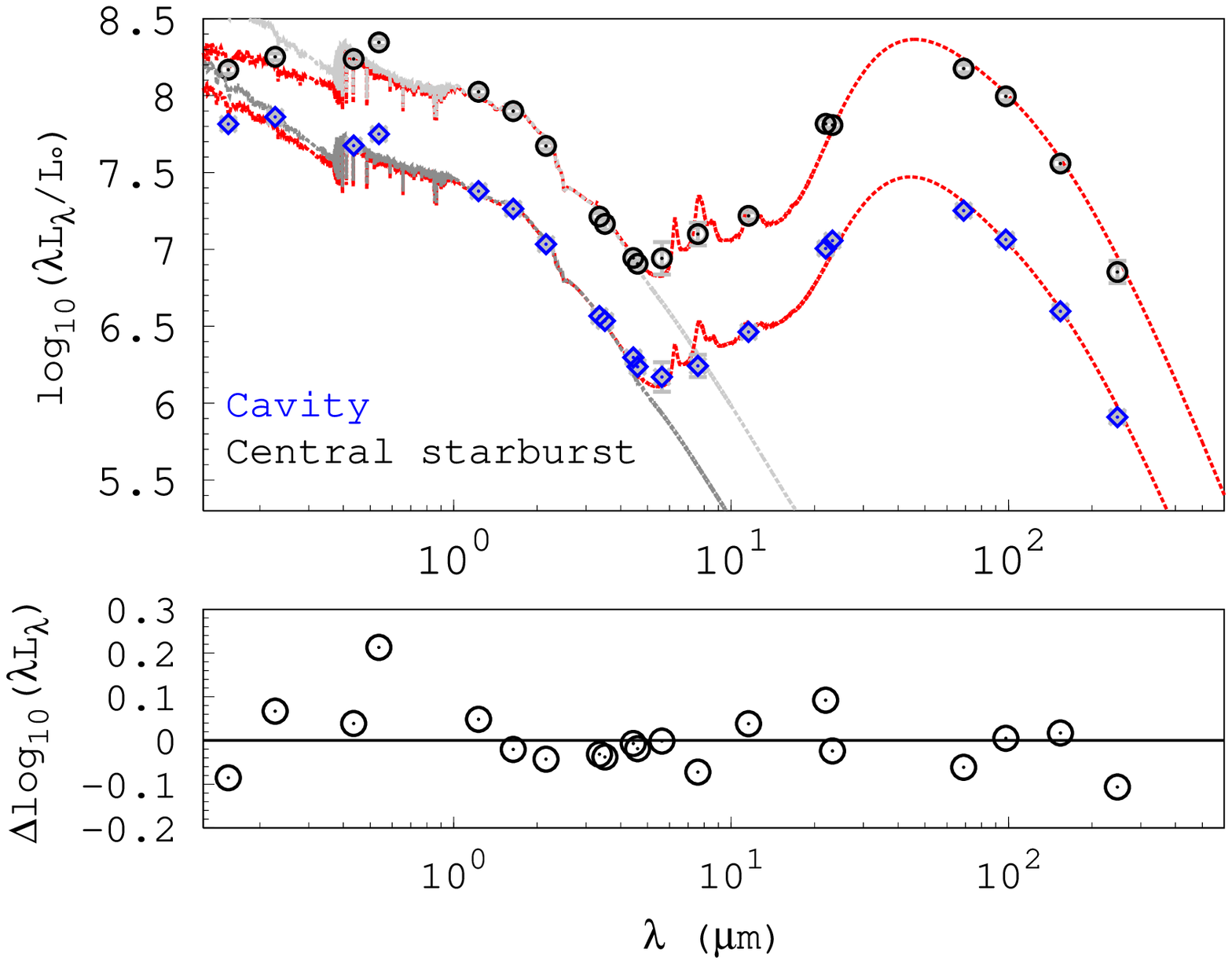}
  \caption{{\it Left panel:} The apertures used for the new SED modelling of the cavity (blue circle) and the central starburst (black ellipse) are overlaid on a three colour image of the galaxy composed of SPIRE\,250\,$\mu$m in red, PACS\,100\,$\mu$m in green, and W3 12\,$\mu$m in blue. The field-of-view covered is 180$\arcsec$ on a side, with a pixel size 21\,$\arcsec$. The images are convolved to the SPIRE\,250\,$\mu$m PSF, which is shown as the white circle on the right bottom.
 {\it Right panel:} Modelled (best--fit) and observed SED for the cavity and the central starburst at the SPIRE\,250\,$\mu$m resolution. The red (darker) curves show the attenuated modelled spectrum, while the grey (lighter) curves show the unattenuated spectrum. Black circles correspond to the central starburst observed spectral luminosity, and blue diamonds to the cavity region one. The lower sub-panel shows with black circles the difference of the observed and modelled SED for the central starburst region.}
  \label{sl_figure13}
 \end{figure*}
%%%%%%%%%%%%%%%%%%%%%%%%%%%%%%%%%%%
%
   For the central starburst, we construct its SED summing the pixels within an ellipse of semi--major axis 82$\arcsec$, major--to--minor axis ratio of 0.55, and position angle of 31$^{\circ}$, centred on the galaxy. The choice of this ellipse area is to maximise the area that contains pixels with S\,/\,N higher than the imposed cutoff. The semi--major axis used corresponds to 76 per cent the D$_{25}$ radius (216$\arcsec$). The ellipse encloses an area of 1.1$\times$10$^{4}$\,sq.~arcsec, which is 31 per cent of the total galaxy area. The apertures used for the cavity and the central starburst are shown overlaid in the three--colour image of Fig.~\ref{sl_figure13}. The right panel in Fig.~\ref{sl_figure13} shows the observed and (best--fit) modelled SEDs for the cavity and the central starburst. 

%
%%%%% TABLE 4 %%%%%%%%%%%%%%%%%%%%%
\begin{table}
     \begin{minipage}[t]{\columnwidth}
      \caption{SED modelling results for the cavity and central starburst regions, using the set of images convolved to the SPIRE\,250\,$\mu$m PSF and with a 21$\arcsec$ pixel size. }
      \label{sl_table4} 
      \centering
      \renewcommand{\footnoterule}{}
      \begin{tabular}{l l l}
\hline\hline
Property                              &Cavity                    &Central starburst     \\
\hline
RA (J2\,000.0)                        &4~30~48.19                &4~30~49.0             \\              
Dec (J2\,000.0)                       &64~50~58.6                &64~50~53              \\              
L$_{D}$ (10$^{8}$\,L$_{\odot}$)           &0.40$^{+0.02}_{-0.02}$       &3.10$^{+0.02}_{-0.04}$    \\          
M$_{D}$ (10$^{4}$\,M$_{\odot}$)           &0.5$^{+0.1}_{-0.1}$          &4.3$^{+1.2}_{-0.6}$      \\            
T$_{C}^{ISM}$ (K)                       &22.6$^{+1.1}_{-1.1}$         &24.0$^{+0.2}_{-0.4}$     \\             
T$_{W}^{ISM}$ (K)                       &59.2$^{+0.6}_{-1.2}$         &59.0$^{+0.8}_{-1.5}$     \\            
SFR (10$^{-2}$\,M$_{\odot}$\,yr$^{-1}$)     &2.52$^{+0.01}_{-0.01}$       &5.78$^{+0.04}_{-0.07}$    \\            
sSFR (10$^{-9}$\,yr$^{-1}$)              &0.9$^{+0.5}_{-0.5}$          &2.4$^{+0.3}_{-0.3}$       \\           
\hline
\end{tabular} 
\footnotetext{Note.-- Units of right ascension are hours, minutes, and seconds, and units of declination are degrees, arcminutes and arcseconds.}% 
\end{minipage}
\end{table}
%%%%%%%%%%%%%%%%%%%%%%%%%%%%%%%%%%%
%
   Table~\ref{sl_table4} summarises the results on the derived properties (median likelihood estimates) for the cavity and central starburst from this new SED modelling with {\it magphys}. These results show that the cavity region contains 13 per cent the L$_{D}$ and 12 per cent the M$_{D}$ of the central starburst region. The SFR and sSFR of the cavity are $\sim$40 per cent that of the central starburst region, signifying the massive star formation in the SSCs. The equilibrium temperatures are statistically equivalent between the cavity and the central starburst region.

   The values listed in Table~\ref{sl_table4} are the result of modelling anew the SEDs of these regions, and are not obtained by summing up the star and dust properties derived in the pixel--by--pixel analysis of Sec.~\ref{sec:pixels}. To compare between the (summed--up) properties derived in Sec.~\ref{sec:pixels} and the properties listed in Table~\ref{sl_table4}, we first normalise each by the area probed.
%%%%% TABLE 5 %%%%%%%%%%%%%%%%%%%%%
\begin{table}
     \begin{minipage}[t]{\columnwidth}
      \caption{Normalised by area properties obtained in this section and in Sec.~\ref{sec:pixels} for the central starburst region.}
      \label{sl_table5} 
      \centering
      \renewcommand{\footnoterule}{}
      \begin{tabular}{l c c}
\hline\hline
Properties                                    &This section            &Sec.~\ref{sec:pixels}  \\
\hline
L$_{D}$ (10$^{8}$\,L$_{\odot}$\,kpc$^{-1}$)         &1.38$^{+0.02}_{-0.04}$     &1.68$^{+0.35}_{-0.60}$    \\
M$_{D}$ (10$^{4}$\,M$_{\odot}$\,kpc$^{-1}$)         &1.9$^{+1.2}_{-0.6}$        &2.3$^{+1.2}_{-0.7}$       \\
SFR (10$^{-2}$\,M$_{\odot}$\,yr$^{-1}$\,kpc$^{-1}$)   &2.57$^{+0.04}_{-0.07}$     &2.87$^{+0.04}_{-0.07}$    \\
sSFR (10$^{-8}$\,yr$^{-1}$\,kpc$^{-1}$)            &0.1$^{+0.3}_{-0.3}$        &1.9$^{+0.3}_{-0.6}$       \\
\hline
\end{tabular} 
\footnotetext{Note.-- To obtain the normalised properties for Sec.~\ref{sec:pixels}, we sum up the properties derived in each individual pixel and then divide by the area covered by the number of pixels used in the sum.}% 
\end{minipage}
\end{table}
%%%%%%%%%%%%%%%%%%%%%%%%%%%%%%%%%%%
% 
The area--normalised properties are listed in Table~\ref{sl_table5}. This comparison shows that the normalised properties for the central starburst region are consistent with each other, with the exception of the sSFR. The sSFR is higher by an order of magnitude in the case of summing up the sSFR found in each pixel in Sec.~\ref{sec:pixels}. The area enclosed in the latter case is larger than the one we use in this section. The sSFR depends on the inverse of the stellar mass, by definition, and the stellar mass is not constant as a function of radius, thus the difference between the two sSFR. 

   As the central starburst region covers a smaller area than that of the whole galaxy, this implies that the fractional dust mass in the cavity versus the whole galaxy is even lower. \citet{Galliano11} find that there is a bias introduced on deriving dust masses according to the spatial resolution probed, i.e., summing up the dust mass on pixel sizes of various spatial resolutions within the LMC versus integrating over a larger area to derive its dust mass results in underestimating the dust mass in the latter case. They find an optimal spatial scale of $\sim$30--50\,pc for the LMC, for which the dust mass estimates become stable over the chosen spatial scale. We cannot probe such small spatial scales. Comparing the spatial scales we probe here for the cavity (and in Sec.~\ref{sec:pixels}; pixel size of 21$\arcsec$ or 294\,pc) with Fig.~6 in \citet[][lower panels]{Galliano11}, the dust mass estimate of the LMC at our spatial resolution is roughly in accord with the one at their optimal spatial resolution element, within the uncertainties. For the central starburst region in this section (spatial scale of 1.2\,kpc in diameter), the findings of \citet{Galliano11} indicate that the dust mass we derive may be underestimated as compared to deriving the dust mass in a finer resolution spatial scale, thus the dust mass we derive here for the central starburst may be considered as a lower limit. All these point to a stronger case of the dust mass within the cavity being a small fraction of that of the central starburst, and further implying that it is even lower than that. This is consistent with the finding of Sec.~\ref{sec:morphology} on the dust emission morphology.

   It is interesting to place our findings on the dust content in NGC\,1569's cavity in the wider context of galaxy evolution and Galactic GCs. Our analysis shows that the dust luminosity and dust mass in the cavity region is low compared to the galaxy's central starburst dust content. The low levels of dust emission in the cavity region compared to the central starburst is in accord with the disrupted nature of the cavity region initially seen in the warm ionised and atomic hydrogen phase. All phases of the ISM thus far studied (warm and hot ionised gas, atomic hydrogen gas, CO--traced molecular hydrogen gas, dust emission) show a deficit of emission in the cavity. The ISM structure in the cavity is a result of the massive starburst, which gave rise to SSC winds and created superbubble kinematics and galactic scale outflows further driving the ISM evolution \citep{Westmoquette08,Martin98}. 

   Even though the SSCs have affected the ISM in the central starburst region of the dwarf galaxy and may have caused the galactic--scale outflows, these have not affected the morphology of the atomic hydrogen gas on global scales or ceased star formation \citep[see also discussion in][]{Holwerda13}. On the contrary, dust knots and ongoing star formation activity is observed in the periphery of the cavity, while the SED modelling implies the need of quiescent cold dust clumps so as to reconcile the G/D mass ratio with an observed flat metallicity gradient \citep{Kobulnicky97,Devost97}. It would require either additional mechanisms external to the galaxy \citep[see, for example,][]{Lianou13b} to completely remove the ISM, or subsequent bursts of massive star formation distributed across the whole body of the galaxy and with intensity similar to the one produced by the SSCs, in order to clear out the ISM in their vicinity. \citet{Bagetakos11} study the distribution of \mbox{H\,{\sc i}} holes in a large sample of galaxies and find that those in dwarf galaxies extend beyond their R$_{25}$ radius. With a stellar mass of 2.8$\times$10$^{8}$~M$_{\sun}$, the feedback from the current massive star formation alone is not enough to affect the atomic hydrogen on global scales in this dwarf galaxy or to cease the star formation activity in the periphery of the cavity and beyond.

   The two SSCs reside within the cavity region and are at an earlier evolutionary phase as compared to their likely old counterparts, i.e. the old massive GCs \citep{deGrijs05}. The majority of the studied Galactic GCs are found to be devoid of dust, which is unexpected considering evolutionary effects on their stellar population \citep{Barmby09,Priestley11}. The evolution of GCs is affected by the geometry and mass of their host galaxy \citep[e.g.,][]{Madrid14}. NGC\,1569 offers a simpler environment than the one in the Galaxy, free from spiral structure, albeit with a complex and disturbed ISM. Assuming the SSCs as the likely progeny of the old GCs seen in our Galaxy, the former possess a low dust content already early in their life, before any complex processes due to the galactic geometry having affected their evolution. 
%______________________________________________________________

\section{Summary \& conclusions}
\label{sec:summary}

   We use a multiwavelength data set from UV to submm to construct the observed SEDs of the central region of NGC\,1569. We model the observed SEDs with {\it magphys} on a pixel--by--pixel basis, as well as the cavity and central starburst regions of the galaxy. We investigate whether the derived physical parameters depend on the choice of which bands to retain for the SED fitting. Results obtained without using the SPIRE 350\,$\mu$m and 500\,$\mu$m bands are consistent with results obtained using those bands. We perform our analysis using images at the SPIRE\,250\,$\mu$m resolution, in order to gain in spatial resolution. We show the variation of the derived properties on a pixel--by--pixel basis and focus on the cavity region versus the central starburst. We derive the total gas mass, considering the warm ionised, atomic and CO--traced molecular hydrogen gas, and investigate its variation as a function of the dust mass on a pixel--by--pixel basis and for the cavity and central starburst regions of the galaxy. 

   Our results are summarised as follows: 
\begin{itemize}

      \item The morphology of the central starburst region of the dwarf galaxy shows low levels of dust emission in the cavity region, as well as several dust knots marking the periphery of the cavity (Fig.~\ref{sl_figure1}). The brightest of the dust knots resolved in most of the {\it Herschel} bands is associated with the onset of an H\,$\alpha$ filament to the west and northwest of SSC\,A (Fig.~\ref{sl_figure1} and Fig.~\ref{sl_figure4}).
      \item The extended emission seen in the warm and hot ionised gas is also seen in the warm and cold dust. All these components of the extended emission spatially correlate, while the emission probed with {\it Herschel} is more extended (Fig.~\ref{sl_figure4} and Fig.~\ref{sl_figure5}).
      \item The dust content within the cavity region is low (12 per cent) compared to the central starburst region (Table~\ref{sl_table4} and Fig.~\ref{sl_figure2}). The spatial variation of the dust mass shows enhanced dust masses in the periphery of the cavity (Fig.~\ref{sl_figure8}). All tracers of the ISM thus far studied show the same picture of the cavity having low emission as compared to its vicinity or central starburst (Fig.~\ref{sl_figure8}, Fig.~\ref{sl_figure11}, and Fig.~\ref{sl_figure12}).  
      \item The G/D mass ratio is higher in the periphery of the cavity compared to that of the cavity itself (Fig.~\ref{sl_figure12}). 
\end{itemize}

   The evolution of NGC\,1569 has been affected by an interaction event \citep{Stil98,Jackson11,Johnson13}, which has triggered the massive starburst \citep{Angeretti05,Anders04,Hunter00}. The massive starburst has affected the surrounding ISM through stellar cluster winds and the ISM there is faint in dust, atomic hydrogen, and warm ionised hydrogen emission. The ISM further presents galactic scale outflows and extended emission beyond the optical body \citep{Westmoquette08,Martin98,Heckman95}. External (i.e., galaxy interactions) and internal (i.e., SSC feedback) mechanisms are linked into driving the dwarf galaxy's evolution. Within this context, dwarf galaxies displaying a starburst nature are now showing evidence of gravitational interactions with a lower-mass satellite companion causing the starburst and peculiar kinematics \citep[e.g.,][]{Martinez12}.
%______________________________________________________________

\section*{Acknowledgements}

The authors would like to thank the referee, Benne W. Holwerda, for the thoughtful comments and suggestions to improve the manuscript. We are grateful to Emmanuel Xilouris for useful suggestions on an earlier draft, to Christine Wilson and Jeroen Stil for useful discussions, to Chris Taylor for making available the CO map, and to Elisabete da Cunha on {\it magphys} specifics. Support for the work of SL and PB is provided by an NSERC Discovery Grant and by the Academic Development Fund of the University of Western Ontario. SL is grateful for an EAS\,/\,Springer grant to attend EWASS 2014 in Geneva, where results of this work have been shown. 

    This research made use of Astropy \citep[http://www.astropy.org]{Astropy13}, a community-developed core Python package for Astronomy; APLpy (http://aplpy.github.com), an open-source plotting package for Python; IRAF, distributed by the National Optical Astronomy Observatory, which is operated by the Association of Universities for Research in Astronomy (AURA) under cooperative agreement with the National Science Foundation; STSDAS and PyRAF, products of the Space Telescope Science Institute, which is operated by AURA for NASA; NASA/IPAC Extragalactic Database (NED), operated by the Jet Propulsion Laboratory, California Institute of Technology, under contract with the National Aeronautics and Space Administration; NASA/IPAC Infrared Science Archive (IRSA), operated by the Jet Propulsion Laboratory, California Institute of Technology, under contract with the National Aeronautics and Space Administration; Aladin; NASA's Astrophysics Data System Bibliographic Services; SAOImage DS9, developed by Smithsonian Astrophysical Observatory. This work made extensive use of the free software GNU Octave and the authors are grateful to the Octave development community for their support. 

    PACS has been developed by MPE (Germany); UVIE (Austria); KU Leuven, CSL, IMEC (Belgium); CEA, LAM (France); MPIA (Germany); INAFIFSI/OAA/OAP/OAT, LENS, SISSA (Italy); IAC (Spain). This development has been supported by BMVIT (Austria), ESA-PRODEX (Belgium), CEA/CNES (France), DLR (Germany), ASI/INAF (Italy), and CICYT/MCYT (Spain). SPIRE has been developed by Cardiff University (UK); Univ. Lethbridge (Canada); NAOC (China); CEA, LAM (France); IFSI, Univ. Padua (Italy); IAC (Spain); SNSB (Sweden); Imperial College London, RAL, UCL-MSSL, UKATC, Univ. Sussex (UK) and Caltech, JPL, NHSC, Univ. Colorado (USA). This development has been supported by CSA (Canada); NAOC (China); CEA, CNES, CNRS (France); ASI (Italy); MCINN (Spain); Stockholm Observatory (Sweden); STFC (UK); and NASA (USA). 

    SPIRE has been developed by a consortium of institutes led by Cardiff Univ. (UK) and including: Univ. Lethbridge (Canada); NAOC (China); CEA, LAM (France); IFSI, Univ. Padua (Italy); IAC (Spain); Stockholm Observatory (Sweden); Imperial College London, RAL, UCL-MSSL, UKATC, Univ. Sussex (UK); and Caltech, JPL, NHSC, Univ. Colorado (USA). This development has been supported by national funding agencies: CSA (Canada); NAOC (China); CEA, CNES, CNRS (France); ASI (Italy); MCINN (Spain); SNSB (Sweden); STFC, UKSA (UK); and NASA (USA).

\bibliographystyle{bibtex/mn2e}
\bibliography{bibtex/biblio}{}

\bsp

\label{lastpage}

\end{document}